\documentclass[12pt, a4paper]{article}
\pdfoutput=1
\usepackage{jheppub}
\usepackage{amsmath, amssymb}

\usepackage{booktabs}
\usepackage{float}
\usepackage{color}

\newcommand{\nn}{\nonumber}

\newcommand{\beq}{\begin{equation}}
\newcommand{\eeq}{\end{equation}}
\newcommand{\beqa}{\begin{eqnarray}}
\newcommand{\eeqa}{\end{eqnarray}}
\newcommand{\beqar}{\begin{eqnarray*}}
\newcommand{\eeqar}{\end{eqnarray*}}

\newcommand{\bea}{\begin{eqnarray}}
\newcommand{\eea}{\end{eqnarray}}

\newcommand{\ie}{{\it i.e.,}\ }
\newcommand{\reef}[1]{(\ref{#1})}

\newcommand\cS{{\cal S}}

\newcommand\cL{{\cal L}}
\newcommand\cG{{\cal G}}
\newcommand\cI{{\cal I}}

\newcommand\cO{{\cal O}}

\title{Black hole chemistry and holography in generalized quasi-topological gravity}

\author[a,b]{Mozhgan Mir,}
\author[c]{Robie A. Hennigar,}
\author[d]{Jamil Ahmed,}
\author[a,e]{and Robert B. Mann}

\affiliation[a]{Department of Physics and Astronomy, University of Waterloo,
Waterloo, Ontario, Canada, N2L 3G1}
\affiliation[b]{Department of Physics, Faculty of Science, Ferdowsi University of Mashhad\\
P.O. Box 1436, Mashhad, Iran}
\affiliation[c]{Department of Mathematics and Statistics, Memorial University of Newfoundland, St. John's, Newfoundland and Labrador, A1C 5S7, Canada}
\affiliation[d]{Department of Mathematics, Quaid-i-Azam University, Islamabad, Pakistan}
\affiliation[e]{Perimeter Institute, 31 Caroline Street North, Waterloo,
ON, N2L 2Y5, Canada}

\emailAdd{mozhganmir@um.ac.ir}
\emailAdd{rhennigar@mun.ca}
\emailAdd{jahmed@student.qau.edu.pk}
\emailAdd{rbmann@uwaterloo.ca}

\abstract{
We investigate the thermodynamics of AdS black holes in Generalized Quasitopological Gravity with and without electric charge, concentrating on the version of the theory that is cubic in curvature. We study new aspects of Hawking-Page transitions that occur for these black holes. Working within the framework of black hole chemistry, we find a variety of familiar and new critical behaviour and phase transitions in four and higher dimensions for the charged black holes.  We also consider some holographic aspects of our work, demonstrating how the ratio of viscosity to entropy is modified by inclusion of these cubic curvature terms.
}

\keywords{Higher Curvature  Gravity, Black Holes, Thermodynamics, AdS/CFT}

\begin{document}
\maketitle

\section{Introduction } \label{intro}

Higher derivative gravity theories play an important role in black hole physics, cosmology, holography, supergravity and string theory.  Efforts to construct a UV complete theory of quantum gravity generically lead to theories that contain a series of terms in the action that are higher-order in curvature in addition to usual Einstein-Hilbert term. For example, this is the case in string theory where an infinite series of terms can be present~\cite{Gross:1986mw}. Higher derivative theories provide a framework for testing which features of gravitational theory are special, and therefore their studies provide a better understanding of Einstein gravity and what type of modifications one can expect due to quantum corrections.

Work in this direction has a long history, dating back to early days of general relativity. Originally, Weyl and Eddington proposed such theories for geometric unification of gravity and electromagnetism \cite{weyl1923allgemeine, carmichael1925eddington}. Somewhat later  the search for higher curvature  theories correcting the Einstein-Hilbert action became motivated by  attempts to construct a quantized theory of gravity. For example, the addition of higher derivative terms to Einstein-Hilbert action can yield a power-counting renormalizable theory~\cite{Stelle:1977ry}.
Further work indicated that  in the low energy effective action of string theory a Gauss-Bonnet term appears \cite{Zwiebach:1985uq}.   Higher curvature gravities have been particularly useful in the context of the AdS/CFT correspondence \cite{Maldacena:1998}, where these terms generically arise when studying the dual theory beyond large $N$, but also have been successfully employed as holographic toy models. The presence of additional couplings in the action allow one to make contact with a larger class of CFTs than those defined by Einstein gravity~\cite{Camanho:2009vw,deBoer:2009pn,Buchel:2008vz,Hofman:2008ar,Hofman:2009ug,Nojiri:1999mh,Blau:1999vz,Buchel:2009sk,Myers:2010jv,mir:1307,Bueno:2018xqc}, which has been used with success to identify universal properties of CFTs, e.g.~\cite{Myers:2010tj,Myers:2010xs,Mezei:2014zla,Bueno1,Bueno2, Cano:2018aqi}. In the context of cosmology, higher curvature gravities have been extensively considered to explain the late-time expansion of the universe, dark matter and inflation \cite{sotiriou2010f, clifton2012modified}.


The most general  higher-curvature theory  yielding second order equations of motion in arbitrary dimensions is known as Lovelock gravity. It is perhaps the most natural generalization of Einstein gravity in higher dimensions~\cite{Lovelock:1971yv}.  Indeed, Einstein gravity can be understood as a special case of Lovelock gravity in dimensions greater than four, with the Einstein-Hilbert term being one of several terms that constitute Lovelock theory in a given dimension. These theories are ghost-free~\cite{Zwiebach:1985uq} and so are candidates for generalizations of Einstein gravity in higher dimensions. However, Lovelock gravity that is $k$th order in  curvature is only non-trivial for spacetime dimensions $d > 2k+1$. Thus, one must go beyond Lovelock gravity to obtain theories that have interesting implications for lower dimensional physics.

Under certain symmetry restrictions, many of the nice properties of Lovelock gravity can be extended to obtain a broader class of \textit{quasi-topological gravity} theories~\cite{Oliva:2010eb, Myers:2010ru}. Quasi-topological theories possess a number of interesting properties.  First, in the context of spherically symmetric metrics, their field equations are second-order (though they would be fourth-order on a generic background). Second, in contrast to Lovelock theory, quasi-topological theories of cubic or higher-order in curvature appear to exist and are non-trivial for any dimension $d \ge 5$.\footnote{Though we note that, at present, explicit examples of five-dimensional quasi-topological theories are known only up to quintic order in curvature~\cite{Dehghani:2013ldu, Cisterna:2017umf}.} Third, the linearized equations of motion of quasi-topological gravity coincide (up to an overall prefactor) with those of Einstein gravity on maximally symmetric spacetime backgrounds
\cite{Myers:2010tj}.  This property ensures  that the theory does not propagate negative energy to asymptotic regions of constant curvature.

These applications have motivated more recent efforts to construct new theories of higher-curvature gravity that are both free of ghosts and interesting both holographically and phenomenologically.  A success in this direction was the construction of \textit{Einsteinian Cubic Gravity} (ECG)~\cite{Bueno:2016xff, Bueno:2016ypa}. ECG was constructed as the unique cubic theory of gravity whose Lagrangian is of the same form in all dimensions and propagates only the usual massless and transverse graviton on maximally symmetric backgrounds. Unlike Lovelock and quasi-topological gravities, ECG is neither trivial nor topological in four dimensions, and admits four-dimensional black hole solutions that possess a number of remarkable properties~\cite{Hennigar:2016gkm, Bueno:2016lrh}: (i) there is a single independent field equation (in the most general case there would be two) that admits an integrating factor, reducing it to a second-order differential equation determining the metric function $f(r)$. (ii) The black hole solutions are ``non-hairy'' in the sense that they are characterized by mass alone. (iii) Despite the lack of an analytic solution to the equations of motion, the thermodynamic properties of black holes can be studied exactly. When evaluated at the horizon, the field equations reduce to two polynomial equations that determine the temperature and mass in terms of the horizon radius. 

It has been realized that it is possible to construct more general theories of gravity in four and higher dimensions that incorporate many of the interesting properties observed for ECG~\cite{Hennigar:2017ego, Bueno:2017sui, Ahmed:2017jod}. These theories, named \textit{generalized quasi-topological gravities}, propagate only the usual massless tranverse graviton in vacuum, admit non-hairy black hole solutions characterized by a single metric function, and allow for non-perturbative studies of black hole thermodynamics.\footnote{It appears that all of these features follow from the requirement that black holes are characterized by single independent field equation, as argued in~\cite{Hennigar:2017ego, Bueno:2017sui}.} The relative simplicity of this class of theories make them ideal for phenomenological purposes (in four dimensions) and
as toy models (in all dimensions).    It was shown that black branes in these theories possess a rich phase structure, contrary to what happens in Lovelock and quasi-topological theories~\cite{Hennigar:2017umz}.  An initial study of holographic aspects of ECG was carried out in~\cite{Bueno:2018xqc}, determining a number of entries in the holographic dictionary for the theory and revealing, for example, that the Kovtun-Son-Starinets bound on the ratio of entropy density to shear viscosity always holds~\cite{Bueno:2018xqc}. In~\cite{Bueno:2018uoy} it was shown that the properties of black hole solutions in these theories extend also to Taub-NUT/Bolt solutions, providing the first examples of explicit solutions of this kind beyond Lovelock theory. Based on that work, a number of universal results were obtained for the free energy of odd-dimensional CFTs on squashed spheres~\cite{Bueno:2018yzo}. See, for example,~\cite{Li:2017ncu, Li:2017txk, Colleaux:2017ibe, Li:2018drw, Carballo-Rubio:2018bmu, Li:2019auk} for a number of other recent developments and applications of these  --- and closely related --- theories.

As we mentioned, there are also reasons for considering these theories in the context of phenomenology. Concerning four-dimensional physics, small asymptotically flat black hole solutions  become stable, a result with possible implications for dark matter and the information loss problem~\cite{Bueno:2017qce}. Recent work has revealed potentially interesting phenomenological signatures of   black holes~\cite{Hennigar:2018hza,Poshteh:2018wqy}. Furthermore, it has been realized that the equations of motion of a subclass of the generalized quasi-topological theories are second-order for FLRW cosmologies (indicating a well-posed initial value problem), with late-time dynamics indistinguishable from $\Lambda$CDM while giving rise to an inflationary epoch~\cite{Arciniega:2018fxj}. It was subsequently realized that this is a generic property of the four-dimensional class of theories~\cite{Cisterna:2018tgx, GeometricInflation}.

In this paper we carry out an extensive study of the thermodynamic properties of charged black holes in cubic generalized quasi-topological gravity. Much of our work is framed in the language of black hole chemistry, in which the cosmological constant is promoted to a thermodynamic variable \cite{Henneaux:1985tv, Creighton:1995au}  interpreted as pressure in the first law of black hole mechanics \cite{Kastor:2010gq, KastorEtal:2010}.  This more general perspective revealed a deep analogy
between charged anti-de Sitter black holes and  Van der Waals fluids \cite{Kubiznak:2012wp}.
A  remarkably rich thermodynamic phase behaviour for black holes has since been discovered, including the examples of  triple points \cite{Altamirano:2013uqa}, re-entrant phase transitions \cite{Altamirano:2013ane}, polymer-like behaviour \cite{Dolan:2014vba}, and even superfluid-like phase transitions \cite{Hennigar:2016xwd,EricksonRobie,Dykaar:2017mba}.
This framework has shown to be particularly fruitful in understanding black holes in higher curvature gravity~\cite{Wei:2012ui, Cai:2013qga, Xu:2013zea, Mo:2014qsa, Wei:2014hba, Mo:2014mba, Zou:2013owa, Belhaj:2014tga, Xu:2014kwa, Frassino:2014pha, Dolan:2014vba, Sherkatghanad:2014hda, Hendi:2015cka, Hendi:2015oqa, Hennigar:2015esa, Hendi:2015psa, Nie:2015zia, Hendi:2015pda, Hendi:2015soe, Zeng:2016aly, Hennigar:2016gkm, EricksonRobie, Hennigar:2016xwd, Cvetic:2010jb, Hennigar:2014cfa, Johnson:2014yja, Karch:2015rpa, Caceres:2015vsa, Dolan:2016jjc, Sinamuli:2017rhp, Li:2017wbi, Dehyadegari:2018pkb, Hendi:2018xuy}; we refer the reader to \cite{Kubiznak:2016qmn} for a detailed survey of this subject.  A study of the thermodynamic behaviour of black holes in the quartic theory
\cite{Ahmed:2017jod} is forthcoming \cite{MirMann2019}.

Our paper is organized as follows.  In Section \ref{sec:bhsolution} we present charged static, spherically symmetric AdS black holes in cubic generalized quasi-topological gravity. In Section \ref{sec:thermo} we collect the thermodynamic properties of the charged black holes. In Section~\ref{sec:HawkingPage} we study the uncharged solutions, discussing their thermodynamics and the Hawking-Page transition in four and five dimensions.  In Section~\ref{sec:thermofpe} we extend our considerations to include charge, working in the grand canonical (fixed potential) ensemble. In Section \ref{sec: thermoce} we discuss the phase structure of the charged black holes in the canonical (fixed charge) ensemble.
In Section~\ref{sec: holog} we begin a holographic study of the theory, focusing on holographic hydrodynamics.  We conclude the paper with a general discussion, and collect some useful results in the appendices.

\section{Charged black hole solutions} \label{sec:bhsolution}

To set up for the thermodynamic analysis in Section~\ref{sec:thermo}, in this section we shall study charged static, spherically symmetric AdS black holes in  generalized quasi-topological gravity.  This includes a more thorough study of the results presented for asymptotically flat solutions and AdS black branes in recent work~\cite{Bueno:2017sui, Bueno:2017qce, Ahmed:2017jod, Hennigar:2017ego, Hennigar:2017umz}, but also includes a study of spherical and hyperbolic black holes for the first time in this context.

\subsection{Full theory and equations of motion}

The most general cubic theory satisfying the condition $g_{tt}g_{rr} = -1$ ensuring dependence on a single metric function  includes the cubic Lovelock and quasi-topological terms, in addition to the generalized quasi-topological term.  Since both Lovelock and quasi-topological terms have been previously studied (see, e.g.~\cite{Frassino:2014pha, Hennigar:2015esa}) here we take Einstein gravity accompanied only by the cubic generalized quasi-topological term and a Maxwell field. In $d$ spacetime dimensions, the action\footnote{Our choice of the coupling here is opposite to that of~\cite{Hennigar:2017ego}, i.e. we choose a positive sign convention for the cubic coupling.} is given by    \cite{Hennigar:2017ego}
\begin{align}
\cI=\frac{1}{16\pi G}\int d^dx& \sqrt{-g}  \bigg[ \frac{(d-1)(d-2)}{L^2}+R -\frac{1}{4}F_{a b}F^{a b}
	\nn\\
	&+\frac{12(2d-1)(d-2) \mu \cS_{3,d}}{(d-3)(4d^4 - 49 d^3 + 291 d^2 - 514 d + 184)} \bigg]
\label{action0}
\end{align}
where the cosmological constant is parameterized in the standard way
\beq
\Lambda = -\frac{(d-1)(d-2)}{2 L^2}
\eeq
and where
\beqa\label{Sd}
\mathcal{S}_{3, d} &=&
14 R_{a}{}^{e}{}_{c}{}^{f} R^{abcd} R_{bedf}+ 2 R^{ab} R_{a}{}^{cde} R_{bcde}- \frac{4 (66 - 35 d + 2 d^2) }{3 (d-2) (2 d-1)} R_{a}{}^{c} R^{ab} R_{bc}
\nonumber\\
&&-  \frac{2 (-30 + 9 d + 4 d^2) }{(d-2) (2d-1)} R^{ab} R^{cd} R_{acbd}
-  \frac{(38 - 29 d + 4 d^2)}{4 (d -2) (2 d - 1)} R R_{abcd} R^{abcd}
\nonumber\\
&&+ \frac{(34 - 21 d + 4 d^2) }{(d-2) ( 2 d - 1)} R_{ab} R^{ab} R -  \frac{(30 - 13 d + 4 d^2)}{12 (d-2) (2 d - 1)}  R^3 \, .
\eeqa
The ansatz for the metric is in the following form
\beqa
\label{eqn:metricAnsatz}
ds^2&=& -N(r)^2f(r)dt^2+\frac{dr^2}{f(r)}+r^2d\Sigma^2_{(d-2),k}
\eeqa
and the field equations permit $N(r)=constant$ \cite{Hennigar:2017ego}; we set $N(r)=1$ for simplicity.\footnote{In general, one can choose $N=1/\sqrt{f_{\infty}}$, to normalize the speed of light on the boundary or in the dual CFT to be $c=1$ \cite{Myers:2010ru}. However we set $N=1$ by time reparametrization of the metric. } In the above,  $d\Sigma^2_{(d-2),k}$ denotes the line element of the $(d-2)$-dimensional transverse space, which we take to be a surface of constant scalar curvature $k=+1,0,-1$, associated with spherical, flat, and hyperbolic topologies, respectively.\footnote{The case   $k=0$ has been previously investigated \cite{Hennigar:2017umz} and so we only concentrate on non-planar black holes.}



A particular case of the metric~\eqref{eqn:metricAnsatz} is a maximally symmetric space, for which the metric function takes the form,
\beqa
f_{\rm AdS}(r) = k+f_{\infty} \frac{r^2}{ L^2} \, . 
\eeqa
Here, $L$ is the length scale associated with the cosmological constant, while $f_\infty$ is a constant that solves the following polynomial equation:
\beqa\label{asympf}
h(f_\infty) := 1 - f_\infty +(d-6) \frac{\mu f_\infty^3}{L^4}     = 0 \, ,
\eeqa
which is insensitive to the value of $k$. With $\mu \neq 0$, $f_\infty$ will differ from unity, indicating that the higher curvature terms contribute to the radius of curvature of the space. In general, the real solutions to this polynomial may be positive or negative --- we discard any negative solutions for $f_\infty$, since these would correspond to dS vacua. Restricting to only $f_\infty > 0$, the effective radius of the AdS space is then given by $L_{\rm eff} = L/\sqrt{f_\infty}$.

The negative of the derivative of Eq.~\eqref{asympf} with respect to $f_\infty$ coincides with the prefactor appearing in the linearized equations of motion~\cite{Hennigar:2017ego}, and therefore must be positive
\beqa\label{asympder}
-h'(f_\infty) =  1 -  3 (d-6) \frac{\mu}{L^4}   f_\infty^2 > 0
\label{PF}
\eeqa
to ensure that the graviton is not a ghost in these backgrounds.



As our aim is to study charged black holes, we introduce a Maxwell field $F_{a b}=\partial_a A_b-\partial_b A_a$, with electromagnetic one form defined as
\beqa
\label{eqn:vecPot}
A &=& q E(r) dt
\eeqa
By substitution of above expression in the Maxwell equation, the unknown function is determined
\beqa
E(r) &=& \sqrt{\frac{2(d-2)}{(d-3)}}\frac{1}{r^{d-3}}
\eeqa
where the specific choice of the prefactor was chosen to simplify the thermodynamic expressions and we have set to zero a constant term in the potential.

The only independent field equation from \eqref{action0} becomes
\beq\label{Feqn}
\frac{d}{dr} F[f, f', f''] = 0
\eeq
with
\beqa
F&=&r^{d-3}\left(k-f(r)+\frac{r^2}{L^2}\right)+\mu F_{\cS_{3,d}}+r^{3-d}q^2 \, .
\eeqa
The term $F_{\cS_{3,d}}$ is the contribution from the cubic generalized quasi-topological term to the field equation and is given by
\beqa
\label{eqn:fullEFE}
F_{\cS_{3,d}}&=& \frac{12}{(4d^4 - 49 d^3 + 291 d^2 - 514 d + 184)}  \Bigl[ (d^2+5d-15)\Bigl(
 \frac{4}{3}  r^{d-4} f'^3- 8 r^{d-5} f f'' \bigl(\frac{r f'}{2} + k - f \bigr) \nonumber\\
&& - 2 r^{d-5} ((d-4)f -2k) f'^2
 + 8(d-5) r^{d-6} ff'( f - k) \Bigr)
 -\frac{1}{3} (d-4) r^{d-7}(k-f)^2 \nonumber\\
 &&\times \Bigl( \bigl(-d^4 + \frac{57}{4} d^3 -  \frac{261}{4} d^2 + 312 d - 489  \bigr)f + k\bigl( 129 - 192 d + \frac{357}{4} d^2 - \frac{57}{4} d^3 + d^4 \bigr) \Bigr)  \Bigr]\,.\nonumber\\
\eeqa

Since the left-hand side of Eq. \eqref{Feqn}  is a total derivative,  direct integration  yields
\beqa
F = m
\eeqa
where $m$ is an integration constant with dimensions of $[{\rm length}]^{d-3}$ and we shall see shortly that it is related to the mass of black hole.
Although exact solutions to these field equations are not possible (except in special cases~\cite{Feng:2017tev}), it is possible to study the asymptotic behaviour and near horizon behaviour of the metric perturbatively. From the near horizon expansion it will be possible to completely characterize the thermodynamics of the black holes.


\subsection{Asymptotic solution}

To begin our solution of the equations of motion, we first focus on the case of large-$r$. In this limit, the solution will consist of a homogeneous and particular part.  For the particular solution, we take the following series ansatz:
\beq 
f_{1/r}(r) = f_\infty \frac{r^2}{L^2} + \sum_{n=-1}^{\infty} \frac{b_n}{r^n} \, ,
\eeq
where we have included a possible linear dependence $b_{-1} r$. Plugging this expansion into Eq.~\eqref{eqn:fullEFE} and solving order-by-order yields the following result:
\beqa
f_{1/r}(r)&=&f_\infty \frac{r^2}{L^2} + k + \frac{m}{h'(f_{\infty}) r^{d-3}}
-\frac{q^2}{h'(f_{\infty}) r^{2d-6}} \nonumber\\
&&\left.+ \mu
\frac{(72 d^5-294 d^4+2358 d^3-11880 d^2+18888 d-6624)}{2(4d^4 - 49 d^3 + 291 d^2 - 514 d + 184)}\frac{h'(f_{\infty})+2}{h'(f_{\infty})^2}\frac{f_{\infty} m^2}{L^2 r^{2 d-4}}
\right.\nonumber\\
&&\left.-\mu \frac{(216 d^5-342 d^4-2442 d^3+5064 d^2-1992 d+2016)}{(4d^4 - 49 d^3 + 291 d^2 - 514 d + 184)} \frac{h'(f_{\infty})+2}{h'(f_{\infty})^2}
\frac{f_{\infty} m q^2}{L^2 r^{3 d-7}}
\right.\nonumber\\
&&\left.+24\mu \frac{(d-2) (d-1)^2 \left(d^2+5 d-15\right)}{(4d^4 - 49 d^3 + 291 d^2 - 514 d + 184)}
\frac{h'(f_{\infty})+2}{h'(f_{\infty})^2}\frac{k  m^2}{ r^{2 d-2}}\right.\nonumber\\
&&\left.
+\cO\left(\frac{g_1(\mu,d,L)m^3}{h'(f_{\infty})^3 r^{3 d-5}},\frac{g_2(\mu,d,L)k q^2 m}{h'(f_{\infty})^2 r^{3 d-5}},\frac{g_3(\mu,d,L)q^4}{h'(f_{\infty})^2 r^{4 d-10}}\right)
\right.
\eeqa
where $h'(f_{\infty})$ is defined in \reef{asympder}. We have written the first five leading terms and have indicated the falloff behaviour of the next corrections to $f_{1/r}(r)$.  It is easy to see that as $\mu \to 0$ $f_{1/r}(r)$ approaches the full solution in Einstein gravity,
\beqa
f^{\rm Ein}(r)=k+\frac{r^2}{L^2}-\frac{m}{r^{d-3}}  +\frac{q^2}{r^{2d-6}} \, .
\eeqa
This is so because, in this limit, $f_\infty \to 1$ and $h'(f_\infty) \to -1$ putting the first four terms into the expected form, while $\mu \to 0$ removes the remaining terms.

To obtain the homogeneous solution, we substitute $f(r) = f_{1/r}(r) + \epsilon f_{\rm h}(r)$ into Eq.~\eqref{eqn:fullEFE}. Here we will work to linear order in $f_{\rm h}(r)$ (which is accomplished by working to linear order in $\epsilon$, then setting $\epsilon = 1$), and to leading order in the large-$r$ limit.  In this case, the equation determining the homogeneous solution reads
\beqa
f_{\rm h}''-\frac{4}{r}f_{\rm h}'-\gamma^2 r^{d-3}f_{\rm h}=0 \, ,
\label{homog}
\eeqa
where
\beqa
\gamma^2&=& -\frac{ 3 (4d^4 - 49 d^3 + 291 d^2 - 514 d + 184) L^2 \left[h'(f_\infty) \right]^2}{144 (d-1) \left(d^2+5 d-15\right) f_{\infty}   \mu \ m}
\, .
\label{gamma2}
\eeqa
Note that, at this order, the homogeneous equation does not care about the value of $k$.


Let us now understand the solutions to the homogeneous equation in the relevant cases. First, consider the case of $\gamma^2>0$. In this case the solution to~\eqref{homog} takes the form,\footnote{Note that the term involving $f_{\rm h}'$ is subleading compared to the other terms in the equation. This justifies neglecting that term in the large-$r$ limit. Doing so leads to identical conclusions concerning the sign of $\gamma^2$ as we obtain here.}
\beqa
f_{\rm h}^{(+)} = A r^{5/2} I_{\frac{5}{d-1}}\left(\frac{2 \gamma r^{\frac{d-1}{2}}}{d-1}\right)+B r^{5/2} K_{\frac{5}{d-1}}\left(\frac{2 \gamma r^{\frac{d-1}{2}}}{d-1}\right) \label{bessel}
\eeqa
where $I$ and $K$ denote the modified Bessel functions of the first and second kinds, respectively and $A$ and $B$ are constants. Schematically, in the limit of large $r$, the behaviour is
\beq
f_{\rm h}^{(+)} \sim  A r^{5/2} \exp\left(\frac{2 \gamma r^{\frac{d-1}{2}}}{d-1}\right)+B r^{5/2}\exp \left(-\frac{2 \gamma r^{\frac{d-1}{2}}}{d-1}\right)\label{exp}
\eeq
which shows that by imposing $A = 0$, the homogenous solution falls off super-exponentially in the asymptotic region --- this can be viewed as a consequence of the fact that the theory does not propagate ghosts on AdS.  The super-exponential falloff of the second term also justifies our dropping of the homogenous solution below.

Consider next $\gamma^2<0$; the homogenous solution at large $r$ becomes
\beqa
f_{\rm h}^{(-)} = C_1 r^{5/2} J_{\frac{5}{d-1}}\left(\frac{2 |\gamma| r^{\frac{d}{2}-\frac{1}{2}}}{d-1}\right)+C_2 r^{5/2} Y_{\frac{5}{d-1}}\left(\frac{2 |\gamma| r^{\frac{d}{2}-\frac{1}{2}}}{d-1}\right) \, ,
\eeqa
where $J$ and $Y$ are the Bessel functions of the first and second kinds, respectively. Note that the radial dependence is such that, in any dimension, we get solutions that oscillate rapidly and grow faster than $r^2/L^2$, and thus do not approach AdS at infinity.  The only consistent possibility would be to impose $C_1 = C_2 = 0$, eliminating the homogenous part of the solution and fixing all of the integration constants characterizing the solution. This appears to be too restrictive, as it seems to be impossible to construct solutions with $\gamma^2 < 0$ numerically while demanding a sensible black hole solution in the bulk. This, combined with the fact that for all other choices of the constants $C_1$ and $C_2$ the solution is not asymptotically AdS, leads us to disregard solutions with $\gamma^2 <0$ in the remainder of the paper --- they do not seem to exist. We henceforth will restrict ourselves to those solutions that have $\gamma^2 > 0$.

Let us note that in all cases of interest here (i.e. $d\ge 4$), the dimension-dependent pre-factors in~\eqref{gamma2} are always positive. Further, the requirement of having asymptotically AdS solutions constrains $f_\infty > 0$. Thus, ensuring the full positivity of $\gamma^2$ reduces to the inequality $m \mu < 0$. As we will see below, the parameter $m$ is related to the mass of the solution. In this work we will restrict ourselves to positive mass solutions, and hence demand that $\mu < 0$.\footnote{Note that negative mass solutions are not necessarily pathological in asymptotically AdS spaces~---~see~\cite{Mann:1997jb} for more details. In this case however, it is not simply the fact that the mass is negative that leads to the exclusion of the solutions, it is the absence of well-behaved asymptotics.}

\subsection{Near horizon solution}\label{nearsol}

Next, we look at the solution near the horizon, which is achieved by performing the following expansion for the metric function:
\beqa\label{eqn:nh_ansatz}
f(r)=4\pi T (r-r_+)+\sum_{i=2}a_n (r-r_+)^n
\eeqa
where $T$ is Hawking temperature of the black hole: 
\beqa
T=\frac{f'(r_+)}{4 \pi} \, ,
\eeqa
which follows from the regularity of the Euclideanized solution.
Inserting the near horizon expansion of the metric function into the field equation and demanding  it satisfy the field equations at each order of $(r-r_+)$ leads to conditions on the series coefficients. The first two equations involve only the mass parameter and the temperature, and read:
\begin{align}
m &=  \frac{\mu r_+^{d-7}}{(4 d^4 - 49 d^3 + 291 d^2 - 514 d + 184)}  \bigg[ 256 \pi^2 (d^2 + 5 d - 15)   (3k + 4 \pi r_+ T) r_+^2 T^2
	\nn\\
	&- (d-4)(4 d^4 - 57 d^3 + 357 d^2 - 768 d + 516) k^3 \bigg] + r_+^{d-3} \left(k + \frac{r_+^2}{L^2} \right)   +\frac{q^2 }{ r_+^{d-3}}  \, ,
	\\
0 &= (d-3) k r_+^{d-4} + (d-1)\frac{r_+^{d-2}}{L^2} - 4 \pi r_+^{d-3} T - (d-3) r_+^{2-d} q^2
	\nn\\
	&+ \frac{\mu r_+^{d-8}}{(4 d^4 - 49 d^3 + 291 d^2 - 514 d + 184)}  \bigg[12 \pi (d-4) (d-6) (4d^3 - 33 d^2 + 127 d - 166) k^2 r_+ T
	\nn\\
	&
	- 512 \pi^3 (d-4)(d^2 + 5d - 15) r_+^3 T^3 - 768 \pi ^2 (d-5)(d^2 + 5d -15) k r_+^2 T^2
	\nn\\
	&
	-(d-4)(d-7)(516 - 768 d + 357 d^2 - 57 d^3 + 4 d^4) k^3   \bigg]\, .	
	\label{MT0}
\end{align}
These two equations determine the mass parameter and temperature (non-perturbatively) as functions of the horizon radius and coupling.
These formulae are enough to determine the thermodynamic properties of the black hole. At higher orders in $(r-r_+)$, the equations are more complicated. However, the general pattern is simple: the next condition fixes $a_3$ in terms of $a_2$. Each successive order then fixes $a_n$ in terms of the previous coefficients. The only free parameter in the series is $a_2$ and, as we will see, the value of $a_2$ ends up being fixed by requiring the solution to be well-behaved asymptotically.

With near horizon and asymptotic solutions in hand, we use numerical methods to verify that these solutions are indeed joined in the intermediate region. In order to do this
we first rescale  the metric function by a factor of $L^2/r^2$ so that  when $r\rightarrow \infty$,  $(L^2/r^2) f(r)\rightarrow f_\infty$. Recall that permissible solutions for $f_\infty$ will be real, positive numbers that solve Eq.~\eqref{asympf}. We then  choose specific values for the coupling, electric charge and mass parameter, finding the corresponding values of $r_+$ and $T$ using \reef{MT0}. To solve the second order differential equation we need to have initial values for the field and its first derivative. We use the near horizon expansion, evaluated at $r = r_+(1+\epsilon)$, to obtain:
\begin{align}
\label{eqn:initial_data}
f\left(r_+(1+\epsilon) \right) &=  4 \pi T r_+ \epsilon + a_2 r_+^2 \epsilon^2 \, ,
\nn\\
f'\left(r_+(1+\epsilon) \right) &= 4 \pi T + 2 a_2 r_+ \epsilon
\end{align}
where $\epsilon$ is some small parameter. Since $a_2$ is not fixed by the field equations, its value must be determined via the shooting method: for given values of the charge, coupling, horizon radius, and $\epsilon$, a value of $a_2$ is selected and then the field equations are integrated using \eqref{eqn:initial_data} as initial data. The result is then compared to the asymptotic solution at some large value of $r$. This process is repeated until satisfactory agreement is obtained, which determines the value of $a_2$. Remarkably, we find a unique value of $a_2$ through this process. Also, owing to the fact that the differential equation is stiff, we are only able to obtain a solution  to a certain precision. With our choice of $a_2$ the asymptotic solution up to $\cO(r^{-12})$ is precise to one part in 1,000 or better.

\begin{figure}
\centering
\includegraphics[width=0.45\textwidth]{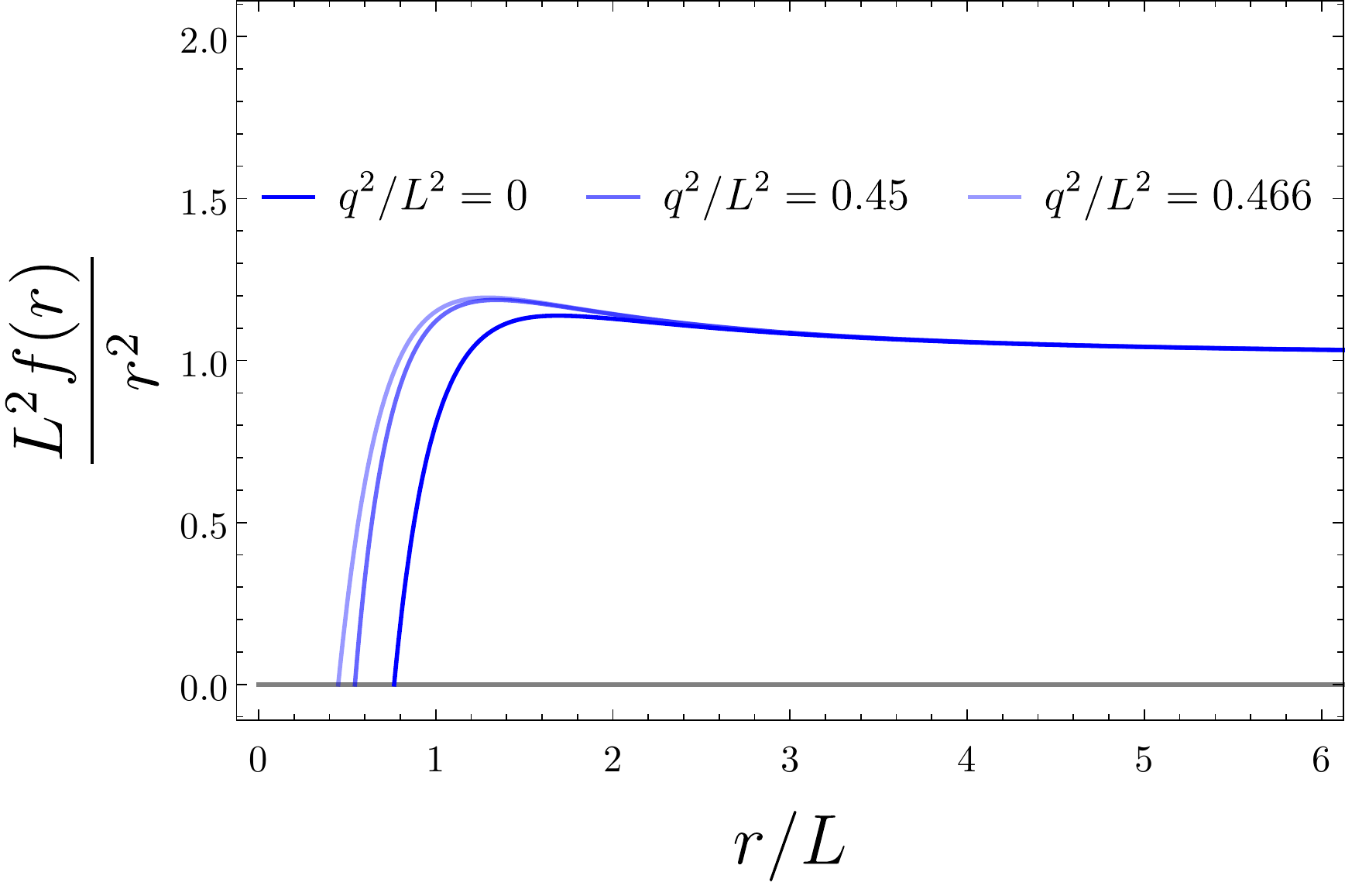}
\quad
\includegraphics[width=0.45\textwidth]{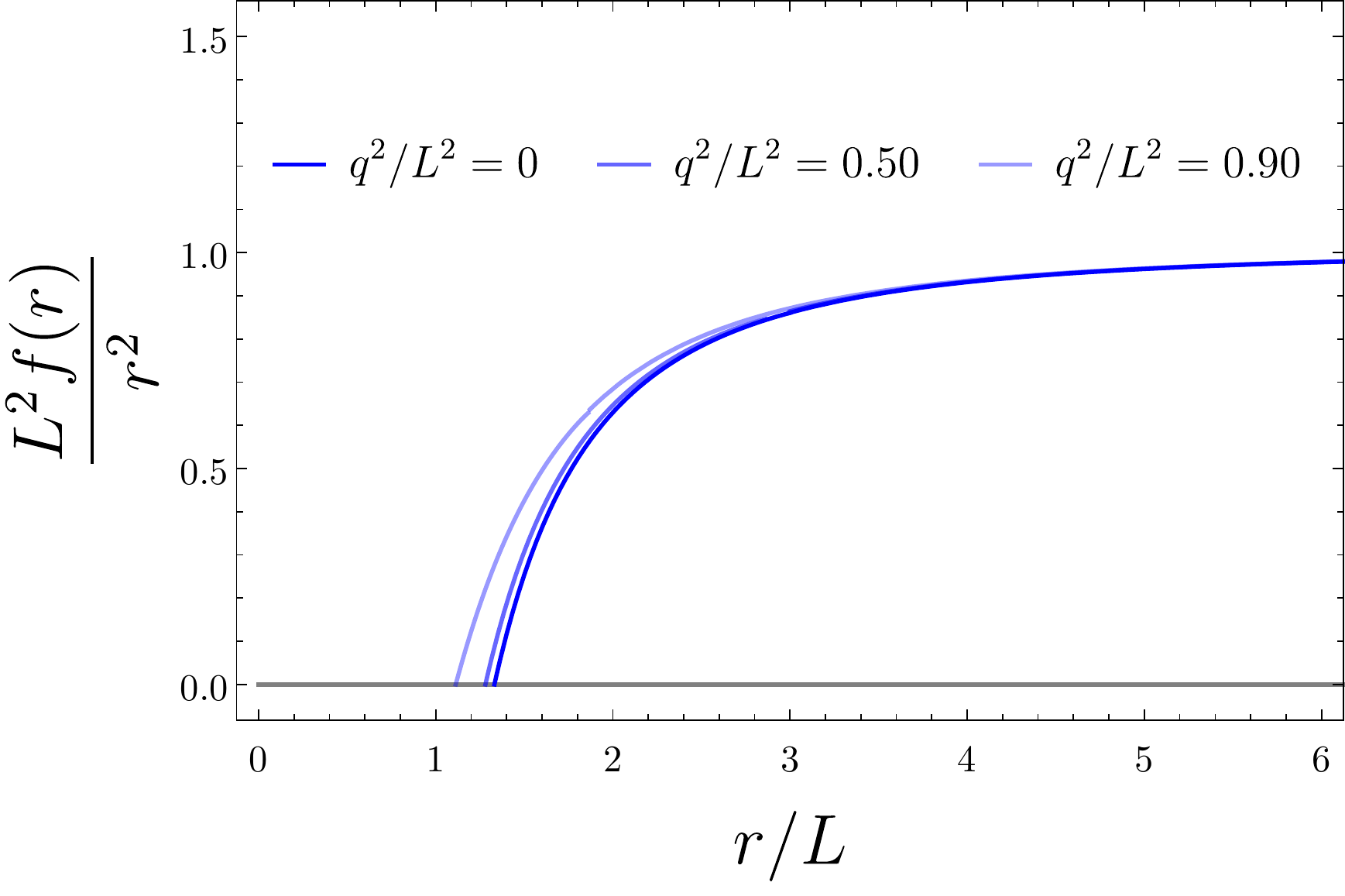}
\\
\includegraphics[width=0.45\textwidth]{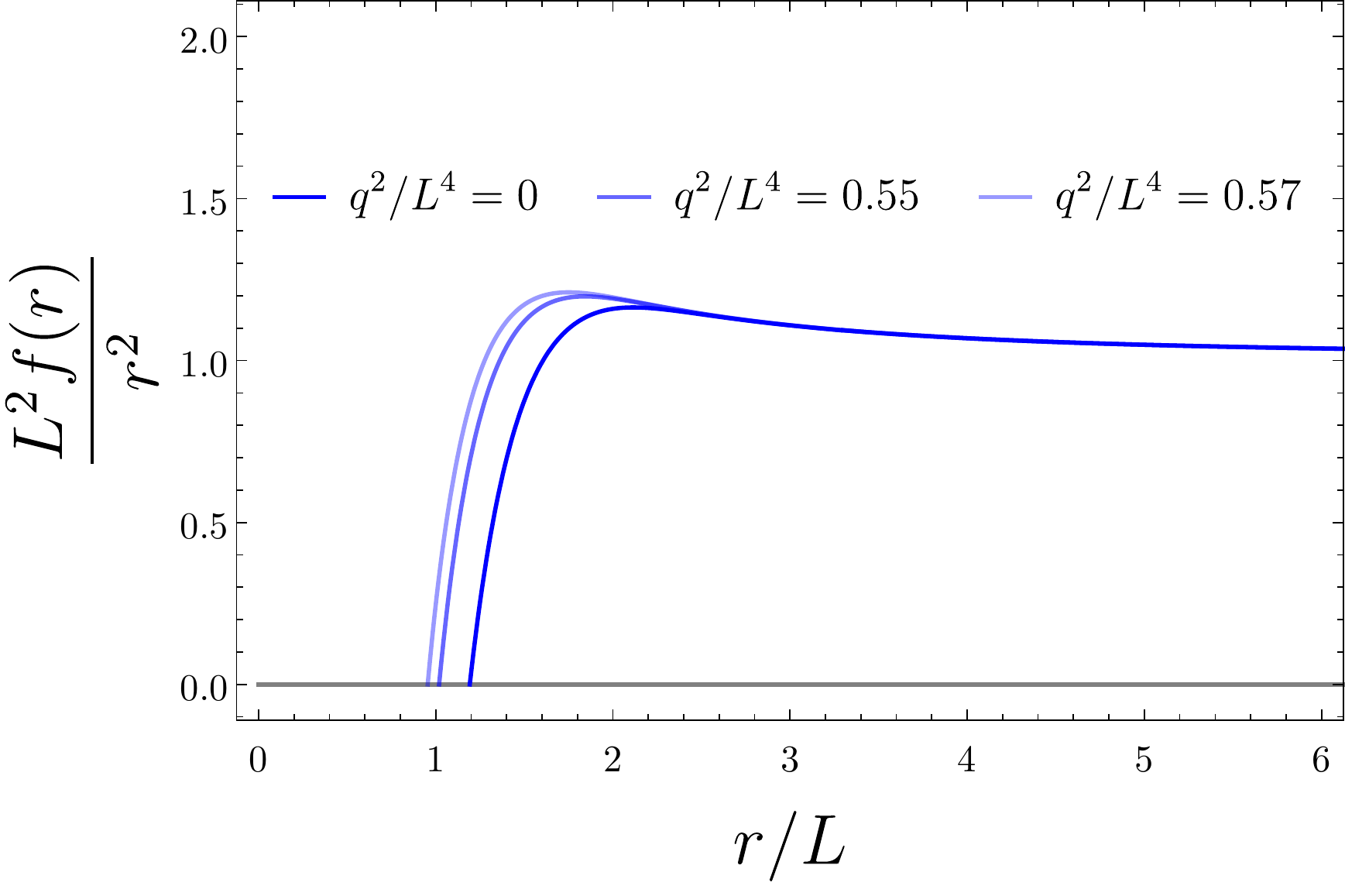}
\quad
\includegraphics[width=0.45\textwidth]{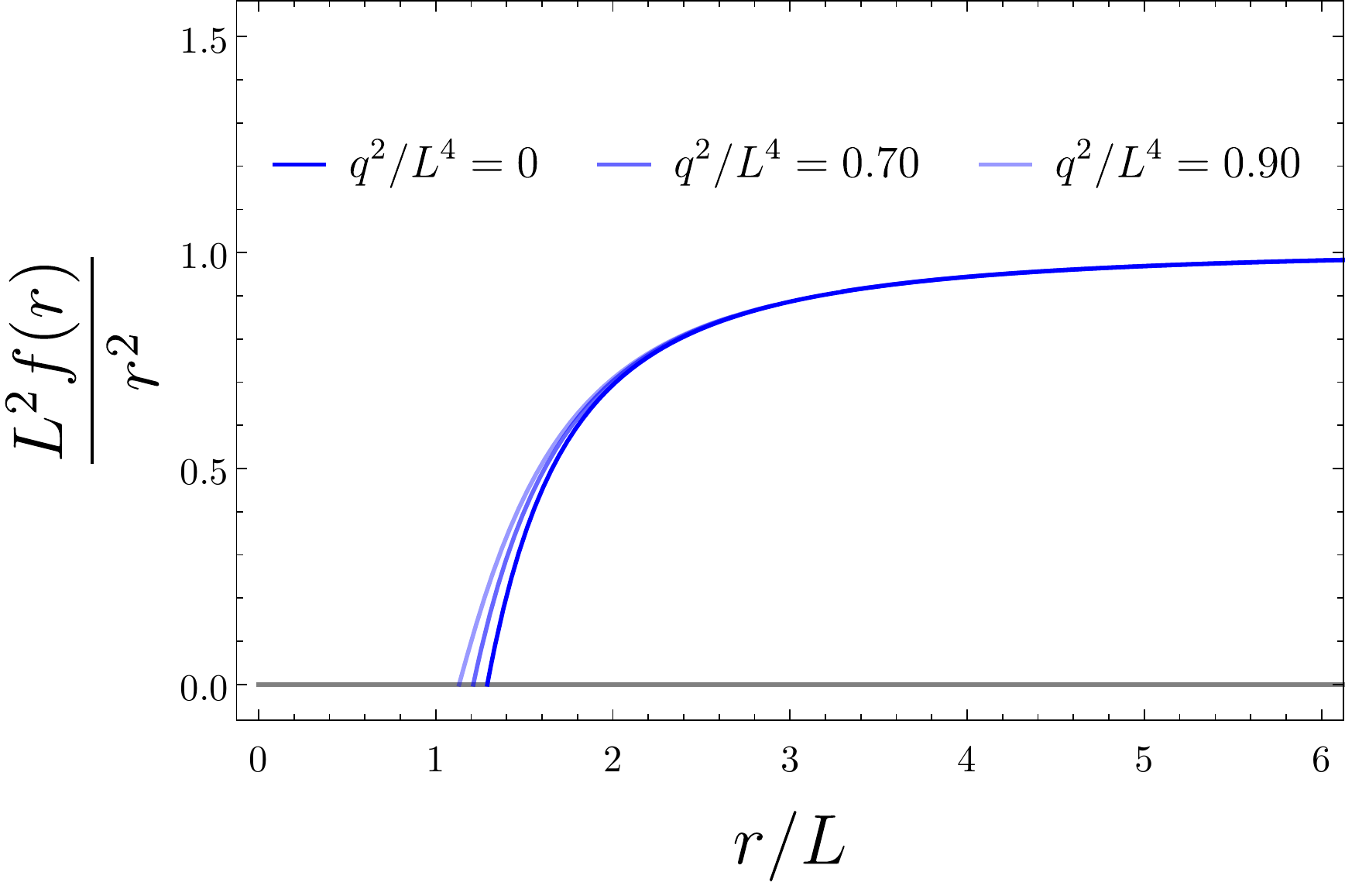}
\caption{{\bf Numerical solutions}. Here we show numerical solutions for the metric function $f(r)$ outside the black hole horizon for the cases: $d = 4$ with $k=1$ (top left), $d=4$ with $k=-1$ (top right), $d=5$ with $k=1$ (bottom left), and $d=5$ with $k=-1$ (bottom right). In the case of four dimensions, we have chosen $\mu/L^4 = -1/50$ and $m/L = 1$, while in five dimensions we set $\mu/L^4 = -1/100$ and $m/L^2 = 1$. In all cases, the value of the electric charge is indicated on the plot.  }
\label{fig:numerics}
\end{figure}

We show in Figure~\ref{fig:numerics} some sample numerical solutions in four (top row) and five dimensions (bottom row) for spherical and hyperbolic black holes with various values of electric charge. At fixed coupling, we observe that increasing the electric charge has the effect of decreasing the horizon radius. As in the uncharged case~\cite{Hennigar:2017ego}, the effect of holding the charge fixed and increasing the coupling is to increase the horizon radius. We have also produced numerical profiles for the metric function $f(r)$ in higher dimensions, but there are no qualitative differences compared to the results displayed in Figure~\ref{fig:numerics}.

Another interesting property of the solutions is their behaviour near the origin $r = 0$, which is sensitive to the spacetime dimension. We consider an expansion near the origin of the form
\beq
f(r) = r^s \left(b_0 + b_1 r + b_2 r^2 + \cdots \right) \, .
\eeq
The most interesting feature is the leading order behaviour, which is governed by $r^s$. To determine the value of $s$ we substitute the above expansion into the field equations and extract the lowest-order in $r$ term in the limit $r \to 0$. In the uncharged case\footnote{A numerical analysis of the interior solutions with $q\neq 0$ is considerably more involved due to the presence of an inner horizon at which the numerical scheme breaks down.} with $k=1$, we find that the vanishing of this term requires that $s$ solve the following cubic equation:
\beq
\label{eq:determineS}
4 s^3 + 3(d-10)s^2 - 12(d-6) s - \frac{(d-4)(4d^4 - 57 d^3 + 261 d^2 - 1248 d + 1956)}{8(d^2 + 5d - 15)} = 0 \, .
\eeq
To be physically admissible, the solution for $s$ must be real. Calculating the discriminant of the cubic reveals that it takes the form $\Delta = (d-6) \times ({\rm positive})$, and so in four and five dimensions there is a single real solution, while in $d \ge 6$ there are three real solutions.

To determine which value of $s$ controls the behaviour of the metric function near the origin we must again resort to numerics. The generalization to construct the interior solution is straightforward. We first construct the exterior solution in manner described above, which allows us to determine the value of $a_2$. With the appropriate value of $a_2$ selected, we then run the numerical scheme once again, this time setting $\epsilon$ to be a small, negative number. The numerical scheme encounters no difficulties inside the horizon. The value of $s$ can then be extracted by plotting $r f'(r)/f(r)$ in the limit $r \to 0$. In all cases that we have explored, we find that it is the smallest (real) root of the cubic~\eqref{eq:determineS} that governs the behaviour of the metric function near the origin; the value of $s$ is shown in table~\ref{tab:nearOrigin} for cases that we have verified numerically. It is interesting that, in six and higher dimensions, there appears to be three admissible solutions based on the small $r$ analysis but the black hole solution (which appears to be unique) selects only one of these possibilities. It would be interesting address what (if any) solutions the additional families of small $r$ solutions represent.

\begin{table}[h]
\centering
\begin{tabular}{ll} \toprule
    {Dimension} & $s$
    \\\midrule
    $d = 4$ & $0$
    \\
    $d=5$ & $-0.43962$
    \\
    $d = 6$ & $-1$
    \\
    $d=7$ &  $-1.62444$
    \\
    $d = 8$ & $-2.26912$
    \\ \bottomrule
\end{tabular}
\caption{{\bf Behaviour of metric function near origin}: Here we display several values of $s$ where $f(r) \sim b_0 r^s$ as $r\to 0$ and $f(r)$ represents a black hole solution. In the cases of five, seven, and eight dimensions we have displayed the result to 5 decimal places. In all cases, we have set $q=0$. }
\label{tab:nearOrigin}
\end{table}

\section{Thermodynamic considerations \label{sec:thermo}} \label{prop}

In this section we investigate the thermodynamic properties of   charged black holes in  cubic generalized quasi-topological gravity. Applying the black hole chemistry formalism \cite{Kubiznak:2016qmn}, we start by investigating the first law and Smarr relation, taking both
$\Lambda$ and $\mu$ to be thermodynamic variables.
 We then look at the physical constraints between the cubic coupling and the charge and present the domain for parameters to get physical critical points. We also illustrate the critical behaviour for the black holes here.

\subsection{First law and Smarr relation}

The near horizon expansion of the metric function discussed in   Section~\ref{nearsol} above allows for the mass and temperature of the black holes to be determined algebraically by \reef{MT0}, despite the lack of an exact solution. However except for $d=4$ an explicit solution for the temperature is complicated, so we shall use the second equation implicitly instead to show that the first law is satisfied.

To calculate the entropy, we use the Iyer-Wald formalism~\cite{Wald:1993nt, Iyer:1994ys},
\beqa
S = -2\pi \oint d^{d-2}x \sqrt{\gamma}P^{a b c d}\hat{\varepsilon}_{a b}\hat{\varepsilon}_{c d}
\eeqa
where
\beqa
P^{a b c d}=\frac{\partial \cL}{\partial R_{a b c d}}
\eeqa
and $\hat{\varepsilon}_{a b}$ is the binormal to the horizon, which is normalized as $\hat{\varepsilon}_{a b}\hat{\varepsilon}^{a b}=-2$. The integration is performed on the horizon with induced metric $\gamma_{a b}$ and $\gamma=\textrm{det} \gamma_{a b}$.  Direct calculation yields the form of the entropy for the action \reef{action0},
\beqa
\label{sratio}
S&=&\frac{\Sigma_{(d-2),k}}{4} r_+^{d-2} \Big[1+\frac{48(d-2)\mu}{(4 d^4 - 49 d^3 + 291 d^2 - 514 d + 184) r_+^4 }\Big(8 \pi  \left(d^2+5 d-15\right)  k r_+ T
	\nn\\
&&\left.  + 8 \pi^2 \left(d^2+5 d-15\right) r_+^2 T^2  -\frac{1}{16} (d-4)\left(4 d^3-33 d^2+127 d-166\right) k^2\Big)
\Big]\right. \, ,
\eeqa
where $\Sigma_{(d-2),k}$ is the volume of the submanifold with line element $d\Sigma_{(k)d-2}$. When $k=1$, this is just the volume of the $(d-2)$-dimensional sphere, while for $k=0$ and $k=-1$ the numeric answer depends on what type of identifications are performed. The pressure is defined in the standard way,
\beqa\label{press}
P=-\frac{\Lambda}{8\pi} = \frac{(d-1)(d-2)}{16 \pi L^2}
\eeqa
with other thermodynamic quantities given by
\begin{align}
V &=\frac{\Sigma_{(d-2),k} r_+^{d-1}}{(d-1)} \, , \quad Q=\Sigma_{(d-2),k} \frac{\sqrt{2(d-2)(d-3)}}{16\pi}q \, ,\quad
\Phi =\sqrt{\frac{2(d-2)}{d-3}}\frac{q}{r_+^{d-3}} ,
\nonumber\\
\Psi_{\mu} =& -\frac{32(d-2)(d^2 + 5d - 15) \Sigma_{(d-2),k}}{(4 d^4 - 49 d^3 + 291 d^2 - 514 d + 184)} \left(\pi^2 r_+^{d-4} T^3 + \frac{3}{2} \pi k T^2 r_+^{d-5} \right)
\nn\\
&+ \frac{(d-2)(d-4)\Sigma_{(d-2),k}}{4 (4 d^4 - 49 d^3 + 291 d^2 - 514 d + 184)} \bigg[3\left(4d^3 - 33 d^2 + 127 d - 166 \right) k^2 T r_+^{d-6}
\nn\\
&- \left( 129 - 192 d + \frac{357}{4} d^2 - \frac{57}{4} d^3 + d^4 \right) \frac{k^3 r_+^{d-7}}{\pi} \bigg]
\end{align}
and the mass is~\cite{Deser:2002jk}
\beqa
\label{eqn:adm_mass}
M =\frac{(d-2) \Sigma_{(d-2),k} m  }{ 16 \pi }  \,.
\eeqa

These quantities satisfy the (extended) first law of black hole thermodynamics
\beqa
d M =T dS+V dP+\Phi dQ+\Psi_{\mu}d\mu
\eeqa 
with $V$ the thermodynamic volume conjugate to the pressure
and $\Psi_\mu$ the potential conjugate to the coupling $\mu$.
The quantities also satisfy the Smarr formula
\beqa
(d-3)M = (d-2)T S-2 P V+(d-3)\Phi Q+4 \mu \Psi_{\mu}
\eeqa
that follows by a scaling argument and the first law. In Appendix~\ref{sec:EucAct}, we show that the same thermodynamic potentials follow from the Euclidean action.

Our aim is to study the critical behaviour of these black holes, and so we must obtain the equation of state. This is constructed by replacing $L^2$ in the second equation in Eq. \reef{MT0} in terms of pressure, yielding
\beqa
P&=&\frac{T}{v}-\frac{(d-3)}{\pi  (d-2)} \frac{k}{v^2}+\frac{e^2}{v^{2 d-4}}
+ \frac{2^8 (d-7)(d-4) (4 d^4-57 d^3+357 d^2-768 d+516 ) \mu k }{\pi  (d-2)^5 (4 d^4 - 49 d^3 + 291 d^2 - 514 d + 184)  v^6}
\nn\\
&&-\frac{3\times 2^8 (d-4)(d-6) \left(4 d^3-33 d^2+127 d-166\right) k^2  \mu T}{(d-2)^4 (4 d^4 - 49 d^3 + 291 d^2 - 514 d + 184) v^5}  \nonumber\\
&&\
+\frac{3\times 2^{12} \pi (d-5)  \left(d^2+5 d-15\right) k \mu T^2  }{(d-2)^3 (4 d^4 - 49 d^3 + 291 d^2 - 514 d + 184) v^4}
\nn\\
&&+ \frac{2^{11} \pi^2 (d-4) \left(d^2+5 d-15\right) \mu T^3 }{(d-2)^2 (4 d^4 - 49 d^3 + 291 d^2 - 514 d + 184) v^3}
\label{eos0}
\eeqa
where, to simplify the resulting expressions we have introduced
\beq
\label{rescaled}
v=\frac{4 r_+}{(d-2)}, \quad \quad
e^2=\frac{16^{d-3}}{\pi } (d-3) (d-2)^{5-2 d} q^2
\eeq
where we refer to $v$ as the specific volume and $e$ is a rescaled electric charge. The non-linear dependence of the equation of state on the temperature in \eqref{eos0} has been observed in previous studies of the generalized quasi-topological theories \cite{Hennigar:2016gkm, Hennigar:2017umz}.

In the bulk of the paper we will study how including cubic generalized quasi-topological terms modify the results for Einstein gravity in various dimensions. To facilitate the study of the thermodynamics, we present the explicit form of the Gibbs free energy valid for arbitrary $d$. In the canonical --- fixed charge --- ensemble the Gibbs free energy is given by $G = M - TS$ and reads
\beqa
\cG &=& \left[\frac{4}{d-2}\right]^{d-1} \frac{G}{\Sigma_{(d-2), k}} =
\frac{ v^{d-1}P}{d-1}+\frac{ v^{d-3}k}{\pi  (d-2)}
+\frac{e^2 }{(d-3)v^{d-3}}
\nn\\
&&-  \frac{2^8 (d-4) \left(4 d^4-57 d^3+357 d^2-768 d+516\right) \mu  k v^{d-7} }{\pi  (d-2)^5 (4 d^4 - 49 d^3 + 291 d^2 - 514 d + 184)}  \nonumber\\
&& - \left( \frac{v^{d-2}}{d-2} -   \frac{3\times 2^8 (d-4) \left(4 d^3-33 d^2+127 d-166\right) k^2 \mu v^{d-6} }{(d-2)^4 (4 d^4 - 49 d^3 + 291 d^2 - 514 d + 184)}\right)T
\nn\\
&&- \frac{3 \times 2^{12} \pi   \left(d^2+5 d-15\right)  v^{d-5} \mu k T^2   }{(d-2)^3 (4 d^4 - 49 d^3 + 291 d^2 - 514 d + 184)}
\nn\\
&&- \frac{2^{11} \pi ^2 \left(d^2+5 d-15\right)  \mu v^{d-4}T^3}{(d-2)^2 (4 d^4 - 49 d^3 + 291 d^2 - 514 d + 184)}  \label{Gibbsd0}
\eeqa
where the overall positive factor is suppressed in the new definition to simplify the expression and other parameters are defined in Eq. \reef{rescaled}. In the grand canonical ensemble this expression is supplemented by an additional $\Phi Q$ term, i.e. $G = M - TS - \Phi Q$. 
In stable equilibrium, the preferred state of the system is that which minimizes the Gibbs free energy at constant temperature and pressure.  In subsequent sections we will denote the free energy as $F$ when considering the cosmological constant as a fixed parameter and $G$ when working explicitly in the black hole chemistry framework. The expressions are identical in either case, only the interpretation differs.

\subsection{Physical constraints}\label{constraints}

Here we discuss the constraints on the cubic coupling that  we impose to ensure the theory is physically reasonable.  Recall first that, As discussed in Section~\ref{sec:bhsolution}, the asymptotic structure of the solutions is problematic when the parameter $\gamma^2$ --- defined in Eq.~\eqref{gamma2} --- is negative. Ensuring that $\gamma^2 > 0$, requires that $m \mu <0$. If we wish to study positive mass solutions, this then means that we must have $\mu < 0$.  We leave consideration of the negative mass solutions for future work, and consider only positive mass solutions with $\mu < 0$ here.

There are constraints on the coupling/pressure that arise due to the existence of stable AdS vacuum solutions to the theory. As described earlier, the AdS vacua of the theory are determined by the roots of the embedding equation $h(f_\infty) = 0$. Naturally, we require that the solutions have  $f_\infty > 0$ --- so that they are AdS --- and $h'(f_\infty) < 0$ --- so that they are stable, with positive effective Newton constant. Combining these requirements yields a bound on the coupling/pressure $|\mu| \le |\mu_c|$ where
\beq
\mu_c = \frac{4 L^4}{27 (d-6) } \, .
\eeq
This actually corresponds to the critical limit of the theory, where both $h(f_\infty)$ and $h'(f_\infty)$ are identically zero. This is a special point in the parameter space of the theory since the linearized equations of motion are identically satisfied. In fact, in the four dimensional version of the theory, it is possible to solve the full equations of motion exactly in this limit --- see~\cite{Feng:2017tev}. We see that the coupling at the critical limit is negative in four and five dimensions, there is no critical limit in six dimensions, and the coupling is positive at the critical limit in $d \ge 7$. When the coupling exceeds (in magnitude) the critical coupling, the theory does not admit AdS vacua. This means that the coupling/pressure is constrained only in four and five dimensions where the constraint reads $\mu > \mu_c$.  The coupling is not constrained by this requirement in higher dimensions, since there $\mu_c > 0$ and the coupling must satisfy the stricter requirement of being negative. If we write the constraint in terms of the pressure, it reads:
\beq
P \le P_{\rm max} := \frac{\sqrt{3}}{72 \pi} \frac{(d-1)(d-2)}{\sqrt{(d-6) \mu } } \, ,
\eeq
where, of course, $P_{\rm max}$ exists only in four and five dimensions.

It turns out that in higher curvature theories black hole entropy  for some regions in  parameter space can be negative.  In the context of Gauss-Bonnet gravity, it has been argued that some of these negative entropy black holes could be unstable~\cite{Cvetic:2001bk, Nojiri:2001pm}.  While it is common to simply discard negative entropy solutions as unphysical, in general the situation requires more careful thought. This is partly because there exist ambiguities in the definition of the black hole entropy --- adding to the Lagrangian a total derivative or a term proportional to the induced metric on the black hole horizon will shift the entropy by an arbitrary constant without having an effect on the other properties of the solution. For example, in even dimensional spacetimes one can add the Euler densities to the action to accomplish such a shift --- we review this in Appendix~\ref{sec:GBent} for the case of Gauss-Bonnet gravity in four dimensions. However, adding an arbitrary constant will be in tension with the expectation that the entropy should vanish when the spacetime does not contain a horizon, and so a judicious choice must be made. It is beyond the scope of this work to completely solve the issue of negative entropy in gravitational thermodynamics, but we shall make a point to elaborate on some of the issues that arise in the sections that follow.  The conditions that determine whether or not the entropy is positive will depend on the spacetime dimension and how the temperature behaves as a function of horizon radius.


\section{Hawking-Page transitions}
\label{sec:HawkingPage}

Let us begin a more thorough study of the thermodynamics of these black holes by revisiting the Hawking-Page transition. That is, we will consider the case of the uncharged black holes with spherical horizon topology. This is not only interesting in its own right, but will allow for some subtleties in the thermodynamic analysis to be discussed in a less complicated setting.  We perform this analysis in four and five dimensions. In this section we regard the cosmological constant as fixed, and hence refer to the free energy (which is then interpreted as the Helmholtz free energy) as $F$. Additionally, we measure the cubic coupling relative to its value in the critical limit, which is $\mu_c/L^4 = -2/27$ in $d = 4$ and $\mu_c/L^4 = -4/27$ in $d = 5$.

\subsection{Four dimensions}

In four dimensions, our considerations become equivalent to those for Einsteinian Cubic Gravity, which were first carried out in~\cite{Bueno:2018xqc}. Here, for the sake of completeness, we review some of these considerations with additional commentary. In this simplest case, the near-horizon equations of motion reduce to
\begin{align}
8 \pi M &= r_+ \left(1 + \frac{r_+^2}{L^2} \right) + \frac{8 \pi^2 T^2 \mu }{r_+}(3 + 4 \pi T r_+ ) \, ,
  \nn\\
0 &= 1 + 3\frac{r_+^2}{L^2} - 4 \pi T r_+ + \frac{24 \pi^2 T^2 \mu}{r_+^2}  \, ,
\end{align}
which can be solved exactly.

\begin{figure}[h]
\centering
\includegraphics[width=0.45\textwidth]{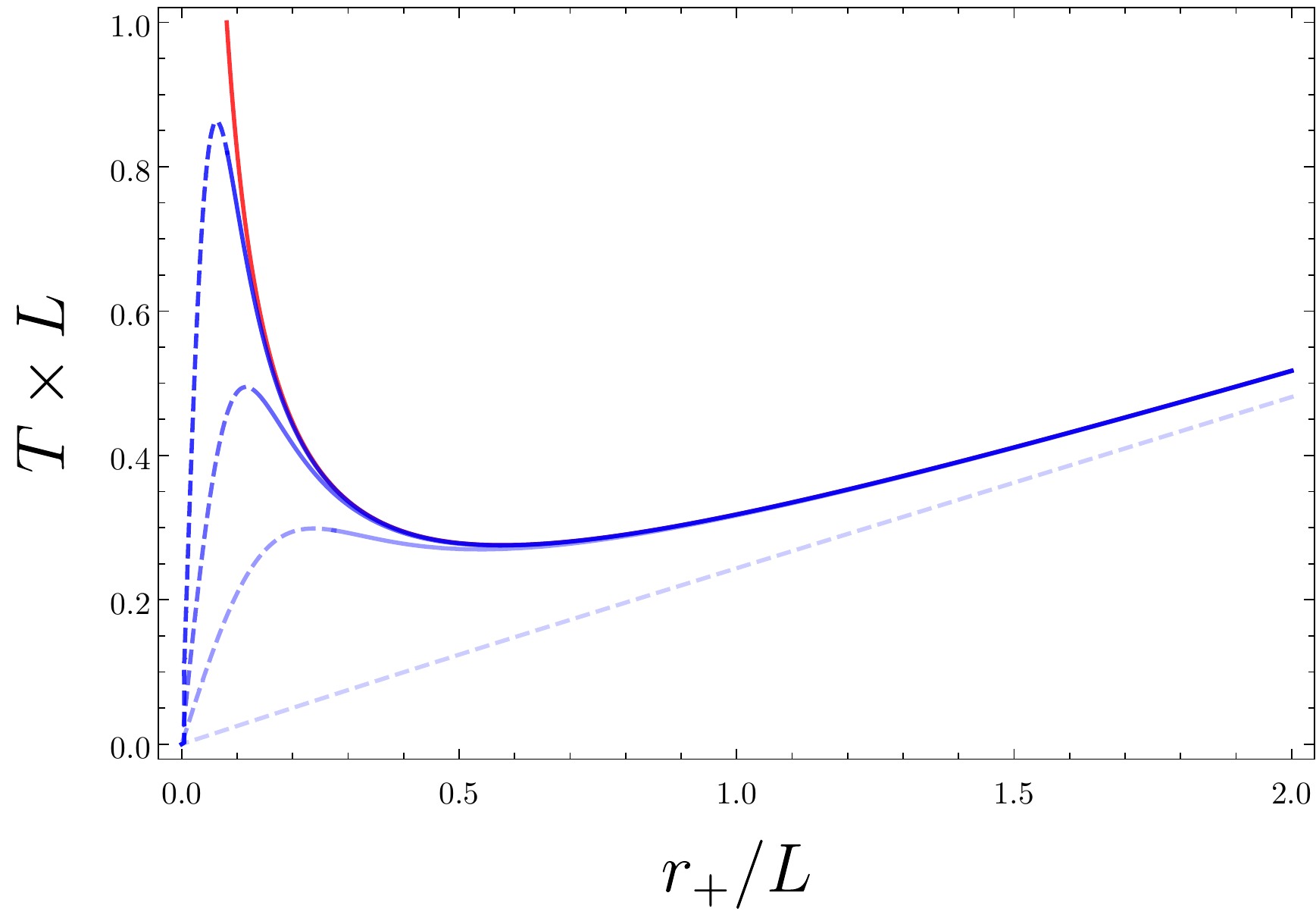}
\quad
\includegraphics[width=0.45\textwidth]{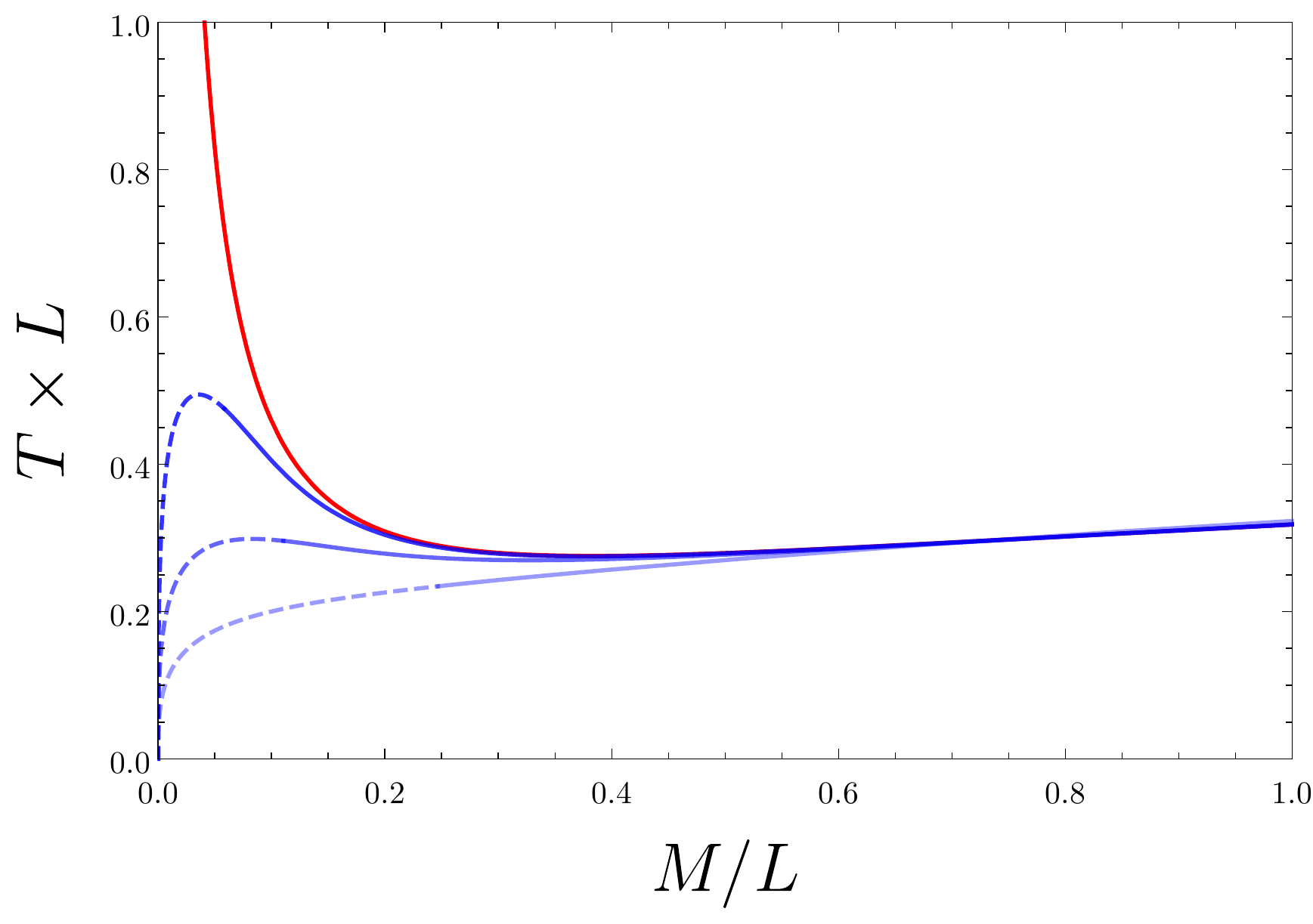}
\quad
\includegraphics[width=0.45\textwidth]{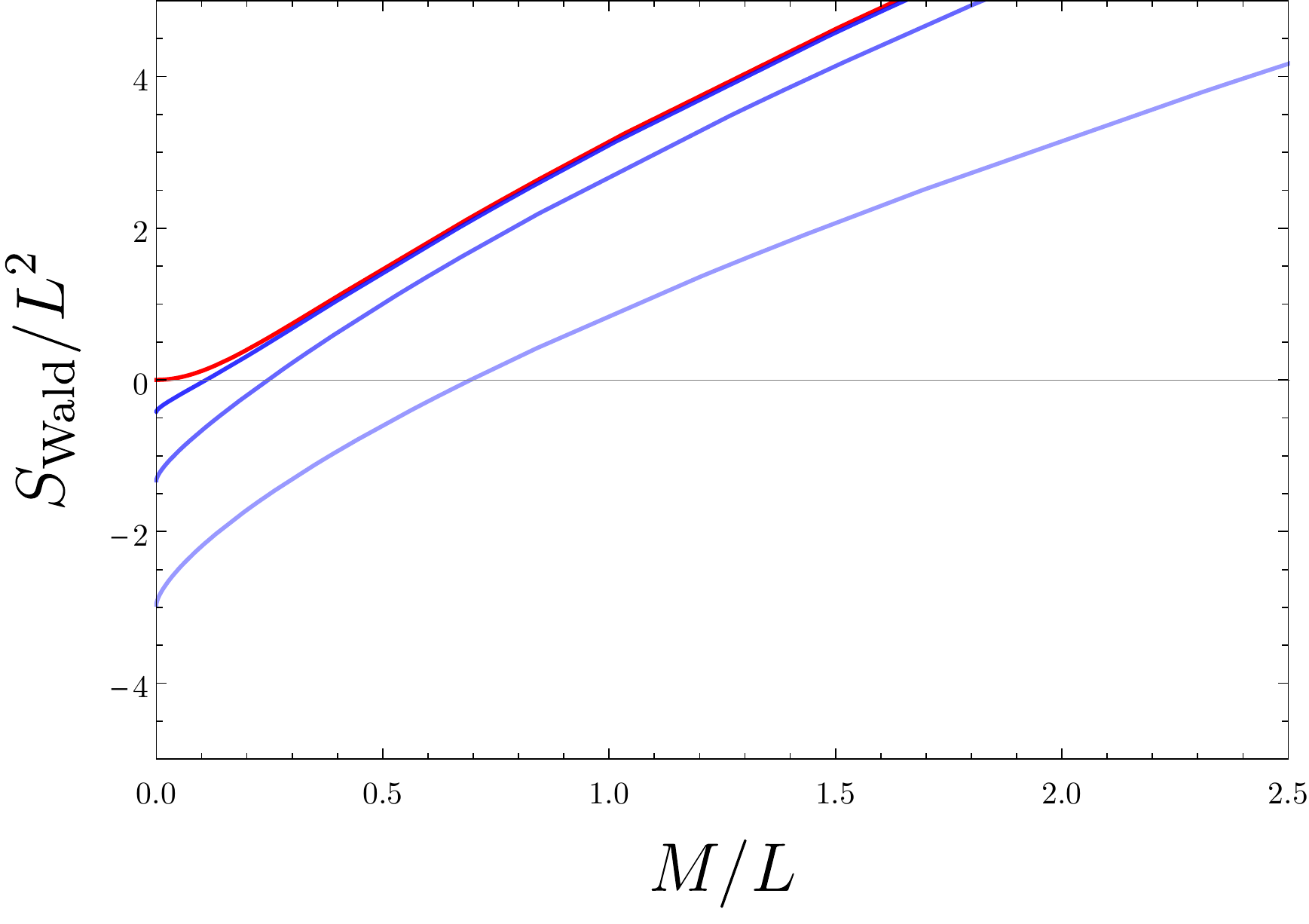}
\caption{{\bf Properties of four-dimensional uncharged black holes}. {\it Top Left}: A plot of temperature vs. horizon radius for the four-dimensional spherical black holes. The red curve represents the Einstein gravity case, while the blue curves correspond to different values of the coupling $\mu$, with curves of lower opacity corresponding to larger values of $\mu$. The dashed portions of the curves indicate that the Wald entropy of the black holes is negative. {\it Top Right}: A similar plot, this time showing the temperature against the mass. {\it Bottom Center}:  Here we plot the Wald entropy against the black hole mass. The red curve corresponds to the Einstein gravity case, while the blue curves correspond to different, non-zero values of the coupling, with curves of lower opacity corresponding to larger values of $\mu$. We see that for any non-zero $\mu$, the Wald entropy is negative as $M \to 0$. }
\label{fig:ECG-TR}
\end{figure}

It is useful to understand the differences and similarities between these solutions and the usual Schwarzschild-AdS solution. To facilitate this comparison, we show in Figure~\ref{fig:ECG-TR} a number of plots. The top left plot shows the temperature against horizon radius for various values of the coupling. For large black holes, the behaviour is very similar to the Schwarzschild AdS solution (which is shown in red), but the behaviour of small black holes is markedly different.\footnote{ Through out this section we will refer to large and small black holes. While our use of this terminology should be clear from the plots displayed, roughly speaking by `small' we mean $r_+/L < 1$ and by large $r_+/L > 1$.} For a given, fixed temperature there can be up to three distinct black hole solutions in the cubic theory, while there are at most two in the Einstein theory. The top right plot, which shows the temperature plotted against the mass, shows very similar behaviour. This plot is particularly useful since we can extract from it the thermal stability of the black holes. Since $C = \partial M/\partial T$, the slope of this plot represents the reciprocal of the heat capacity. We conclude that in the higher-curvature theory the large black holes ($M/L \gtrsim 0.3$)
are thermally stable (as they are in Einstein gravity) and the small black holes  ($M/L  \lesssim 0.03$) are as well (whereas they are not in Einstein gravity).   In the cubic theory, it is only the intermediate sized black holes that are thermodynamically unstable.

The plots also reveal initially puzzling behaviour: The Wald entropy computed for the black holes can become negative, as indicated by the dashed portions of the blue curves, and shown explicitly for a few examples in the bottom center plot of Figure~\ref{fig:ECG-TR}. In this simple setting we can compute the Wald entropy of the small black holes perturbatively in $r_+$ finding:
\beq
S = - 2 \pi  \sqrt{- 6 \mu} + {\cal O}(r_+^2) \, .
\eeq
All of the small black holes, therefore, possess negative entropy. While negative entropy certainly makes no sense from a statistical mechanics perspectice, there do not appear to be any other pathologies associated with these classical solutions, and so there is no obvious reason to outright reject these negative entropy solutions\footnote{Let us note that the issue of negative gravitational entropy is not only a problem for higher-curvature theories of gravity. For example, AdS Taub-NUT and Taub-Bolt solutions in Einstein gravity can possess negative entropy for certain parameter
values~\cite{Mann:1999pc,Emparan:1999pm,Mann:1999bt}.}. Further, let us recall that ambiguities in the definitions of entropy can allow for the shift of the entropy by an arbitrary constant. Such a shift could be accomplished via a number of ways, e.g. by adding an explicit Gauss-Bonnet contribution to the action, as in~\cite{Castro:2013pqa, Bueno:2016lrh}, or by adding to the Lagrangian a term proportional to the volume form of the induced metric on the horizon, as in~\cite{Clunan:2004tb}. Note that these methods only shift the entropy when a horizon is present, leaving the entropy of the vacuum unchanged. Further, these techniques \textit{only} change the entropy --- the solutions themselves are left unaffected. The most natural way to adjust the entropy would be to ensure that $S \to 0$ as $M \to 0$, thereby avoiding any order of limits issues. In the present case this would amount to adding $2 \pi \sqrt{-6 \mu}$ to the Wald entropy, using either of the methods described above.

The numerical value of the entropy will not have any implications when we consider thermodynamics in the fixed charge ensemble, since there we will be comparing the free energy of different branches of the black hole solutions which would all be shifted by the same amount. However, whether or not one chooses to shift the entropy can have significant implications when comparing the free energy to the vacuum. This is the case both for the Hawking-Page transition, which we consider here, and the thermodynamics in the fixed potential ensemble, which we will consider below. To illustrate these differences, we plot the results one would obtain by taking the Wald entropy to be the ``correct'' thermodynamic entropy versus those obtained using the shifted entropy satisfying $S \to 0$ as $M \to 0$.

\begin{figure}[H]
\centering
\includegraphics[width=0.45\textwidth]{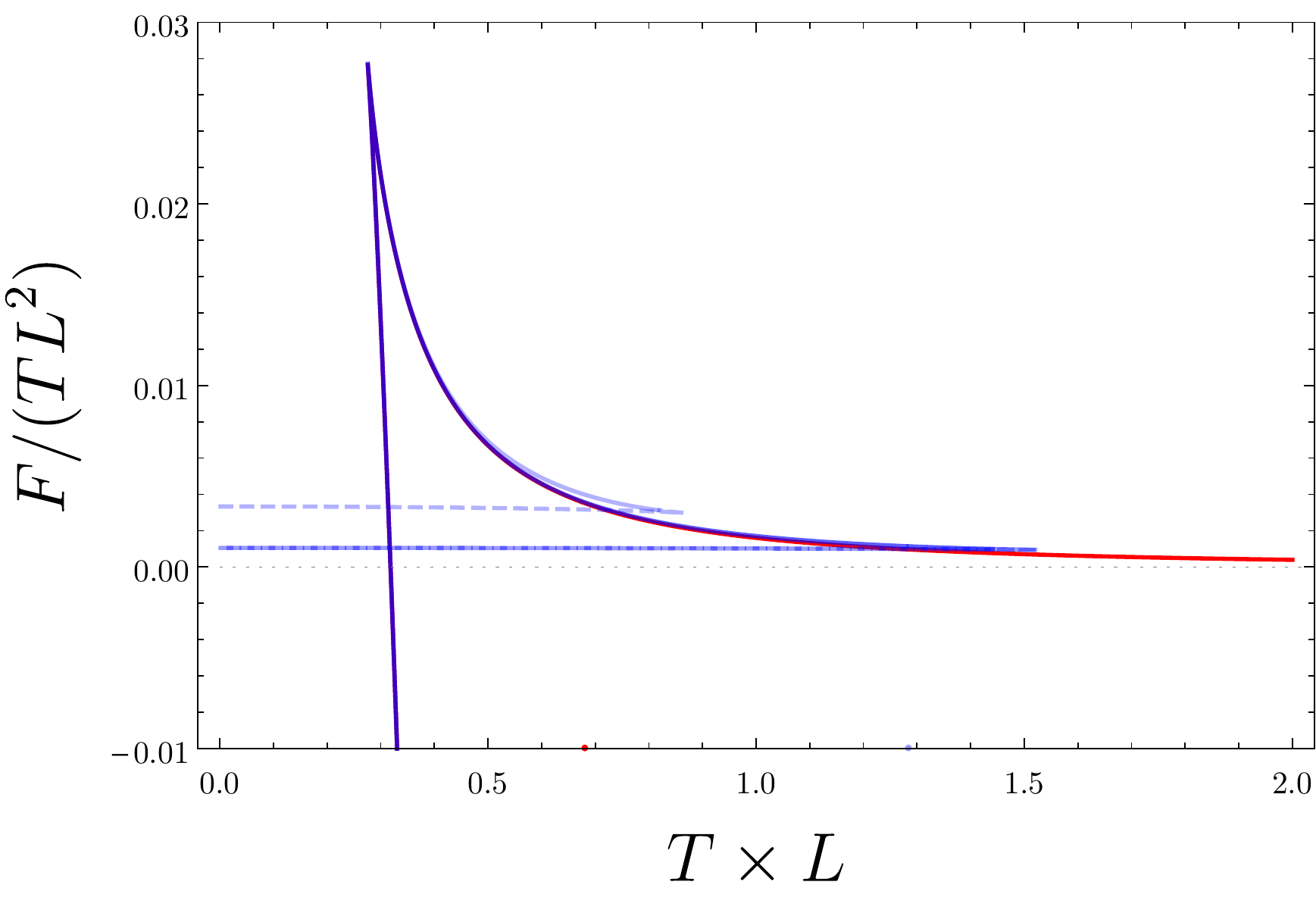}
\quad
\includegraphics[width=0.45\textwidth]{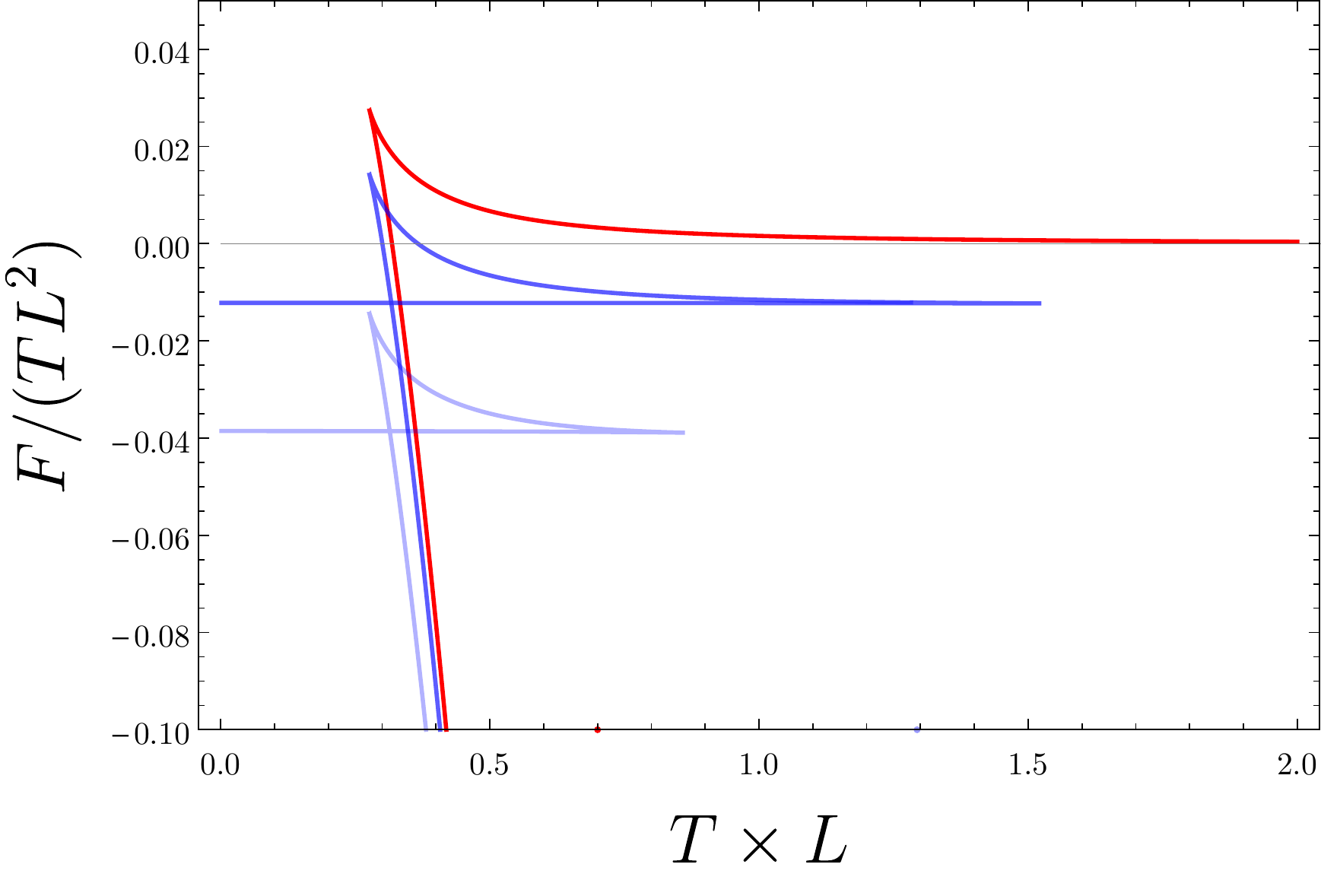}
\caption{{\bf Hawking-Page transitions in four dimensions}. {\it Left}: A plot of the free energy using the Wald entropy. Here we indicate negative entropy regions with a dashed curve. {\it Right}: A plot of the free energy using the shifted Wald entropy satisfying $S \to 0$ as $M \to 0$. In both cases, the red curve corresponds corresponds to the Einstein gravity case with $\mu = 0$, the light-blue curve corresponds to  $\mu = -10^{-4} |\mu_c|$, and the dark-blue curve corresponds to $\mu = -10^{-5} |\mu_c|$. Note that in these plots we have plotted $F/T$, since the free energy itself has a steep slope that makes it difficult to showcase the results.
}
\label{fig:HP-4d}
\end{figure}

We show in Figure~\ref{fig:HP-4d} plots of the free energy in the two scenarios.
On the left, the plots are constructed using the Wald entropy, while on the right the plots are constructed using the shifted Wald entropy.  Note in both cases the existence of a third branch of solutions that exist for any non-vanishing cubic coupling. These appear in the figure as near horizontal lines that extend all the way to $T = 0$. These correspond to the small, thermally stable black holes described above.  The situation portrayed in the left plot is very similar to the Einstein gravity situation: At low temperatures, the dominant contribution to the partition function arises from thermal radiation, and at higher temperatures the dominant contribution is a large AdS black hole.  Although it is hard to see in the diagram, the  temperature at which the transition takes place $T_{\rm HP}$ is larger in the cubic theory.  Performing a series expansion for small $\mu$ near the zero of the free energy makes this more apparent:
\beq
T_{\rm HP} = \sqrt{\frac{8 P}{3 \pi}}\left[1 - \frac{1280}{9}   \pi^2 P^2 \mu + {\cal O}(\mu^2) \right] \, .
\eeq

\begin{figure}[H]
\centering
\includegraphics[scale=0.6]{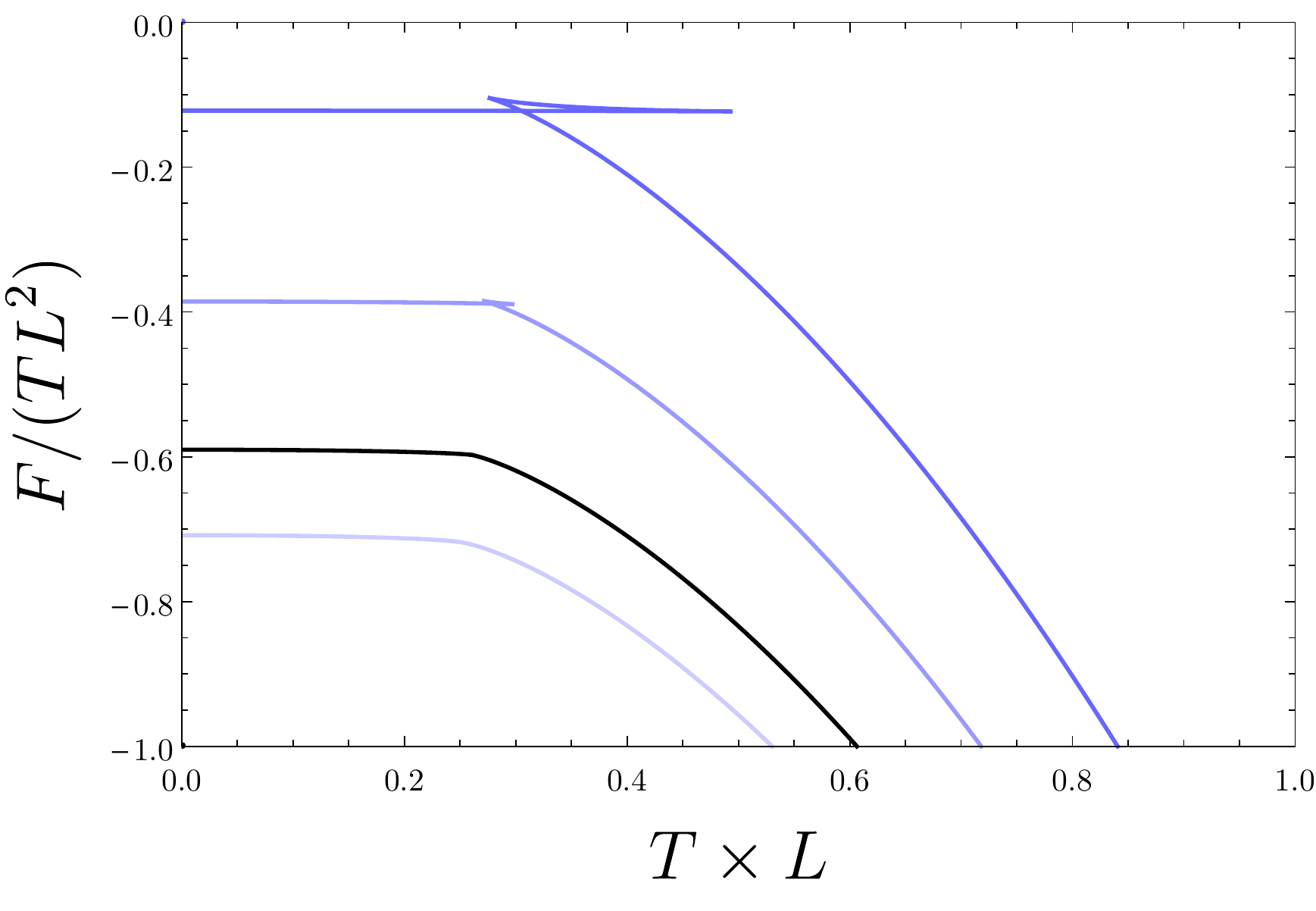}
\caption{{\bf Free energy in four-dimensions}. Here we use an additional plot of the free energy for larger values of the coupling, using the shifted entropy. The various curves correspond to increasing magnitudes of the coupling from top to bottom. The swallowtail present on the curves shrinks, eventually terminating at a cusp for $\mu = - L^4/576$, which corresponds to a second-order phase transition. For smaller larger magnitudes of the coupling, the curve is smooth with only a single branch. }
\label{fig:HP-FT2}
\end{figure}

The right plot tells a very different story. In this case, at low temperatures, the dominant contribution is a small, thermally stable black hole. As the temperature increases, there is a point at which a first order small/large black hole phase transition occurs. For small values of the coupling, the temperature at which this transition occurs is very close to the usual Hawking-Page temperature.  As the magnitude of the coupling is increased, the swallow-tail structure shrinks --- see Figure~\ref{fig:HP-FT2} --- eventually disappearing at $\mu = -L^4/576$. This corresponds to a critical point, i.e. a second order small/large black hole phase transition. The critical exponents that characterize this point are given by the usual mean field theory values --- see, for example,~\cite{Kubiznak:2012wp}.

\subsection{Five dimensions}

With the four-dimensional case illustrating some of the interesting --- and puzzling --- behaviour of these solutions, let us now move on to consider the five-dimensional case. This case is already quite a lot more complicated, with the near horizon equations being cubic polynomials in the temperature:
\begin{align}
M &= \frac{8 r_+^2}{3 \pi} \left(k + \frac{r_+^2}{L^2} \right) + \frac{\mu}{474 \pi r_+^2} \left[-976 k^3 + 8960 \pi^2 r_+^2 T^2 \left(3k + 4 \pi r_+ T \right) \right] \, ,
	\nn\\
0 &= 2 r_+ \left(k + \frac{r_+^2}{L^2}\right) + r_+^2 \left( \frac{2 r_+}{L^2} - 4 \pi T \right)  	 
	\nn\\
	&- \frac{\mu}{1264 r_+^3} \left[-1952 k^3 + 1728 k^2 \pi r_+ T + 17920 \pi^3 r_+^3 T^3 \right]	\, .
\end{align}

Since the equation determining the temperature as a function of $r_+$ is cubic it can be solved exactly. Although the resulting expressions are too messy to be illuminating, we can gain some important information by considering the discriminant of this equation, $\Delta$. Again the full expression is not particularly illuminating, but in the limit of large $r_+$ it takes the following form:
\beq
\Delta = -(79 L^4 + 1890 \mu ) \frac{286 720 \pi^6 r_+^6 \mu}{6241 L^4} + {\cal O}(r_+^4) \, .
\eeq
This means that this discriminant changes sign from positive to negative when $\mu = - 79 L^4/1890$.  Consequently for $ \mu \in (-79 L^4/1890,0)$ the temperature as a function of $r_+$ has three real solutions at large $r_+$, while for $\mu \in (\mu_c, -79 L^4/1890)$ there is only a single solution. Looking directly at the explicit solutions to the cubic equation (and discarding those for which $T < 0$), we find that for $ \mu \in (-79 L^4/1890,0)$, $T(r_+)$ is double-valued at large $r_+$, while for $\mu \in (\mu_c, -79 L^4/1890)$, large black holes with positive temperature do not exist.  A similar analysis as that just described applied to small black holes reveals that the discriminant behaves like
\beq
\Delta = - \frac{6452490240 \pi^6 \mu^4}{493039 r_+^6} + {\cal O}\left( r_+^{-2} \right) \, .
\eeq
This means that, regardless of the value of $\mu$, $T(r_+)$ will always be single-valued at small $r_+$. By explicitly examining the solution, we find that $T(r_+)$ is positive for small $r_+$: small black holes exist over the full range $\mu \in (0, \mu_c)$.

\begin{figure}[H]
\centering
\includegraphics[width=0.45\textwidth]{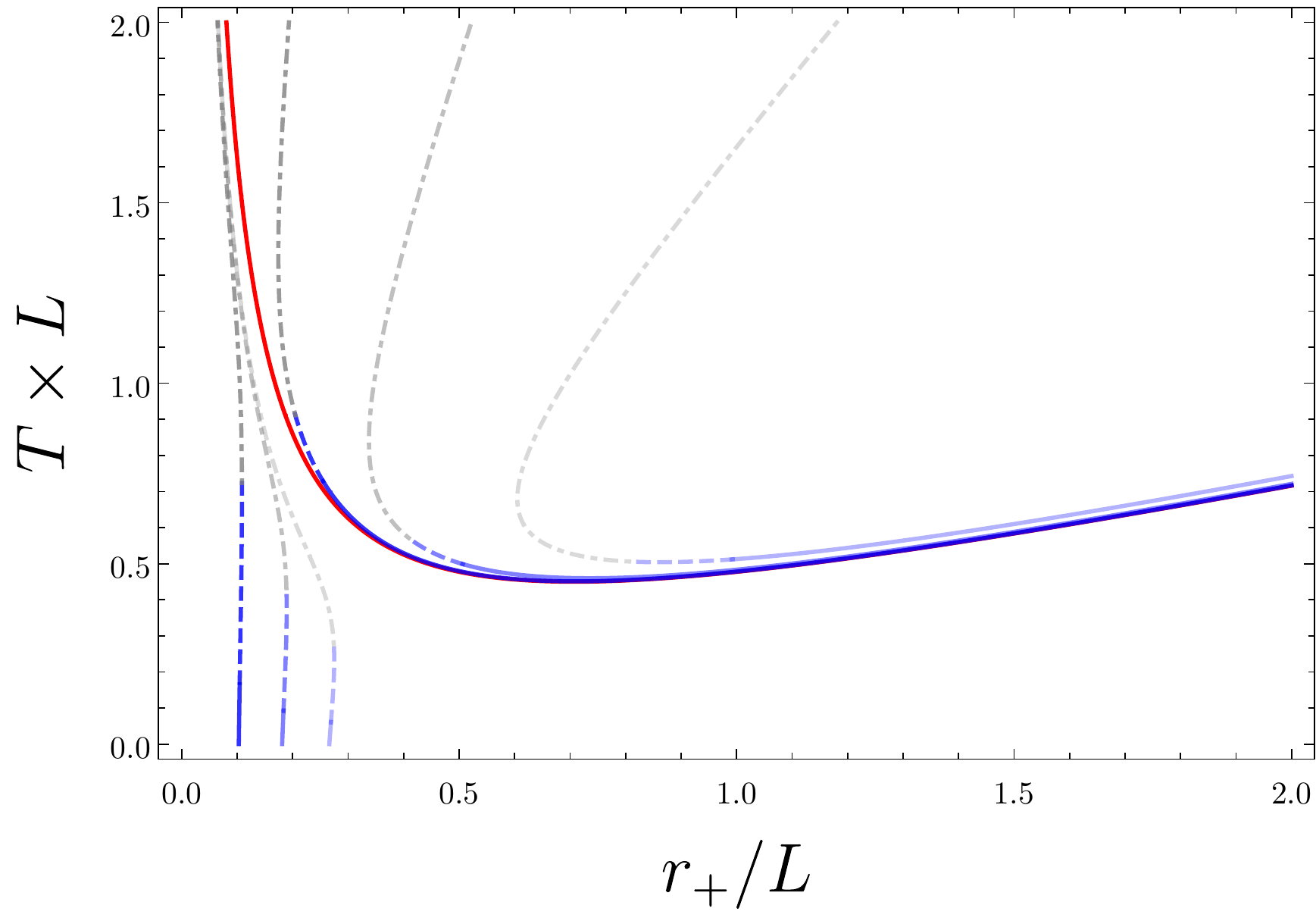}
\quad
\includegraphics[width=0.45\textwidth]{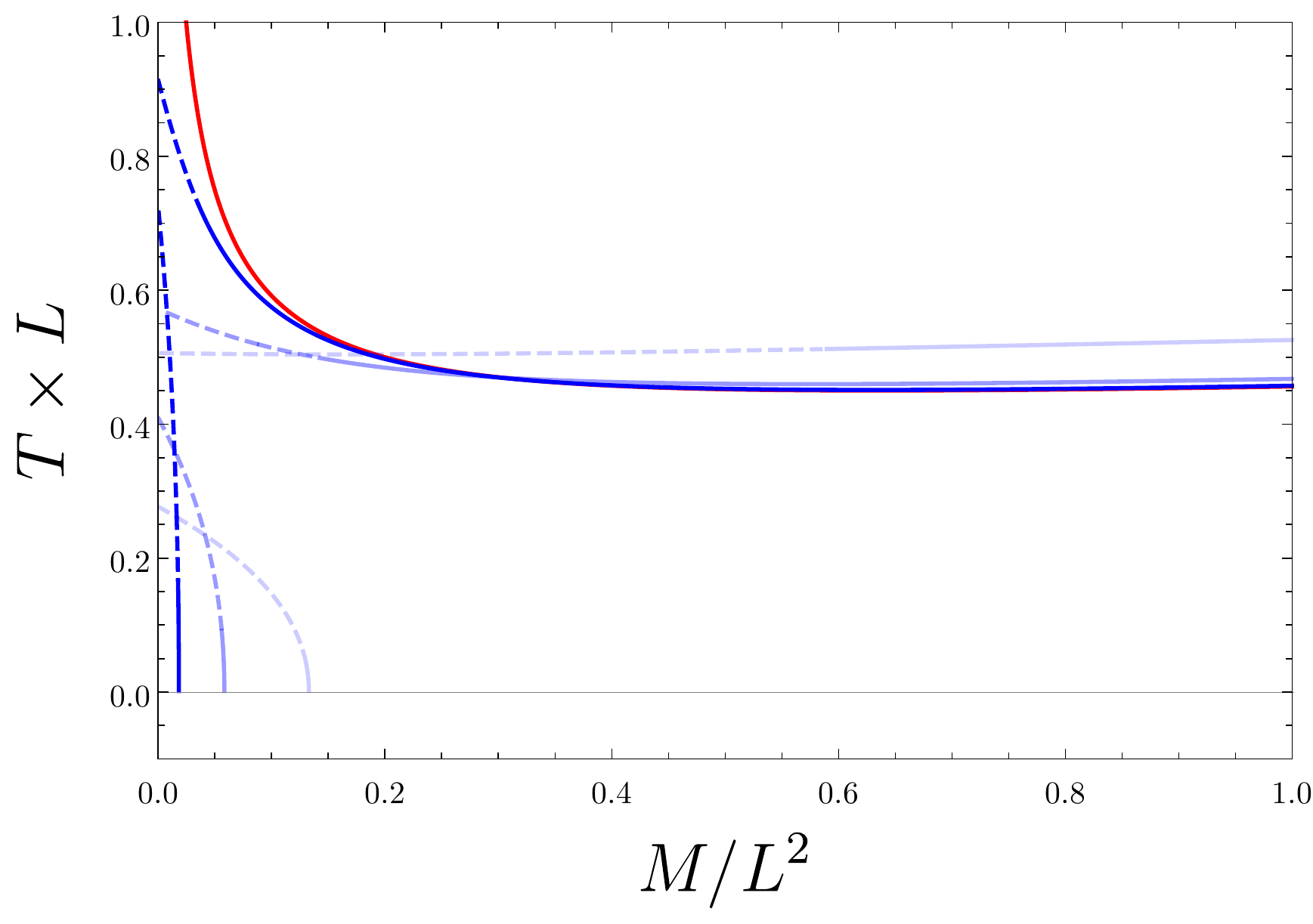}
\caption{{\bf Properties of five-dimensional uncharged black holes I}. {\it Left}: A plot of the temperature vs.~horizon radius for the five-dimensional black holes. The red curve represents the Einstein gravity result, while the blue curves correspond to $\mu/\mu_c = 10^{-3}, \, 10^{-2}, \, 10^{-1}/2$ in order of decreasing opacity. In each case, the dashed portion of the curve indicates negative Wald entropy, while the gray dot-dashed portions indicate that the mass is negative, and hence the solutions do not exist. {\it Right}: A plot of the temperature vs.~the mass; the curves are the same as in the left plot. For large values of the mass (or, equivalently, large values of the horizon radius) the solutions with the cubic correction hug closely the Einstein gravity curve, while significant differences begin to appear for small values of the mass/horizon radius. Though it is a bit hard to see in the figures, note that  the curves that touch $T = 0$ (those on the bottom left of each figure) always have positive entropy.}
\label{fig:5d-HP-properties}
\end{figure}
\begin{figure}[H]
\centering
\includegraphics[width=0.45\textwidth]{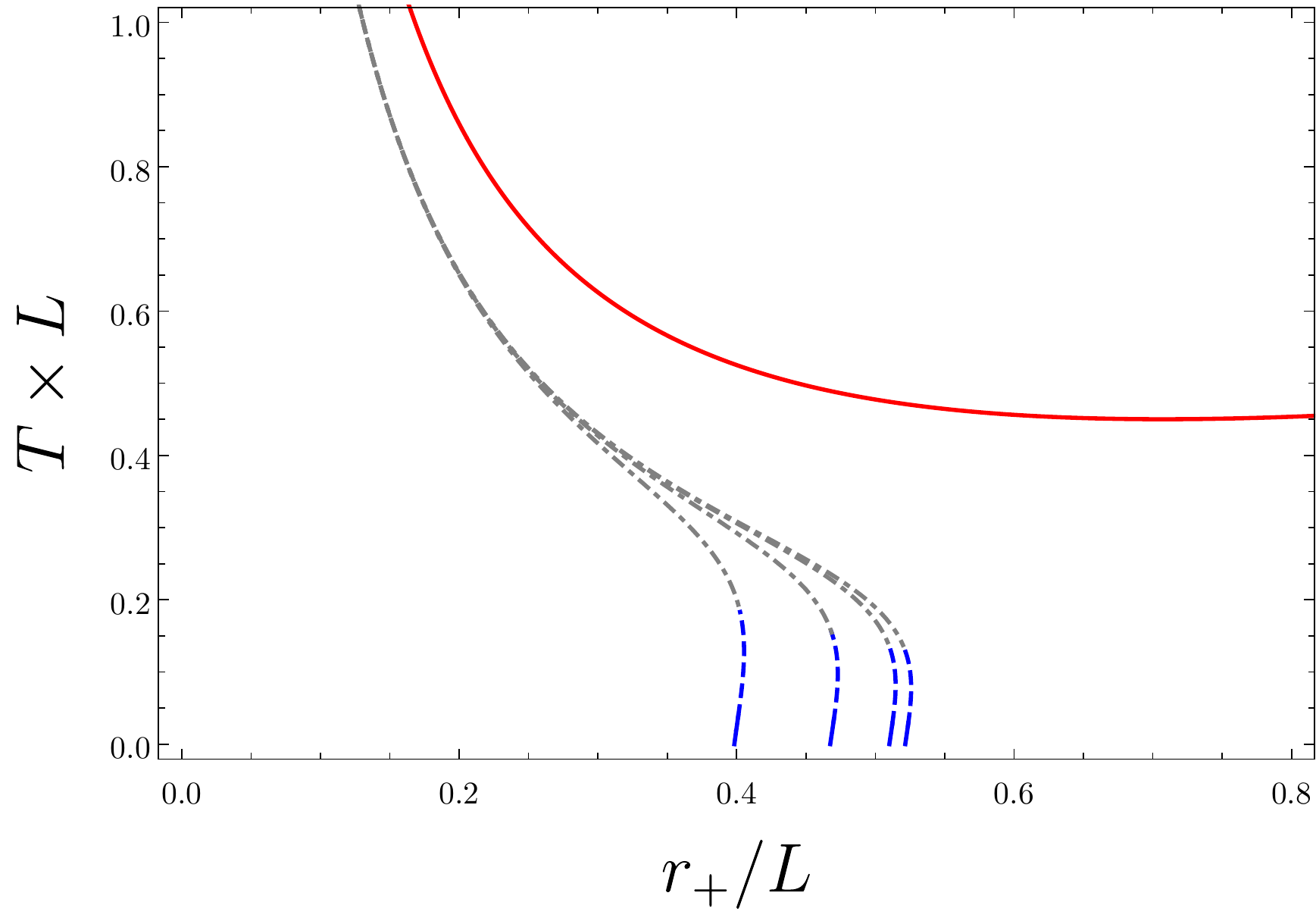}
\quad
\includegraphics[width=0.45\textwidth]{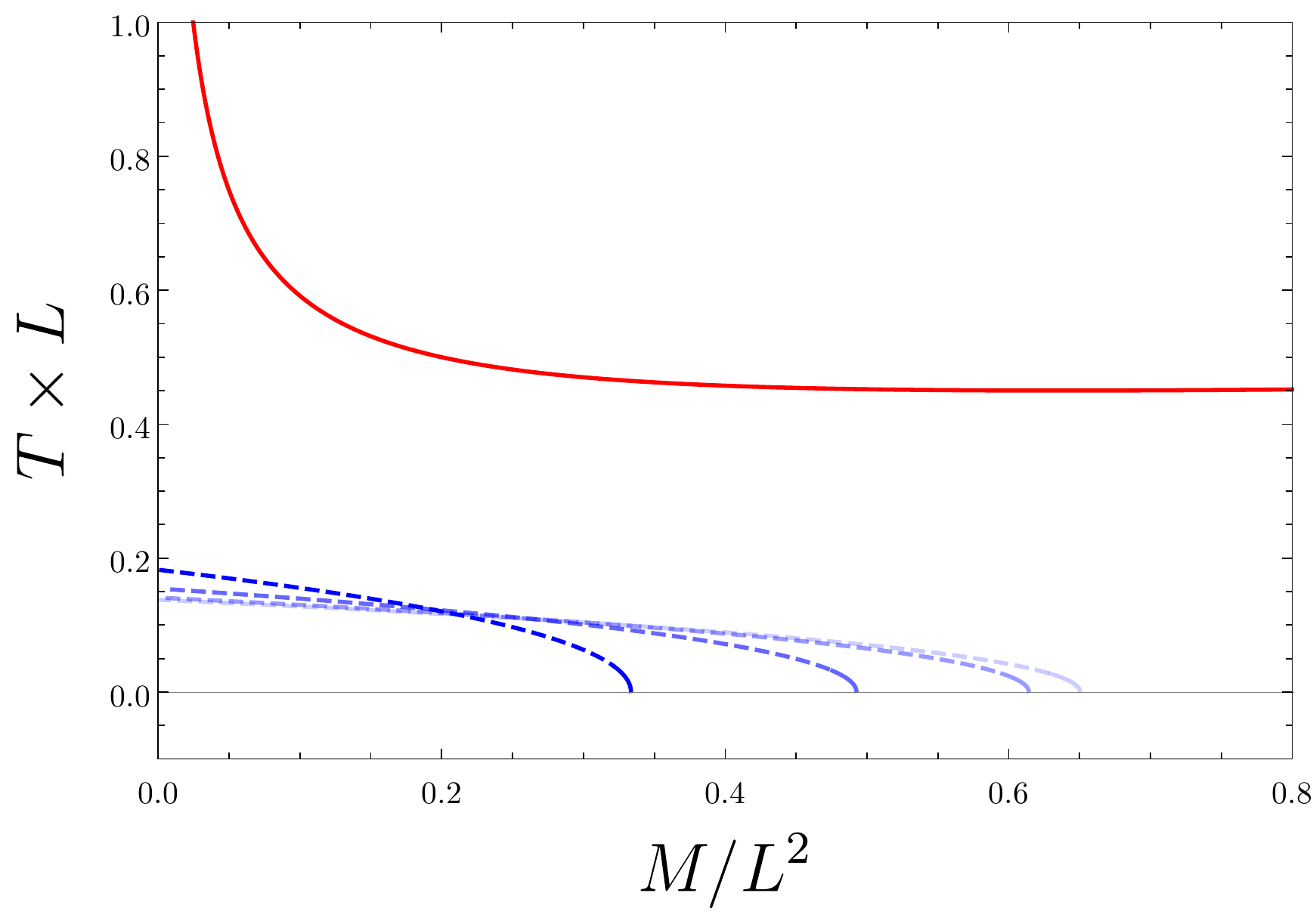}
\caption{{\bf Properties of five-dimensional uncharged black holes II}.  Here we show plots of the temperature vs.~horizon radius and temperature vs.~mass for $\mu = 0$ (red) and $\mu/\mu_c = 0.29, 0.6, 0.9, 0.99$ (blue curves). Each non-zero value of $\mu$ here is larger in magnitude than the special value of $\mu = - 79 L^4/1890$ which marks the point at which the large black holes no longer exist. The dashed blue curves indicate negative Wald entropy, while the dot-dashed grey curves indicate that the mass is negative --- these solutions do not exist. Note that for small temperatures the entropy is positive.
}
\label{fig:5d-HP-small}
\end{figure}
Let us now consider the temperature vs.~horizon radius profiles directly, taking into account various complications like the positivity of mass and entropy. We divide our study into two cases corresponding to $\mu > -79 L^4/1890$ and $\mu < -79 L^4/1890$, with the first case shown in Figure~\ref{fig:5d-HP-properties}. Here we see that two branches of black holes emerge, which ``hug'' the Einstein gravity temperature vs. horizon radius curve on opposite sides. In the limit $\mu \to 0$, it is the upper curve that converges to the Einstein gravity result, while the lower curve disappears. At any given value of the temperature, there can be up to three black hole solutions (opposed to the two present in the Einstein case), though for  most values of the coupling at least one of these possible solutions will have negative mass (and hence the solutions does not exist) or negative entropy (and hence the solution needs more careful attention). Contrary to the four-dimensional case, the small black holes are not thermally stable, as can be deduced from the negative slope in the temperature vs.~mass plot. Similar to the Einstein case, the large black holes are thermally stable. Now, let us move on to consider what happens when we push $\mu$ beyond $-79 L^4/1890$. Plots for this situation are shown in Figure~\ref{fig:5d-HP-small}. Despite the absence of the large black holes, the profiles for the small black holes remain largely the same.

\begin{figure}[H]
\centering
\includegraphics[width=0.45\textwidth]{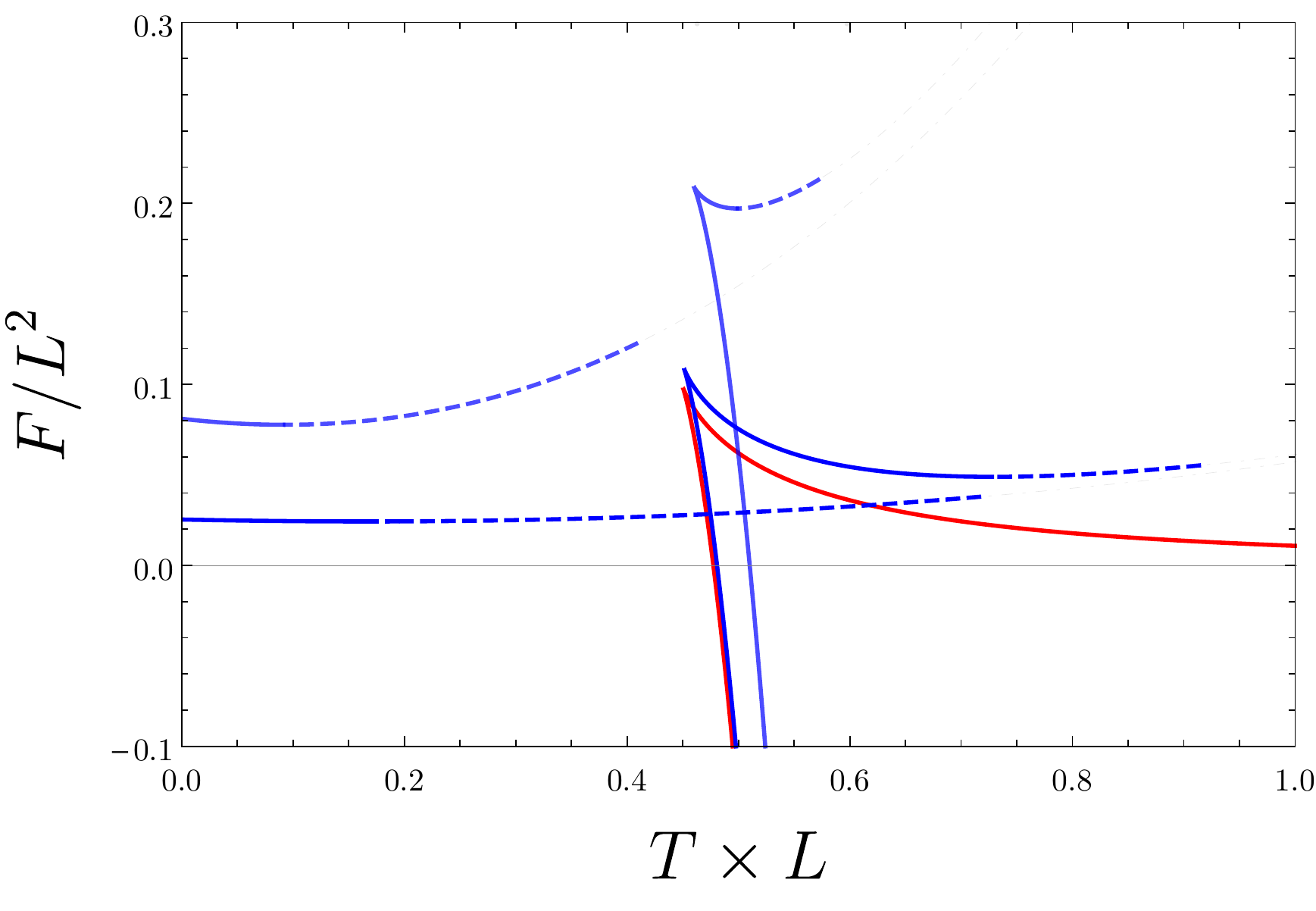}
\quad
\includegraphics[width=0.45\textwidth]{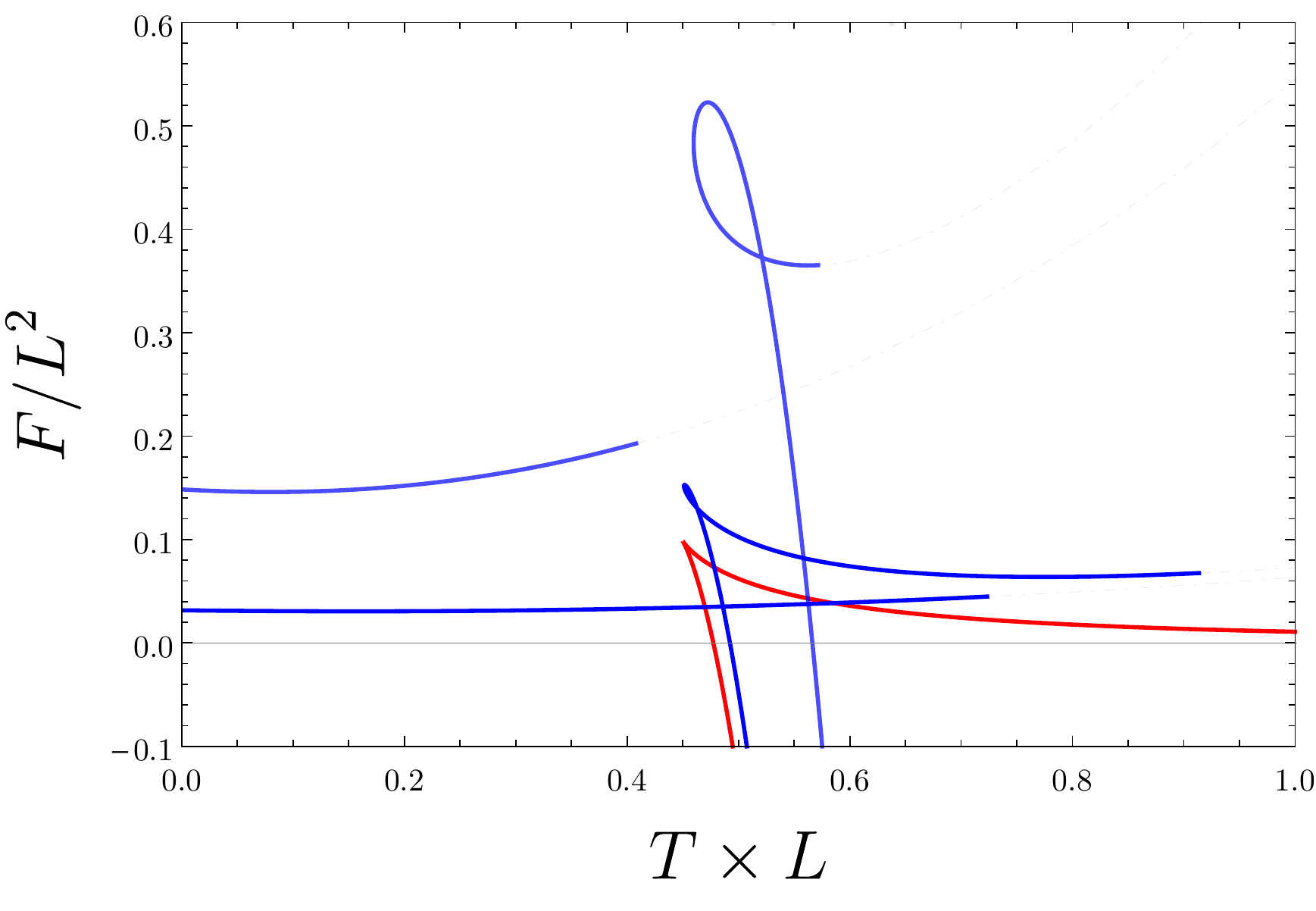}
\caption{{\bf Hawking-Page transition in five dimensions}. {\it Left}: Here we show a plot of the free energy vs.~temperature for the five dimensional uncharged black holes. The red curve corresponds to the Einstein gravity case, the dark blue curve corresponds to $\mu = -10^{-3} |\mu_c|$, and the light blue curve corresponds to $\mu = -10^{-2} |\mu_c|$. In each case, solid curves indicate that both the mass and entropy are positive, while a dashed curve indicates that the Wald-Entropy is negative. The blue curves terminate when the mass becomes negative. {\it Right}: The same plot as on the left, but now using the shifted entropy as described in the text.   }
\label{fig:5d-HP-FvT}
\end{figure}

Both plots in Figure~\ref{fig:5d-HP-properties} indicate regions of negative Wald entropy. At first glance, the situation here is actually more complicated than in four dimensions, since instead of one there are now two branches of black hole solutions. This means that we cannot simply add a universal constant to the entropy to ensure that $S \to 0$ as $M \to 0$. However, it turns out that in this case the effects are not relevant for the Hawking-Page transition. As shown in Figure~\ref{fig:5d-HP-FvT}, which is a plot of the free energy vs.~temperature in the five dimensional case,   the free energy and mass are positive at $T \to 0$, indicating that it is thermal AdS that dominates the partition function at small temperatures. The regions with negative entropy and negative mass (the latter corresponding to solutions that do not exist) are actually excluded by the Hawking-Page transition, since they have positive free energy.  This persists even when a constant is added to the entropy\footnote{Here we have shifted the entropy $S \to S_{\rm Wald} - S_{\rm min}$ with $S_{\rm min} = \min \left\{S^1_{M \to 0} , S^2_{M \to 0} \right\}$, where the superscripts denote the two branches of black holes.} to ensure that $S > 0$ for all $M \ge 0$.  The precise temperature at which the Hawking-Page transition occurs is larger than in the equivalent set up for Einstein gravity, similar to the four-dimensional case. Let us close by noting that when $\mu < - 79 L^4/1890$ (and so large black holes no longer exist), then a Hawking-Page transition does not occur, and thermal AdS is thermodynamically preferred for all temperatures.

\subsection{Remarks on higher dimensions}

 Before moving on to consider the charged solutions, let us pause here to present a few comments on the higher dimensional solutions. In many aspects, the higher dimensional solutions are similar to the five dimensional solutions. One feature that continues into higher dimensions is a limit on the coupling for the existence of large black holes. We saw above that in five dimensions there is a special coupling $\mu^* = -79 L^4/1890$ such that for $\mu < \mu^*$ there are no large black holes. In higher dimensions the value of $\mu^*$ can be determined in the same way by examining the large $r_+$ behaviour of the discriminant of equation~\eqref{MT0}. The result reads
\beq
\mu^* = -\frac{(4 d^4 - 49 d^3 + 291 d^2 - 514 d + 184)L^4}{54(d-4)(d-1)^2(d^2 + 5d - 15)} \, ,
\eeq
and we emphasize that no such bound exists in four dimensions. Noting this, the structure of the temperature vs.~horizon radius profiles are qualitatively similar to the discussion presented above for five dimensions.

\section{Charged black holes: Grand canonical ensemble} \label{sec:thermofpe}

Next we consider the thermodynamics of the cubic corrected black holes in the fixed potential --- or grand canonical --- ensemble. This means we consider the difference of electric potential between the horizon and infinity to be a fixed quantity. From the perspective of holography this setup amounts to a fixed chemical potential in the field theory. Once again we restrict ourselves to the spherical black holes  and present the analysis in four and five dimensions. Further, we emphasize that the four-dimensional results would coincide with those for Einsteinian Cubic Gravity, though in this case there is no precedent for this study and the results here are novel.  Recall that in the grand canonical ensemble the free energy is given by $F = M - TS - \Phi Q$.

\subsection{Four dimensions}

In four dimensions, a number of expressions are quadratic (rather than cubic) in the temperature, allowing for analytic results to be presented. Working in the fixed potential ensemble, we have the following expressions that determine the mass and temperature in terms of the coupling and $r_+$:
\begin{align}
 2 M &=  k r_+ + \frac{r_+ \Phi^2}{4} + \frac{ r_+^3}{L^2 }  + \frac{8 \mu \pi^2 T^2 \left(4 \pi r_+ T + 3 k \right)}{ r_+} 	 \, ,
	\nn\\
	0 &=  k +  \frac{ 3 r_+^2}{L^2} - 4 \pi r_+ T - \frac{\Phi^2}{4} + \frac{24 \mu k \pi^2 T^2}{ r_+^2} \, .
	\label{eqn:GCE-4d-nheqs}
\end{align}
 From the above, we can obtain the equation of state by solving the second expression for the pressure:
\beq
\label{eqn:gce_eos_4d}
P = \frac{T}{v} - \frac{k}{2 \pi v^2} + \frac{\Phi^2}{8 \pi v^2} - 48 \mu \frac{ k \pi T^2}{ v^4} \, ,
\eeq
where we have identified $P = 3/(8\pi L^2)$ and $v = 2 r_+$ is the specific volume. In the following we will remark on the cases where both $P$ is constant and considered a thermodynamic variable. From now on we take $k = +1$ to focus on the spherical black holes.

Let us begin by discussing some of the properties of the black holes when the potential is fixed at the boundary. In this four dimensional case, we note that the terms that arise due to the higher-order curvature terms are all proportional to at least one power of the temperature. This means that the properties of the extremal black holes are in fact the same as in Einstein gravity.  The black holes will be extremal when the following constraint is satisfied:
\beq
\left( L r_+^{\rm ext} \right)^2 = \frac{\Phi^2 - 4 }{32 \pi} \, .
\eeq
In the case of spherical black holes, this means that extremal black holes will exist in the fixed potential ensemble only if the potential satisfies $\Phi^2 > 4$, just as in the Einstein gravity case~\cite{ChamblinEtal:1999a}. As we will see, it turns out that this value of the electric potential also controls other aspects of the behaviour of the black holes and leads to a variety of interesting structures.

\begin{figure}[h]
\centering
\includegraphics[width=0.45\textwidth]{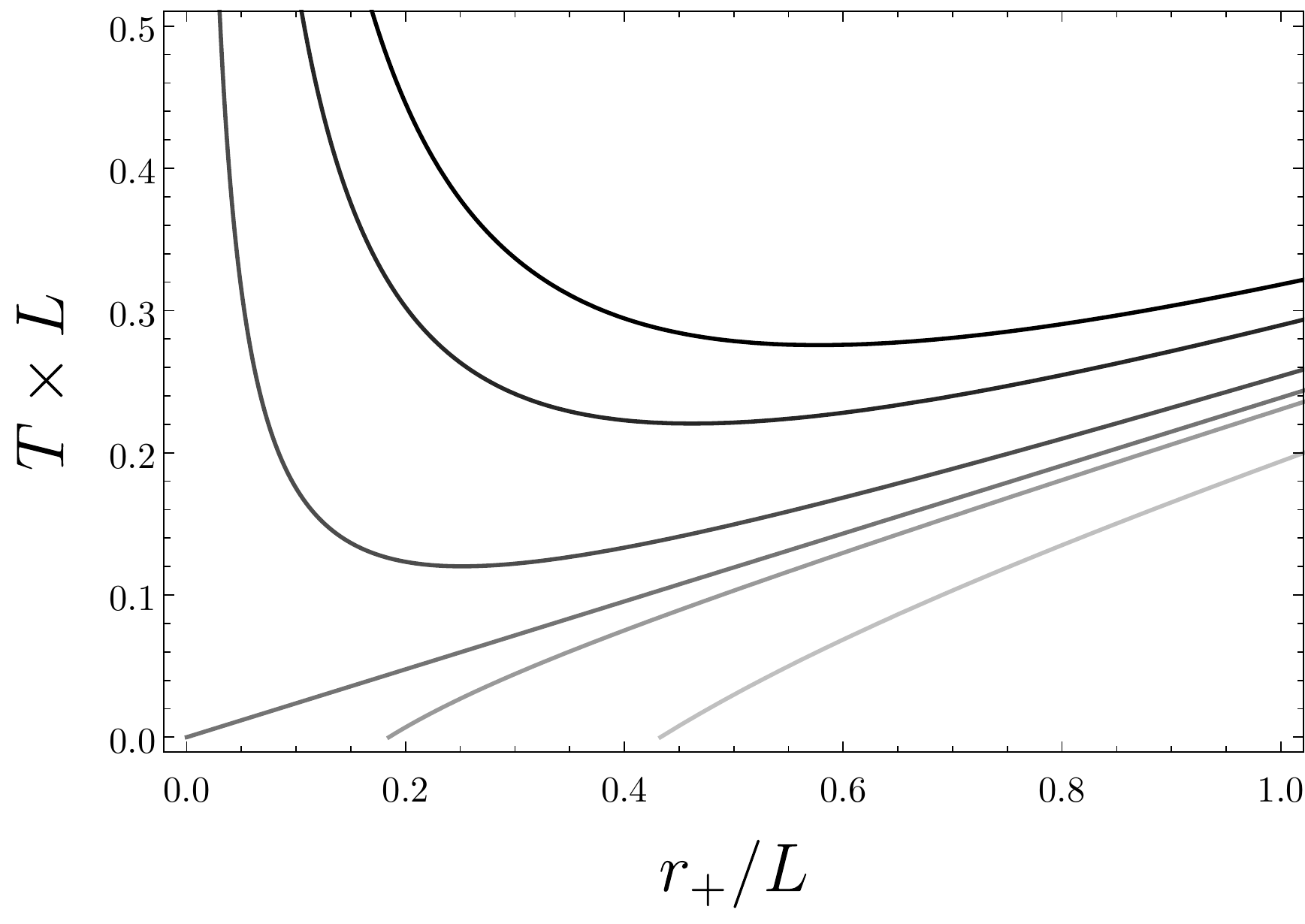}
\quad
\includegraphics[width=0.45\textwidth]{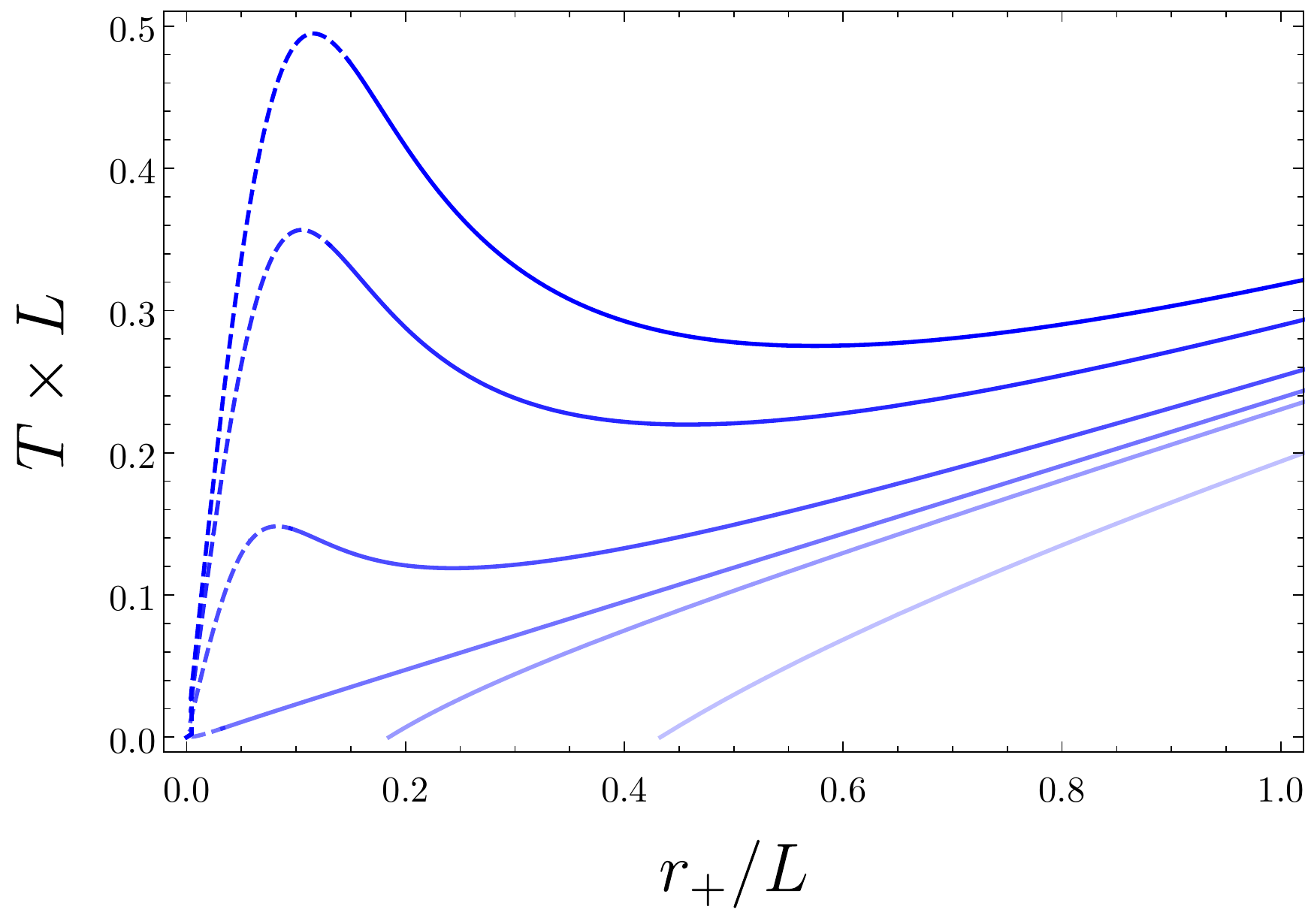}
\quad
\includegraphics[width=0.45\textwidth]{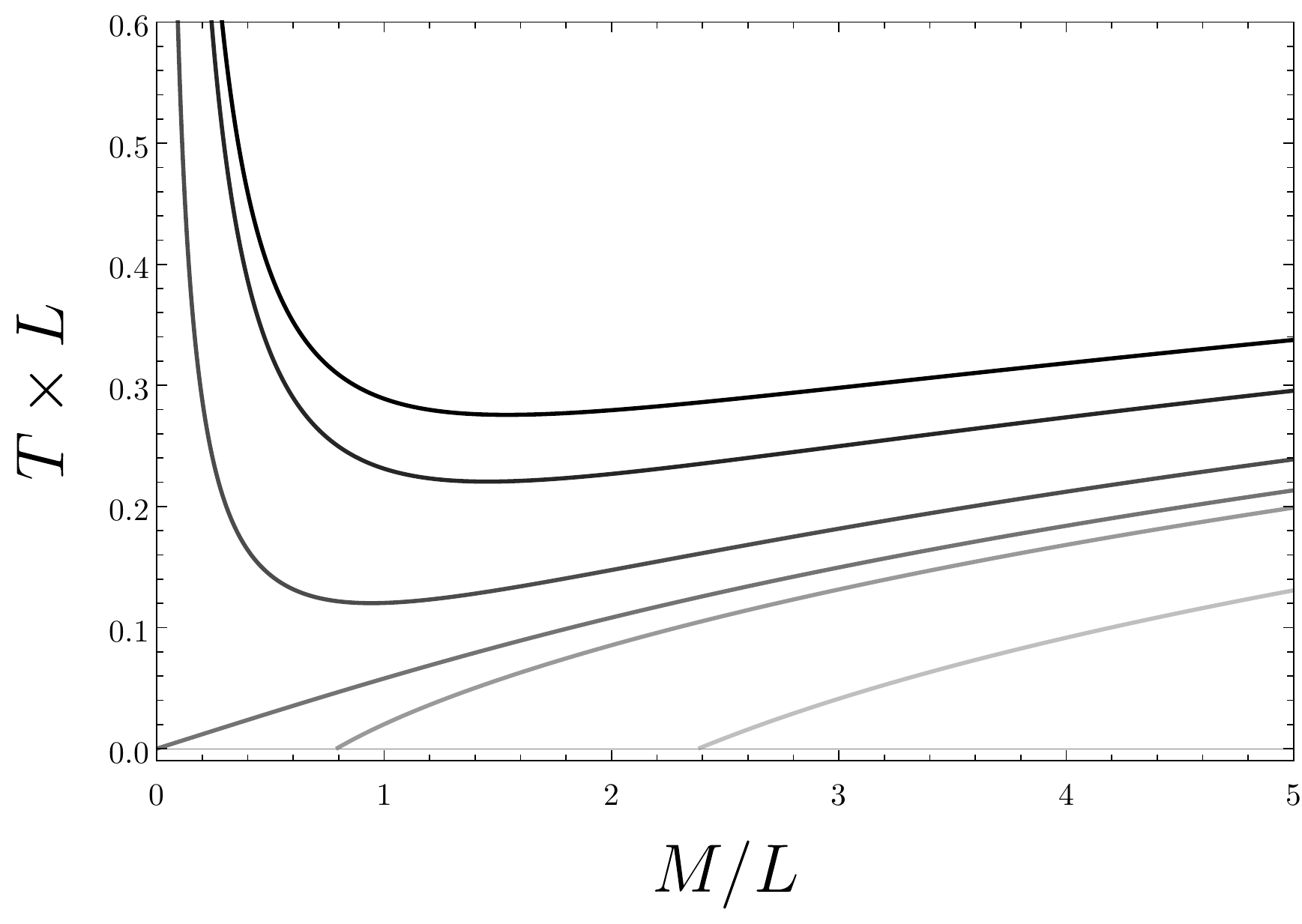}
\quad
\includegraphics[width=0.45\textwidth]{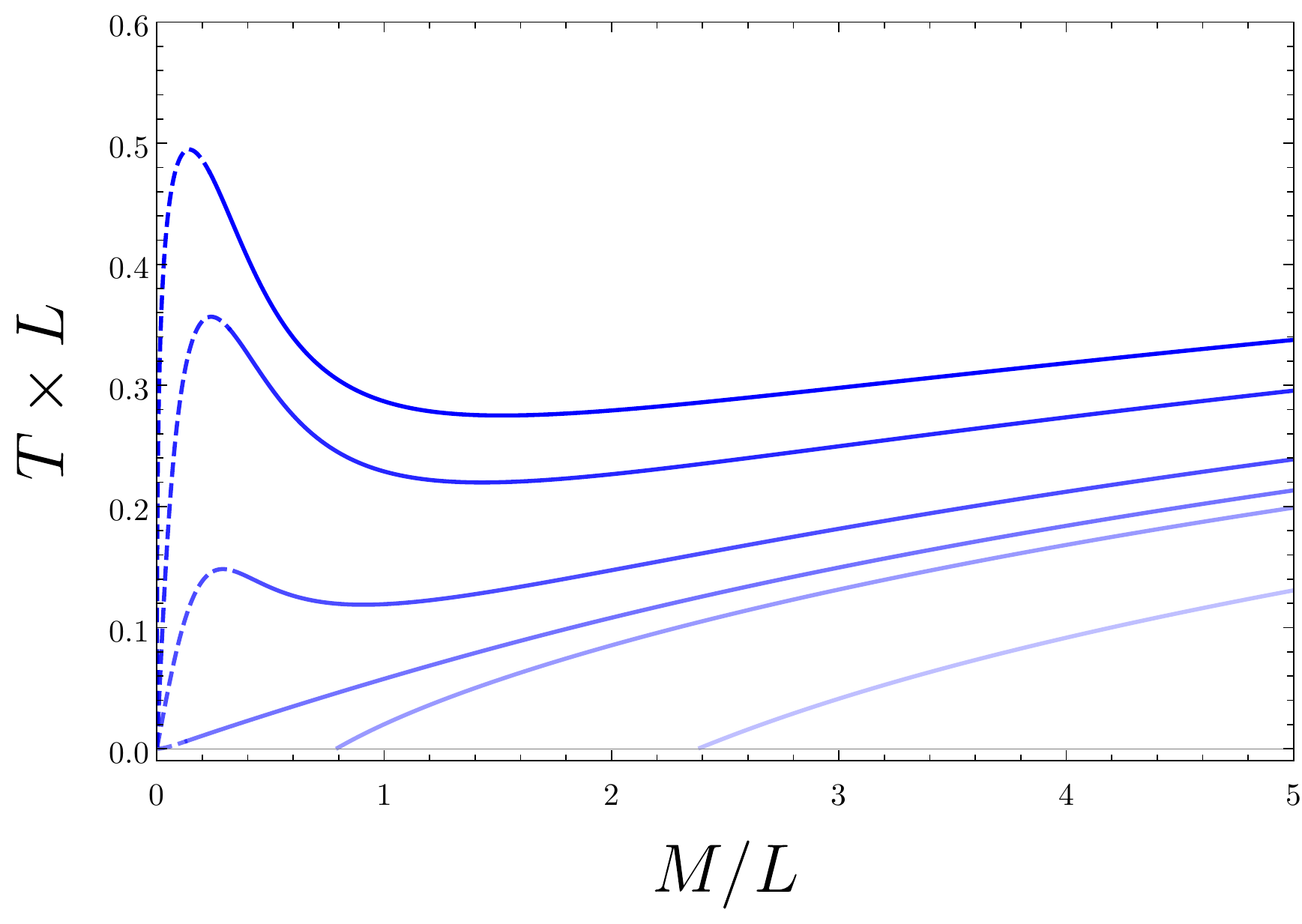}
\caption{{\bf Four-dimensional black holes in the fixed potential ensemble}. {\it Top Row}: Here we display plots of the black hole temperature vs. horizon radius in Einstein gravity (left) and four dimensional generalized quasi-topological gravity (right). The various curves correspond to different values of the potential: $\Phi = 0, 1.2, 1.8, 2, 2.1, 2.5$ from top to bottom (greatest to least opacity). In the right plot, the dashed portion of the curve indicate negative Wald entropy, and the higher-curvature coupling has been set to $\mu/\mu_c = 10^{-3}$. {\it Bottom Row}: Here, for exactly the same parameters, we display the temperature plotted against the mass in Einstein gravity (left) and the cubic theory (right).}
\label{fig:4dPhi_TR}
\end{figure}

To gain a better understanding of the black hole solutions under consideration, it is again helpful to consider plots of the temperature against the horizon radius (and mass), as shown in Figure~\ref{fig:4dPhi_TR}.  First, note that when the electric potential obeys $\Phi^2 < 4$, extremal black holes cannot exist and for both theories the behaviour is qualitatively similar to the uncharged solutions. This means that, for the cubic theory, so long as the potential satisfies this bound the small black holes are thermally stable, opposed to thermally unstable as is the case in Einstein gravity. Let us examine the behaviour of the small black holes in more quantitative detail for the cubic theory.

 The temperature of small black holes is proportional to the horizon radius, rather than inversely proportional:
\beq
T = \sqrt{\frac{4 - \Phi^2}{-96 \pi^2 \mu}} r_+ + {\cal O}(r_+^3) \, .
\eeq
This feature leads to the thermal stability of small black holes with the specific heat taking the following form:
\begin{align}
C_P = \frac{\pi \Phi^2}{\sqrt{-96 \mu \left(4 - \Phi^2 \right)}} +& \frac{\pi r_+^2}{L^2 (4-\Phi^2)^2} \left[L^2(\Phi^6 - 12 \Phi^4 + 128) \right.
	\nn\\
	 &\left.- 2 \sqrt{-96 \mu (4 - \Phi^2)}(\Phi^2 + 8) \right] + {\cal O}( r_+ ^3) \, .
\end{align}
The expression for the specific heat makes manifest the fact that small black holes will always have a positive heat capacity.  Expanding the expression for the mass in the limit of small black holes we see that
\beq
M = \frac{\Phi^2 r_+}{2} + {\cal O}(r_+^3) \, ,
\eeq
indicating that the mass is positive for small black holes and vanishes in the limit that the spacetime does not contain a horizon. However, performing a similar analysis for the entropy we see that
\beq
S = - \pi \sqrt{-6 \mu (4 - \Phi^2)} + {\cal O}(r_+^2) \, .
\eeq
Once again we can add a constant to the entropy to ensure that it is positive as $M \to 0$, but the situation is a bit trickier than in the uncharged case. Here, the limiting value of the entropy cares not only about the coupling $\mu$, but also the value of the electric potential $\Phi$. This is troubling because the methods we introduced in the previous section for shifting the entropy essentially amount to adding a non-dynamical term to the Lagrangian. In this case, if we add precisely the contribution to ensure $S \to 0$ as $M \to 0$, this would require modifying the action in a way that depends on the particular solution. A compromise of sorts can be reached by adding the same constant as in the uncharged case. This would ensure that the entropy is always positive, but would mean that only when $\Phi = 0$ would $S \to 0$ as $M \to  0$, otherwise $S$ would limit to a (positive) constant. Here we will be somewhat agnostic, presenting the results obtained when using the Wald entropy directly, and those obtained when shifting the entropy as just described.

Next let us consider the behaviour of the black holes when $\Phi^2 > 4$, which marks the transition between the two types of behaviour evident in Figure~\ref{fig:4dPhi_TR}.  Recall that, because all instances of $\mu$ in Eqs.~\eqref{eqn:GCE-4d-nheqs} multiply the temperature, the properties of extremal black holes are identical to those in Einstein gravity. In particular, this implies that when $\Phi^2 > 4$, the extremal (and near extremal) black holes will possess positive entropy.  In this case, to see this explicitly, it is helpful to expand the quantities in a small temperature series. We find that the horizon radius goes like
\beq
r_+ = \frac{1}{4} \sqrt{\frac{\Phi^2 - 4}{2 \pi P}} + \frac{T}{4 P} + \sqrt{\frac{\pi}{2}} \left(\frac{-4 - 1536 P^2 \pi^2 \mu + \Phi^2}{4 P^{3/2} (\Phi^2 - 4)^{3/2}} \right) T^2 + {\cal O}(T^3) \, ,
\eeq
which in turn implies the entropy behaves in the following way:
\beq
S = \frac{\Phi^2 - 4}{32 P} +  \sqrt{\frac{\pi P}{2 (\Phi^2 - 4)}} \left(\frac{-4 + 1536 P^2 \pi^2 \mu + \Phi^2}{8 P^2} \right) T + {\cal O}(T^2) \, .
\eeq
Thus we see that the Wald entropy for the near extremal solutions will be positive. We also know that the solutions have positive entropy in the high temperature limit since, in that case, the solutions also reduce to the Einstein gravity results. Thus, provided $\Phi^2 > 4$, only solutions at intermediate temperatures can possess negative entropy, if any do at all.

\begin{figure}[h]
\centering
\includegraphics[width=0.45\textwidth]{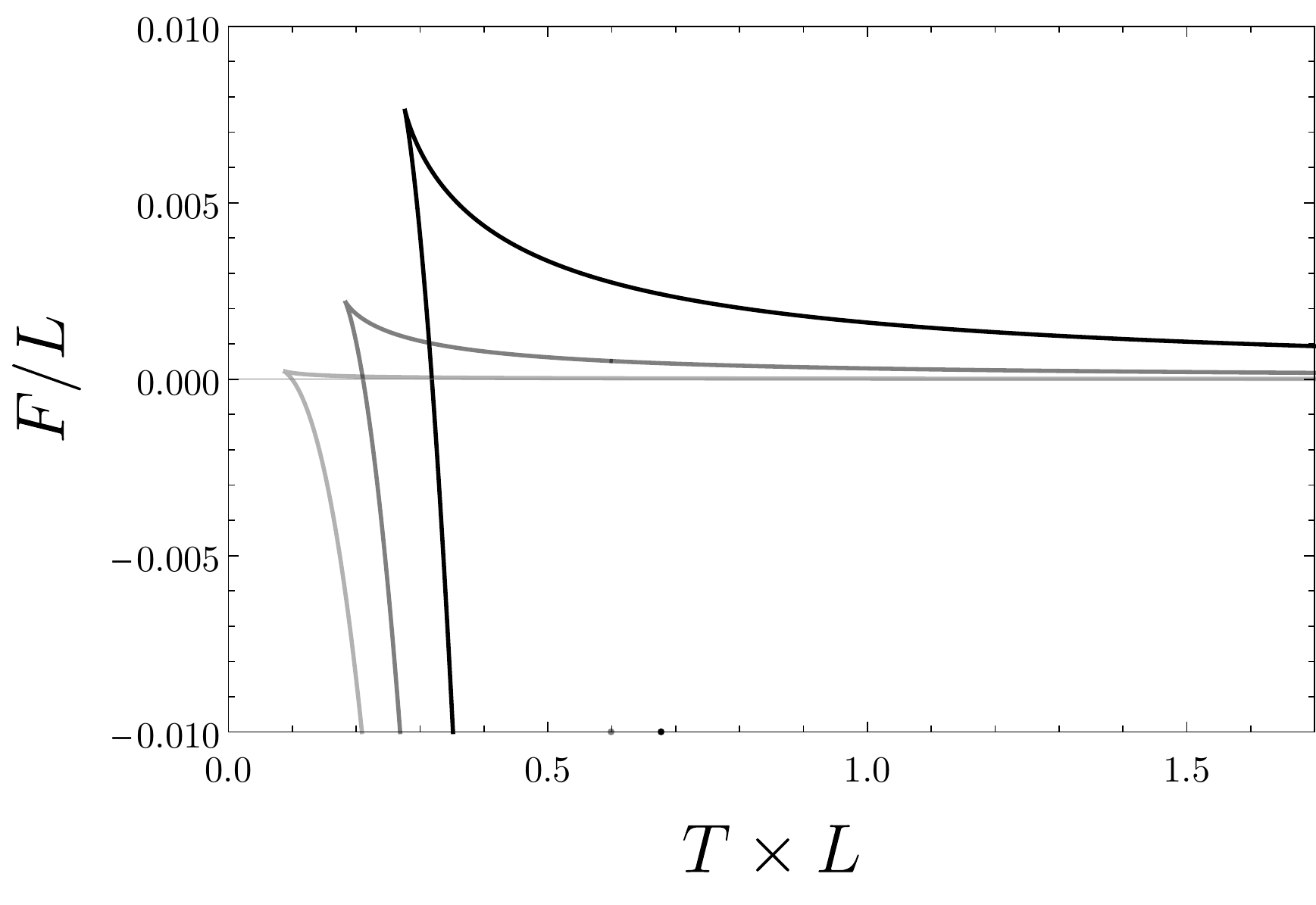}
\quad
\includegraphics[width=0.45\textwidth]{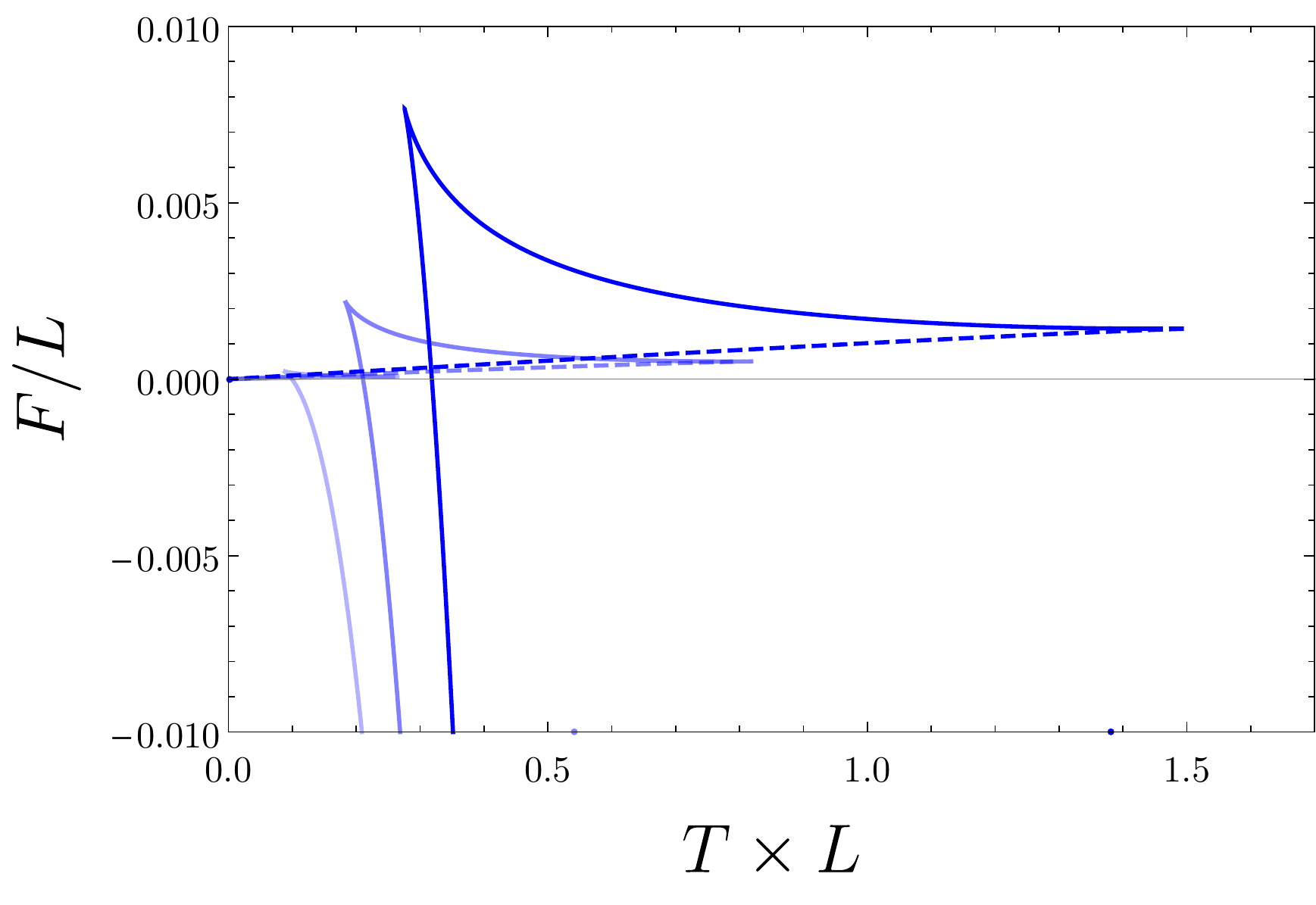}
\quad
\includegraphics[width=0.45\textwidth]{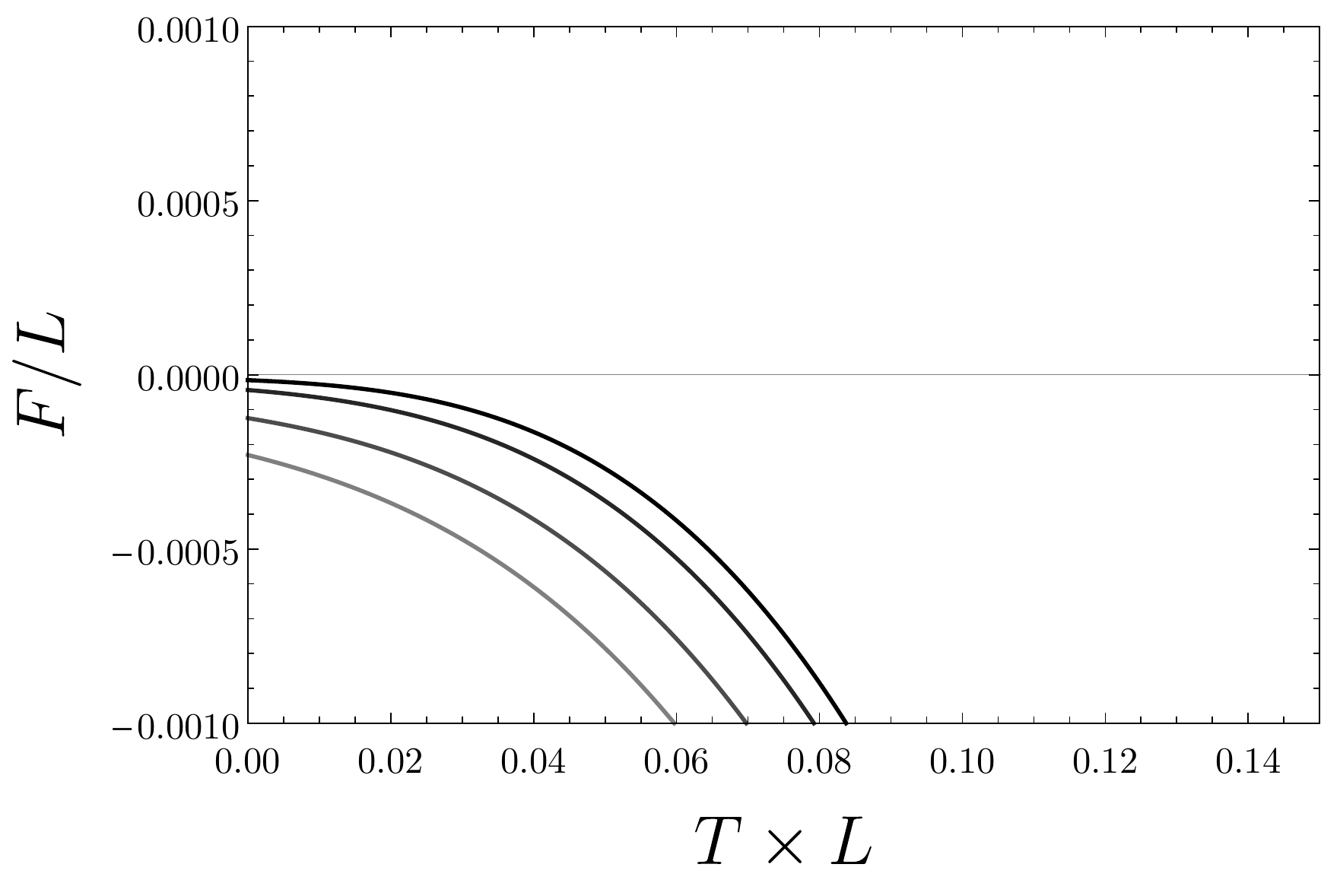}
\quad
\includegraphics[width=0.45\textwidth]{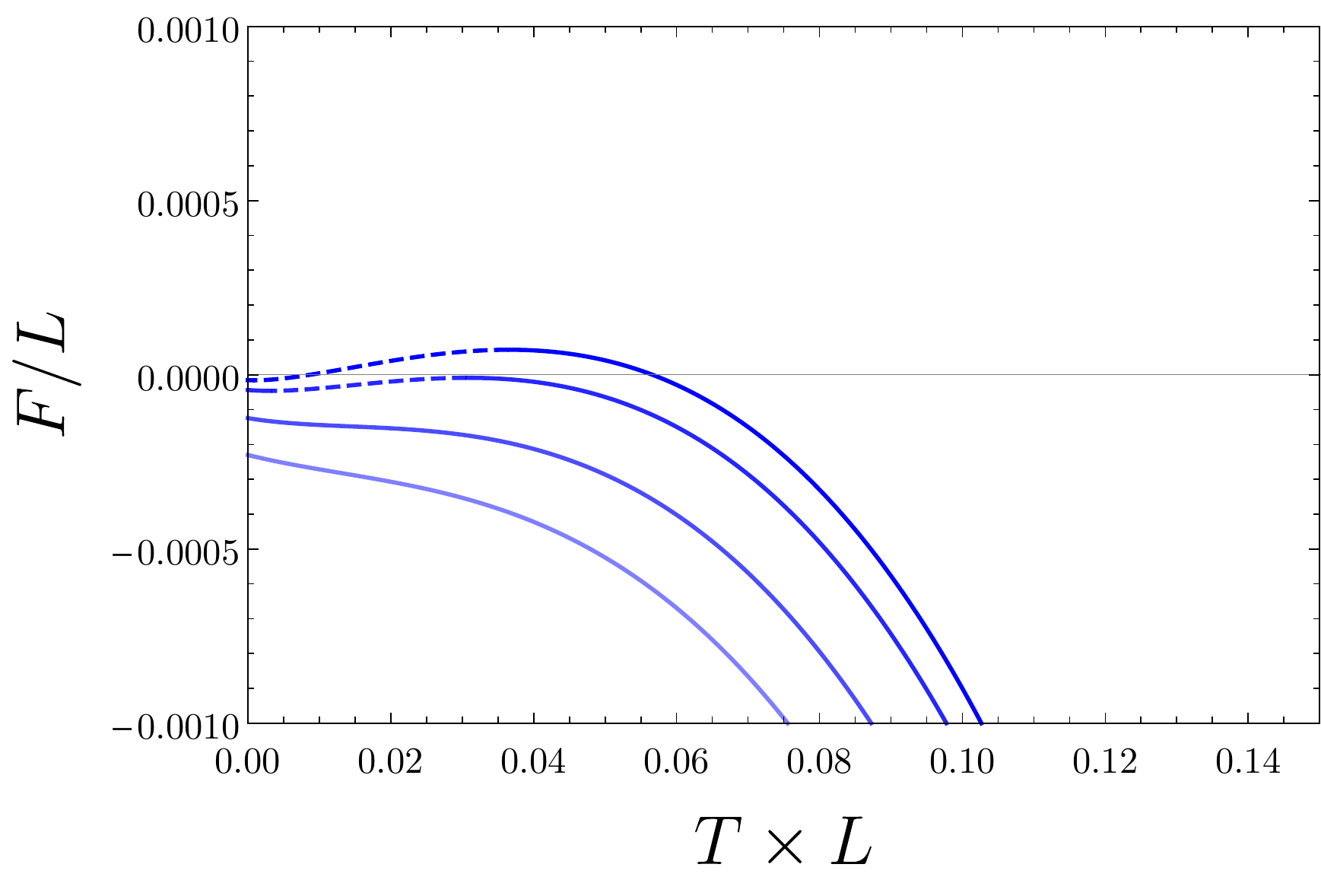}
\quad
\includegraphics[width=0.45\textwidth]{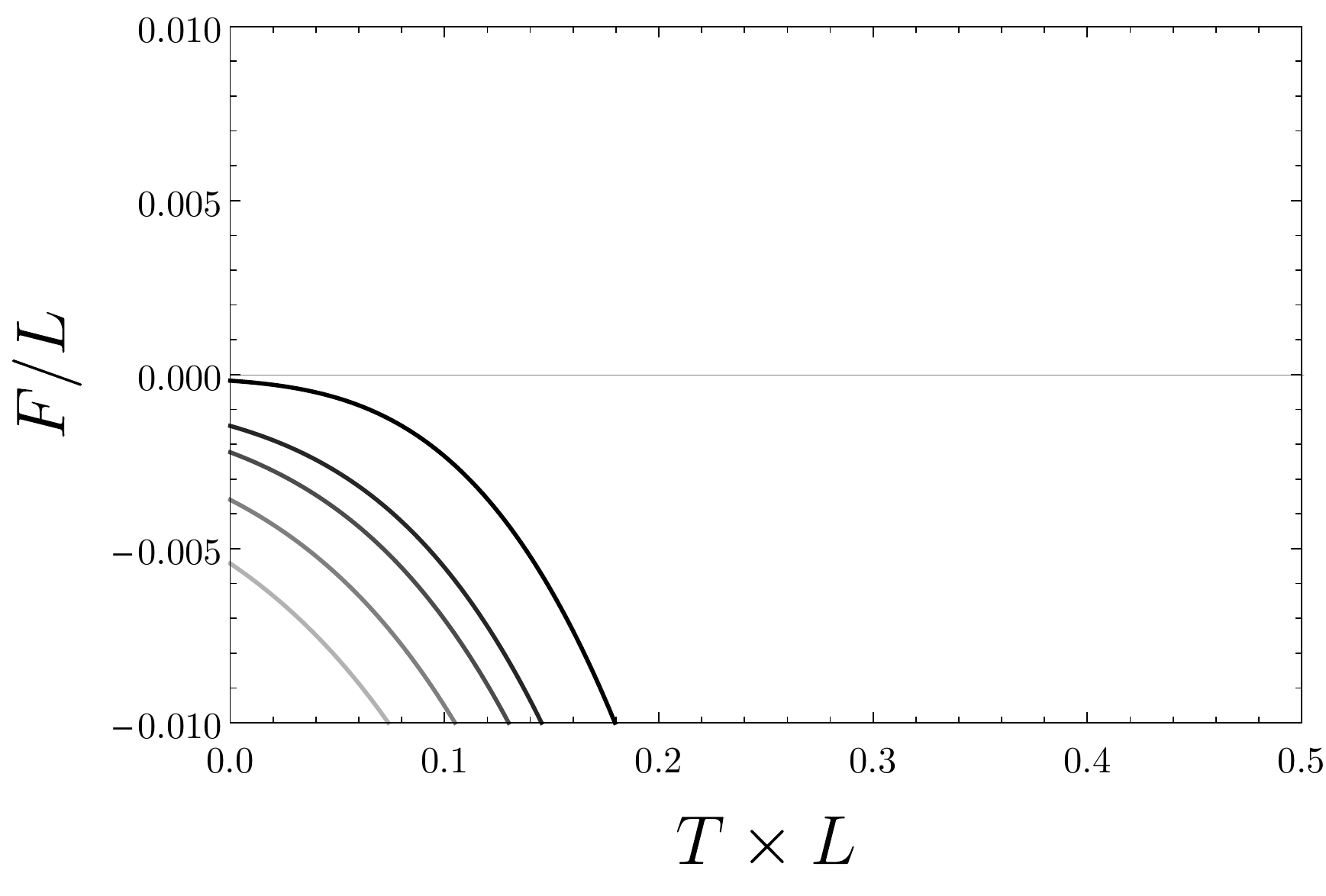}
\quad
\includegraphics[width=0.45\textwidth]{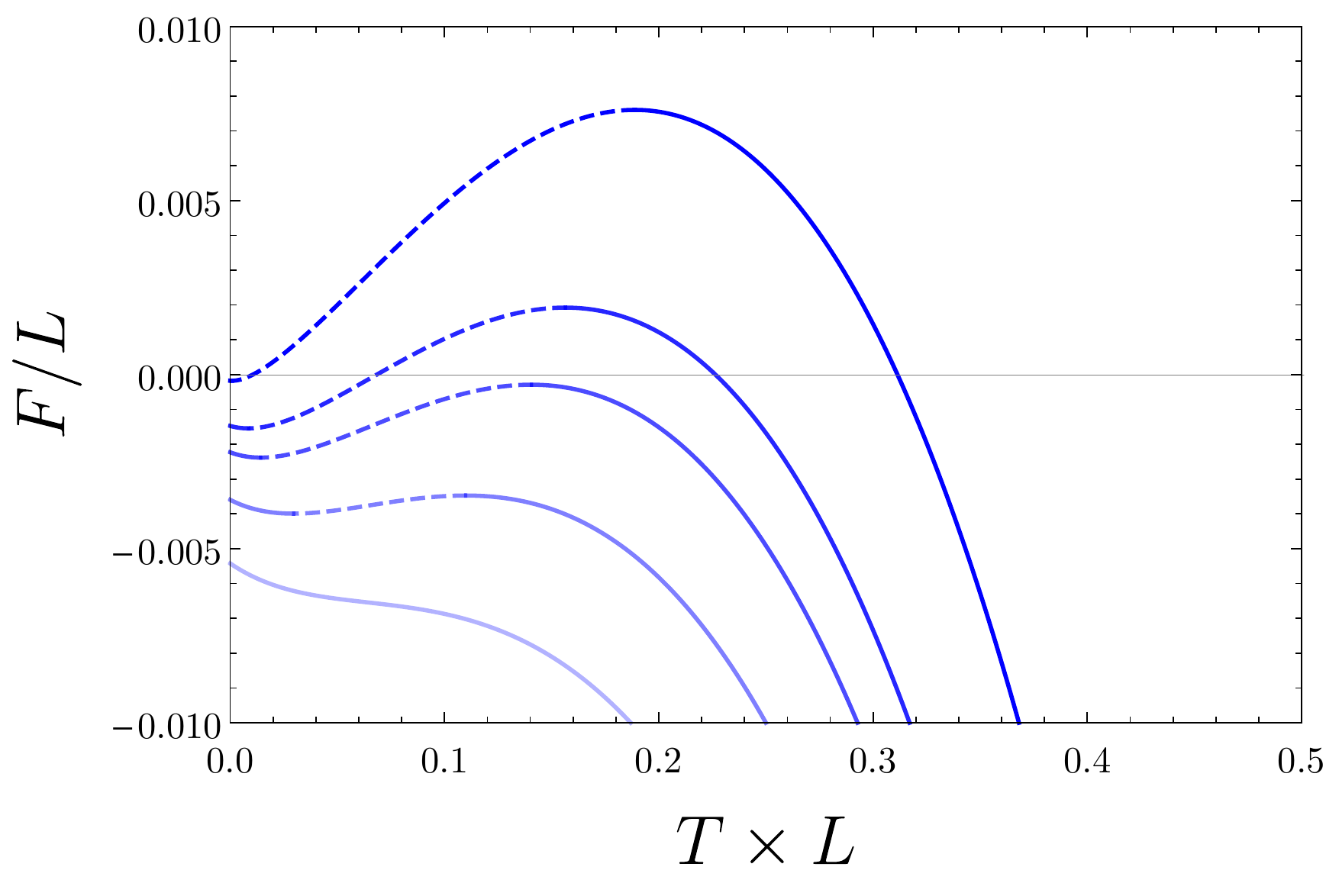}
\caption{{\bf Free energy vs.~temperature for grand canonical ensemble in four dimensions}. {\it Top Row}: Here we plot examples of the free energy for $\Phi = 0, 1.5, 1.9$ (more to less opacity) for Einstein gravity (left) and the cubic theory with $\mu/\mu_c = 10^{-5}$.  {\it Middle Row}: Here we plot the free energy for $\Phi = 2.01, 2.02, 2.04, 2.06$ (more to less opacity) for Einstein gravity (left) and the cubic theory with $\mu/\mu_c  = 100 489/32157432$.  {\it Bottom Row}: Here we plot the free energy for $\Phi = 2.05, 2.2, 2.26, 2.35, 2.45$ (more to less opacity) for Einstein gravity (left) and the cubic theory with $\mu/\mu_c  = 1/2 489/32157432$. In all cases, the dashed portions of the curves indicate negative Wald entropy.}
\label{fig:4d-gce-freeEnergy}
\end{figure}

To understand better the phase structure of the solutions, we display a few relevant free energy curves in Figure~\ref{fig:4d-gce-freeEnergy} taking the entropy to be the Wald entropy.  The top row shows a few relevant examples when the potential satisfies $\Phi^2 < 4$. In this case, the free energy is qualitatively similar to the uncharged results for both Einstein gravity and the cubic theory. In both cases, increasing the electric potential has the effect of decreasing the temperature at which the free energy crosses zero, i.e. the temperature at which the Hawking-Page transition occurs is reduced. In the case of the cubic theory, increasing the potential has the additional effect of shrinking the swallowtail, and so can push the system toward a critical point. The middle and bottom row show examples of what occurs when $\Phi^2 > 4$. In this circumstance, there is only a single branch of solutions in both the Einstein gravity case and the cubic case. In the Einstein gravity case, the free energy of the charged solutions is now always less than zero. This means that the charged black hole always makes the dominant contribution to the partition function provided $\Phi^2 > 4$.

However, in the cubic case the situation is more subtle. Using the Wald entropy, as done in Figure~\ref{fig:4d-gce-freeEnergy}, the interpretation would be the following: For $\Phi^2 < 4$, the situation would be qualitatively similar to that of Einstein gravity, with thermal AdS at fixed potential dominating the partition function at low temperatures, and a large AdS black hole at higher temperatures.  The situation is also similar to Einstein gravity provided $\Phi^2$ is much larger than $4$: then there is a single branch of black holes, always with positive entropy, and with free energy always less than zero --- the dominant contribution to the partition function is a black hole for all values of temperature.  The real differences emerge for $\Phi^2 > 4$, but close to $4$. Here, at low temperatures, the thermodynamically preferred phase is a black hole. As the temperature is increased a region of parameter space is entered where the entropy is negative; if these black holes are considered unphysical, at this point there would be a zeroth-order phase transition to thermal AdS space. As the temperature is further increased, there comes a point where the entropy is positive and the free energy dips below zero again --- at this point there will be a Hawking-Page transition between thermal AdS and the black hole. However, in this case, the Hawking-Page transition need not be first order, but can in fact be a second order transition, akin to those that occur at a critical point. The reason this can happen here is because of the fact that the free energy is ``peaked'' --- if the peak of the free energy occurs exactly when $F = 0$, then both $F$ and its first derivative vanish at that point and hence the transition will be of second order.

To summarize, from this perspective, the interpretation of the thermodynamics would be the following: there is a zeroth-order black hole/thermal AdS transition, followed up a first- or second-order thermal AdS/black hole phase transition as the temperature is monotonically increased. In other words, there is an intermediate regime of $\Phi$ for which we have a re-entrant Hawking-Page transition for the fixed potential ensemble.

\begin{figure}[H]
\centering
\includegraphics[width=0.45\textwidth]{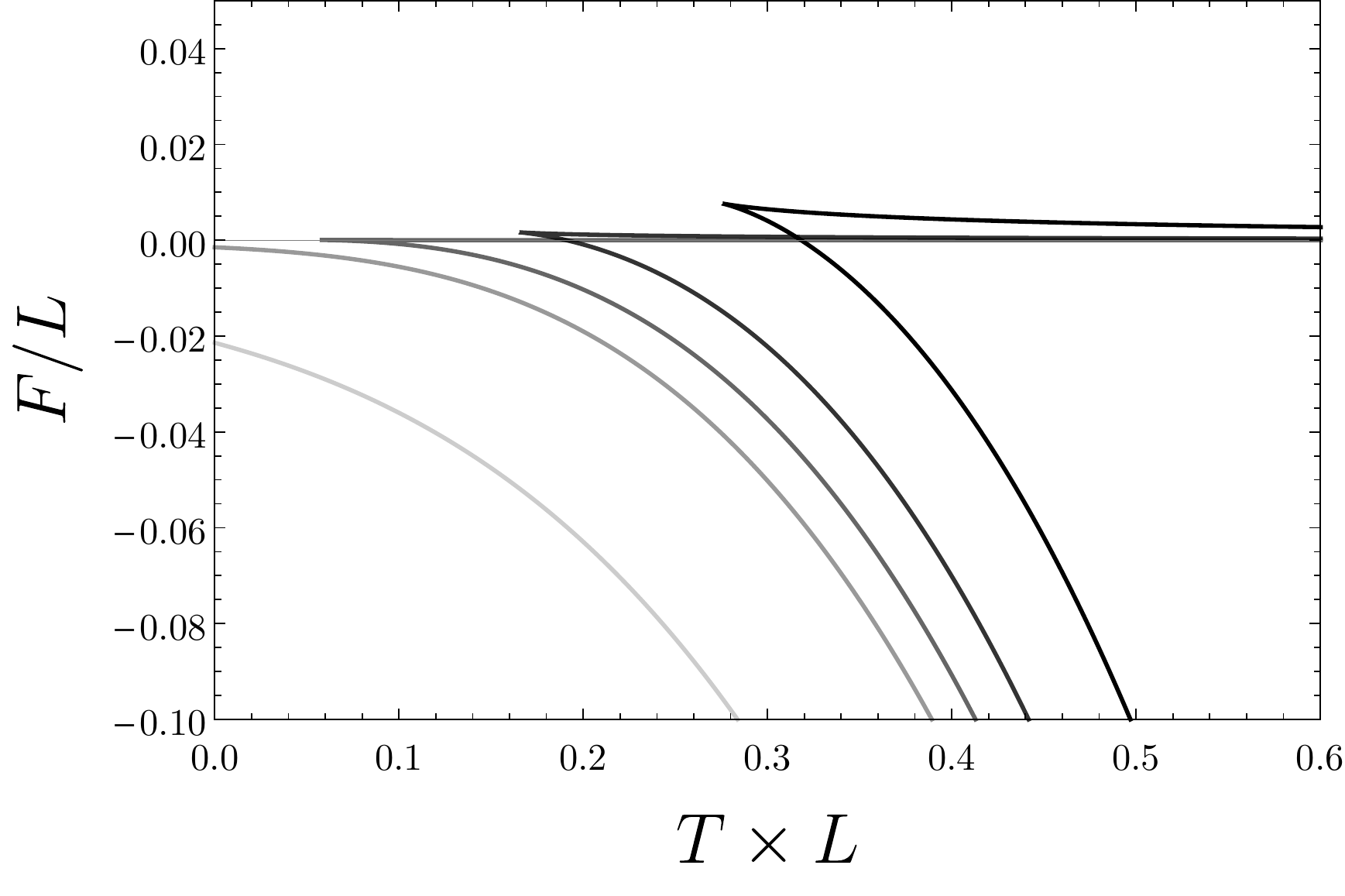}
\quad
\includegraphics[width=0.45\textwidth]{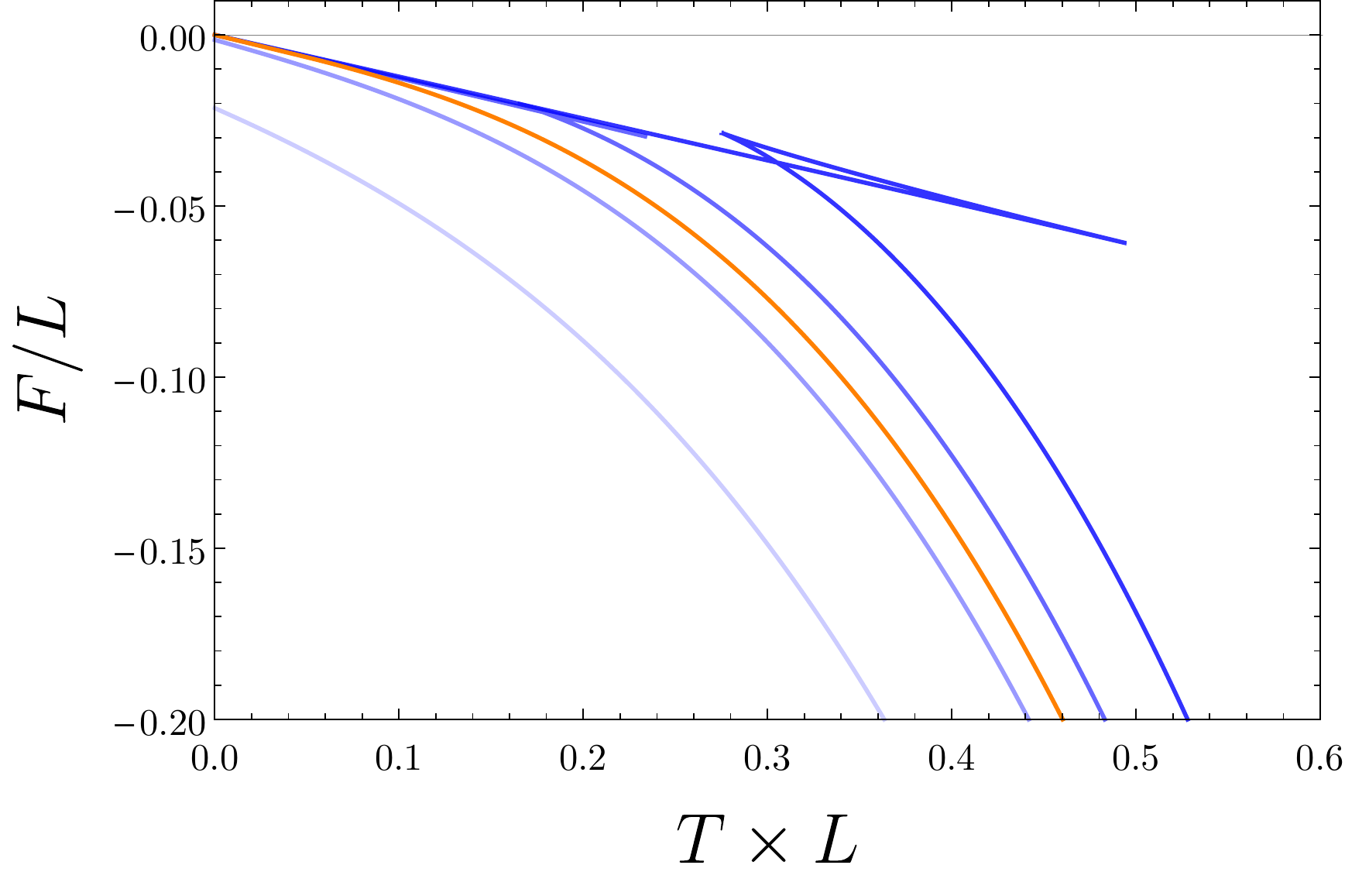}
\caption{{\bf Free energy vs. temperature in four dimensions using shifted entropy}. Here we plot the free energy vs. temperature for $\Phi = 0, 1.6, 1.957, 2.2, 3$ (more to less opacity) for Einstein gravity (left) and the cubic theory with $\mu/\mu_c = 10^{-4}$ (right). In this plot, the entropy has been shifted by the constant $S = S_{\rm Wald} + 2 \pi \sqrt{-6 \mu}$ which ensures that the entropy is always positive. The orange curve on the right corresponds to $\Phi = 1.957$, which results in a critical point. }
\label{fig:4d-gce-freeEnergyPos}
\end{figure}

For completeness, let us also discuss the interpretation of the thermodynamics using the shifted Wald entropy that is always positive. For this case, representative free energy diagrams are shown in Figure~\ref{fig:4d-gce-freeEnergyPos}.  Here the interpretation is a bit different. In this case, regardless of the value of $\Phi$ the free energy of the black holes in the cubic theory is always negative and thermal AdS is never the thermodynamically favoured solution. For $\Phi^2 < 4$, the free energy possesses three branches, and there is a first-order small/large black hole phase transition. As the value of $\Phi$ is increased (or, equivalently, as the pressure is increased at constant $\Phi$), the swallowtail shrinks, eventually terminating at a critical point. In other words, the free energy displays standard van der Waals behaviour. When $\Phi^2$ is larger than the critical value (to be discussed explicitly below), there is only a single branch of black holes, and these are thermodynamically favoured at all temperatures.

Let us discuss the critical point in more detail.   It is easy to check that  \eqref{eqn:gce_eos_4d} admits a critical point with the values
\begin{align}
T_c &= \frac{\sqrt{2} }{12 \pi } \left(\frac{3 (4  - \Phi^2)^3}{-48 \mu } \right)^{1/4}
	\, ,\quad
	v_c = \frac{\sqrt{2}}{3} \left(3^3 (-48) \mu  (4  -\Phi^2) \right)^{1/4}
	\,, \nn\\
	P_c &= \frac{1}{32 \pi} \sqrt{\frac{3 ( 4 - \Phi^2)}{ -48 \mu  }} \, .
	\label{eqn:critPoint}
\end{align}
Note that these expressions for the critical values are valid in the case where $P$ is constant as well as when $P$ is a thermodynamic variable.  In the former case, the expression for the critical pressure can be solved to obtain the value of $\Phi$ that yields a critical point for a given fixed pressure. In the latter case, specifying a value of $\Phi$ then gives a critical pressure, as is the standard in black hole chemistry. The critical values satisfy the following universal relationship,
\beq
\frac{P_c v_c}{T_c} = \frac{3}{8} \, ,
\eeq
which is identical to the van der Waals ratio \cite{Kubiznak:2012wp}, and is the same in both the canonical and grand canonical ensembles (see below). It is worth noting that critical points only exist for a range of potentials: if $\Phi^2 > 4$, then there is no critical point.

To determine the critical exponents, we expand the equation of state near the critical point in terms of the dimensionless variables $\rho, \tau$ and $\phi$ defined by
\beq
P = P_c(\rho + 1) \, , \quad T = T_c(\tau + 1) \, , \quad v = v_c \left(\phi + 1 \right) \, .
\eeq
This yields
\beq
\rho = \frac{10}{3} \tau - \frac{16}{3} \phi \tau + \frac{1}{3}\tau^2 - \frac{4}{3} \phi^3 - \frac{4}{3} \tau^2 \phi + \frac{28}{3} \phi^2 \tau + \cdots \, .
\eeq
Using well-established techniques~\cite{GunasekaranEtal:2012} the critical exponents can be read off from this expansions and are given by the mean field theory values:
\beq
\label{eqn:mft}
\alpha = 0 \, , \quad \beta = \frac{1}{2} \, , \quad  \gamma = 1 \, , \quad \delta = 3 \, .
\eeq

\subsection{Five dimensions}

Let us now consider the differences that arise when considering the grand canonical ensemble in five dimensions. In five dimensions, the near-horizon equations take the following form:
\begin{align}
M &= \frac{8 r_+^2}{3 \pi} \left(k + \frac{r_+^2}{L^2} \right) + \frac{8 r_+^2 \Phi^2}{9 \pi} + \frac{\mu}{474 \pi r_+^2} \left[-976 k^3 + 8960 \pi^2 r_+^2 T^2 \left(3k + 4 \pi r_+ T \right) \right] \, ,
	\nn\\
0 &= 2 r_+ \left(k + \frac{r_+^2}{L^2}\right) + r_+^2 \left( \frac{2 r_+}{L^2} - 4 \pi T \right) - \frac{2 r_+ \Phi^2}{3} 	
	\nn\\
	&- \frac{\mu}{1264 r_+^3} \left[-1952 k^3 + 1728 k^2 \pi r_+ T + 17920 \pi^3 r_+^3 T^3 \right]	
\end{align}
In this case we would identify the pressure as $P = 3/(4 \pi L^2)$ and the specific volume as $v = 4 r_+/3$, leading to the equation of state
\beq
P = \frac{T}{v} - \frac{2 k}{3 \pi v^2} + \frac{2 \Phi^2}{9 \pi v^2} - \frac{31232 k^3 \mu }{19197 \pi v^6}  + \frac{256 k^2 T \mu}{237 v^5} + \frac{4480 \pi^2 T^3 \mu }{711 v^3} \, .
\eeq
Once again, from this point we will set $k=1$ to focus on the spherical black holes.

In the four-dimensional case, we saw that all instances of the cubic coupling multiply powers of the temperature in the near horizon equation. This led to the interesting result that the extremal black holes in the cubic theory are the same as in Einstein gravity. This property is no longer true in five dimensions. It is easy to see that even in the uncharged case extremal solutions can exist --- see, for example, those branches of solutions that intersect $T=0$ at finite $r_+$ in Figure~\ref{fig:5d-HP-properties}. The condition for the existence of extremal solutions is a solution of the following equation:
\beq
0 = 2 r_+\left(1 - \frac{\Phi^2}{3} \right)  + \frac{4r_+^3}{L^2}  	+ \frac{981\mu}{632 r_+^3} \, .
\eeq
It is obvious that this has solutions for $\Phi = 0$ as well as for non-zero $\Phi$ (recall that $\mu < 0$ for the existence of positive mass solutions). In this four-dimensional case, the distinct types of thermodynamic behaviour corresponded to whether or not extremal black holes existed. While that is still true here for the Einstein case, it is no longer the case for the cubic theory where things now become more interesting.

\begin{figure}[h]
\centering
\includegraphics[width=0.45\textwidth]{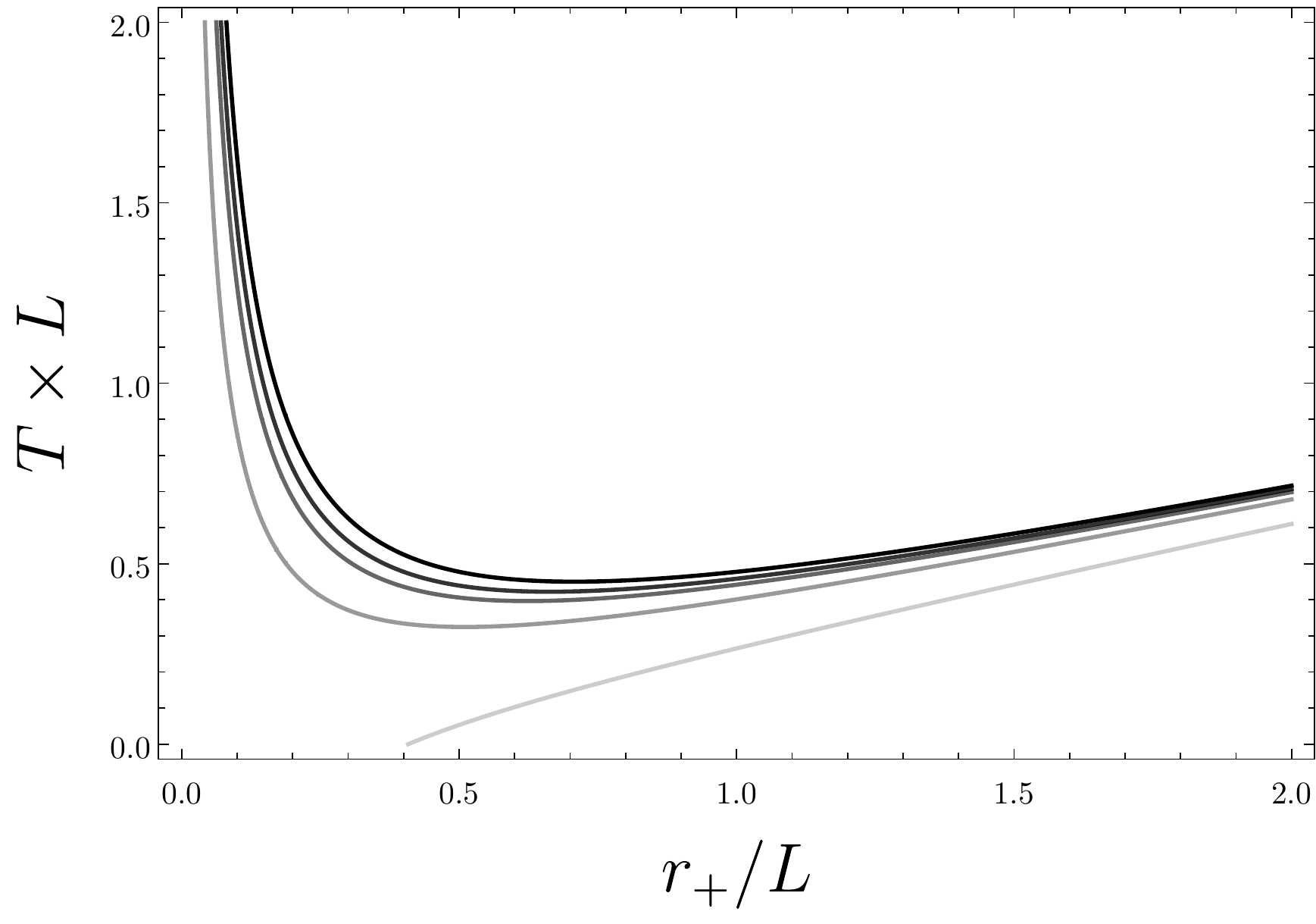}
\quad
\includegraphics[width=0.45\textwidth]{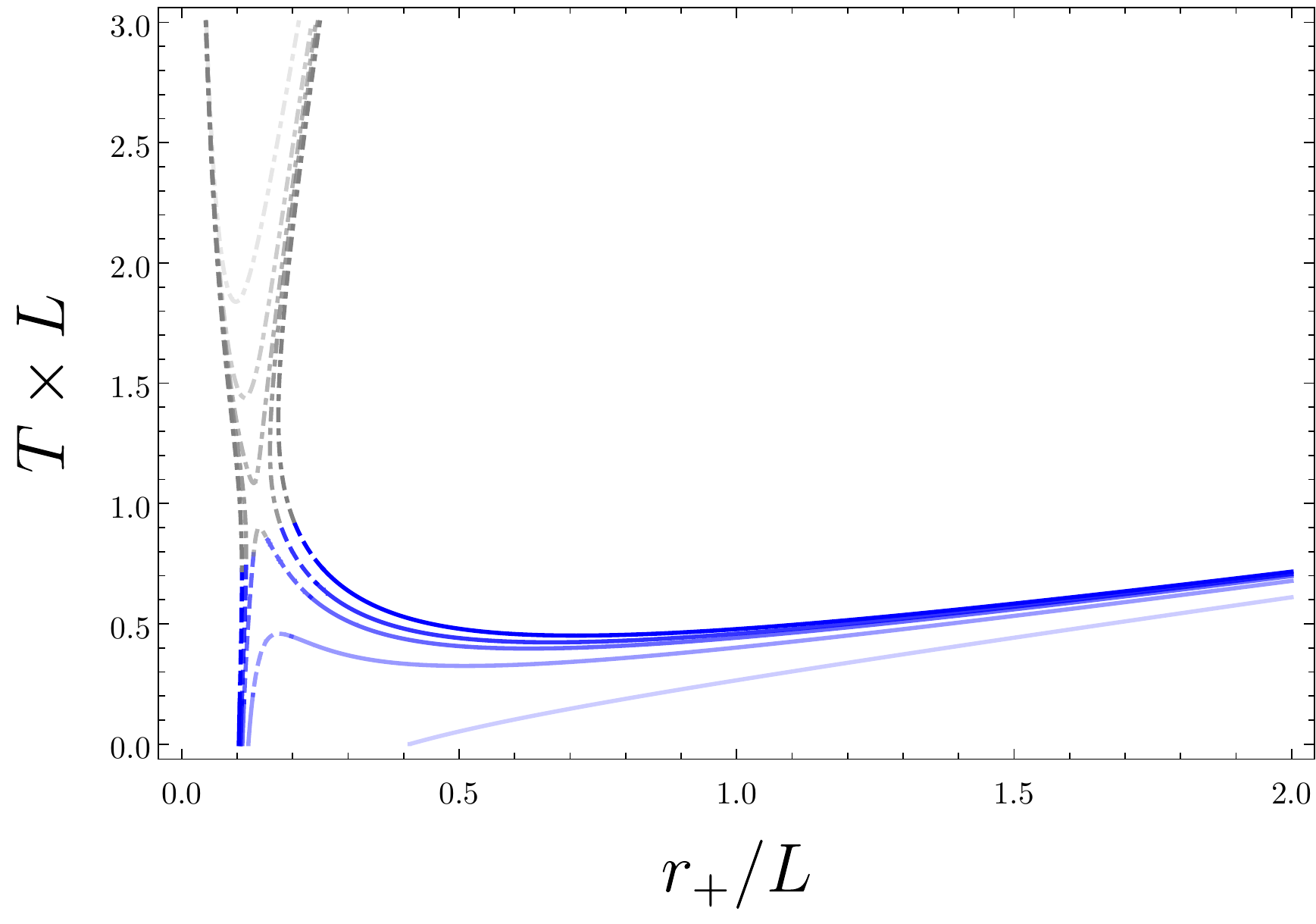}
\quad
\includegraphics[width=0.45\textwidth]{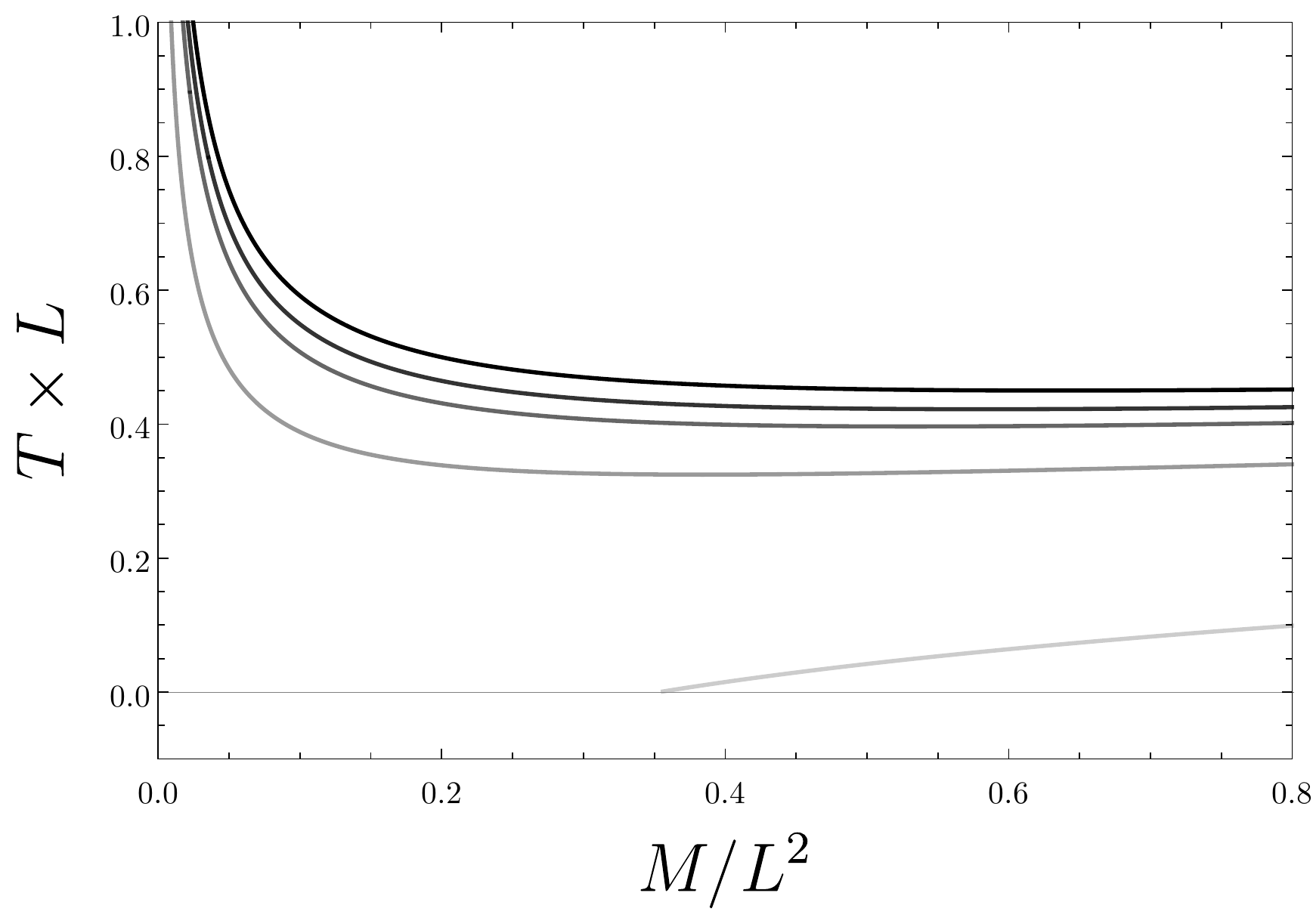}
\quad
\includegraphics[width=0.45\textwidth]{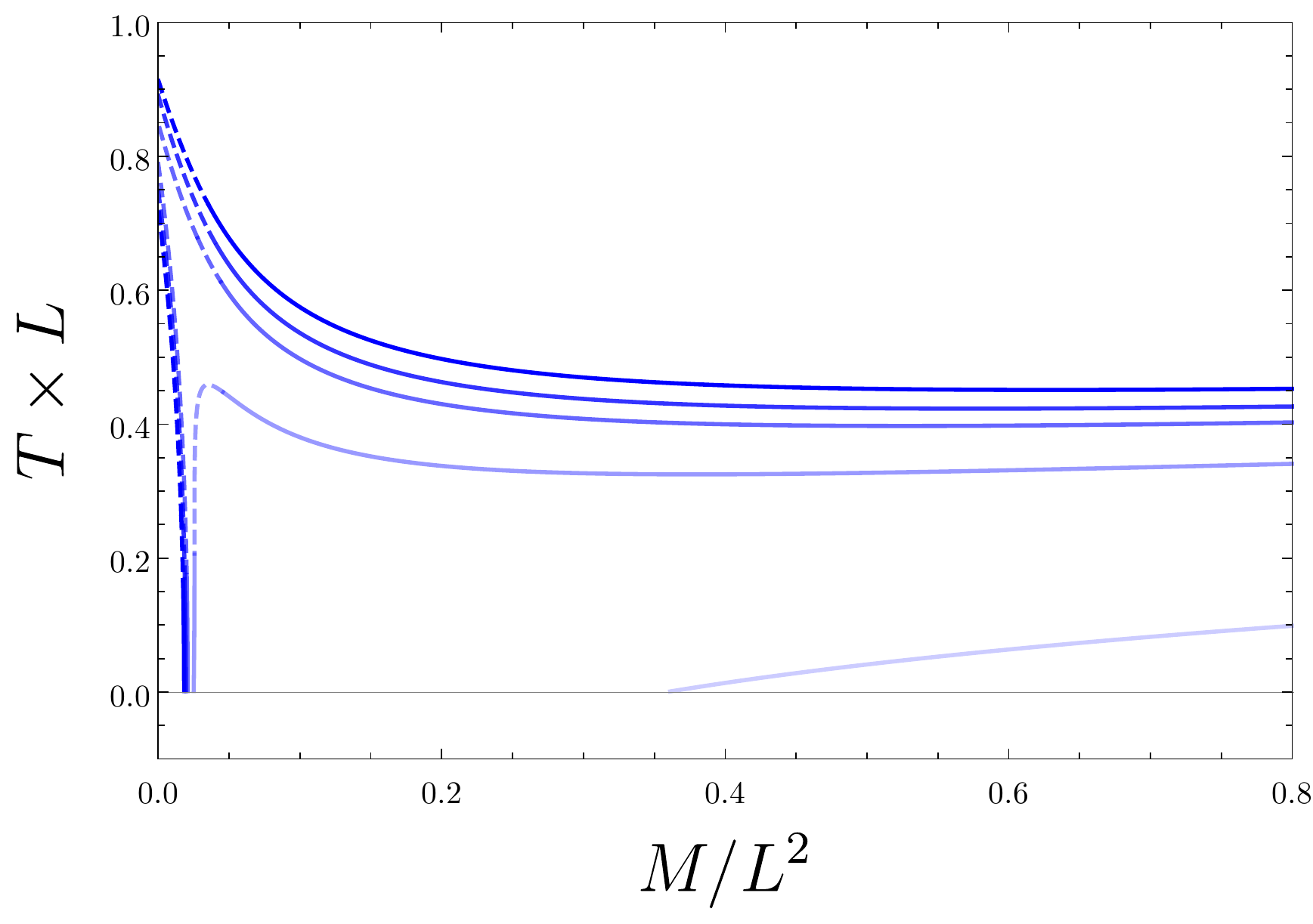}
\caption{{\bf Five-dimensional black hole properties in grand canonical ensemble I}. {\it Top Left}: A plot of the temperature vs.~horizon radius in Einstein gravity for $\Phi = 0, 0.6, 0.82, 1.2, 2.0$ in order of decreasing opacity (or top to bottom through a vertical slice of the plot). {\it Top Right}: Temperature vs.~horizon radius in the cubic theory with $\mu/\mu_c = 10^{-3}$ for the same values of the potential. In this case, the dashed blue lines indicate negative Wald entropy, while the dot-dashed grey lines indicate negative mass (and hence the full solution does not exist). {\it Bottom Left}: The same situation as the top left plot, but now we replace the horizon radius with the mass. {\it Bottom Right}: The same situation as the top right plot, but now we replace the horizon radius with the mass.  }
\label{fig:TR-5d-GC}
\end{figure}

To gain a better understanding of the situation, we once again consider plots of the temperature against the horizon radius for fixed values of the potential and the AdS radius. As in the uncharged case, we divide our study into two parts: first for $\mu \in (0, -79 L^4/1890)$ and then for $\mu \in (-79 L^4/1890, \mu_c)$. For the five dimensional case, the first plots are shown in Figure~\ref{fig:TR-5d-GC}.  The behaviour in the Einstein gravity case (shown on the left) is qualitatively similar to the four-dimensional analysis: For $\Phi^2 < 3$ the structure of the curves is qualitatively identical to the uncharged solutions with up to two black holes at a given temperature, while for $\Phi^2 > 3$ there is only ever a single black hole. In the cubic case (shown on the right) the situation is  quite different. For small values of $\Phi$, the structure of the curves is again qualitatively similar to the uncharged case ---  namely, there are two disconnected branches of the temperature. However, as the value of $\Phi$ is increased there is a point where there is a significant change in the structure of the curves. There are still two disconnected branches of the temperature, but one now consists of purely negative mass black holes (see the curves in the upper left of the plot), while the other somewhat resembles the profiles shown in Figure~\ref{fig:HP-4d} --- for a given temperature there can be up to three black holes. As the value of $\Phi$ is further increased, the hump on this curve flattens out, and the profile resembles that of Einstein gravity for $\Phi^2 > 3$. For large enough $\Phi$, the entropy and mass will be positive along the entire curve.

\begin{figure}[h]
\centering
\includegraphics[scale=0.65]{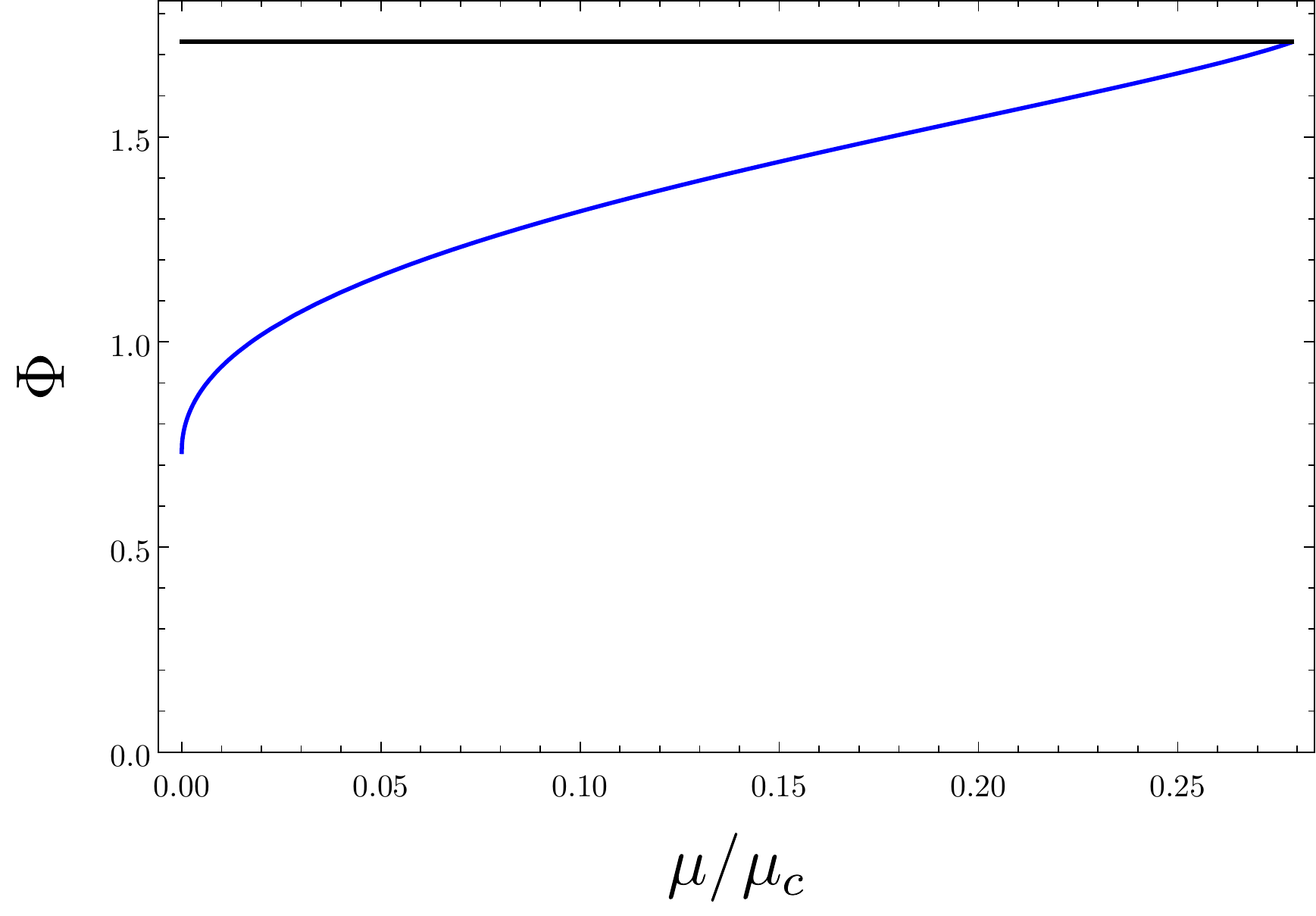}
\caption{{\bf Special values of potential in five dimensions}.  Here we show a plot of the value of potential at which the structure of the $T-r_+$ profiles change from resembling the uncharged case to exhibiting a closed curve. The black curve represents the value at which this occurs in Einstein gravity $\Phi = \sqrt{3}$. Note that the limit between the two cases is discontinuous. The blue curve terminates at $\mu = 79 \mu_c /280 = -79 L^4/1890$, since after this point the theory does not admit large black hole solutions.
}
\label{fig:PhiSpecial}
\end{figure}

The value of $\Phi$ for which the dramatic change just described occurs depends on the value of cubic coupling.  Its precise value can be determined in the following way. Note that when $\Phi$ is less than this value, the temperature vs.~horizon radius profiles have only a single extremum. However, just above this value, the profiles have three extrema (one corresponding to the negative mass branch, and two corresponding to the other branch). Determining when the number of extrema jumps provides a way to determine this value of $\Phi$. In practice, this means solving a complicated polynomial equation, and so here we simply provide a plot of the result in Figure~\ref{fig:PhiSpecial}.  Note that the coupling only runs to $\mu = - 79 L^4/1890 = 79/280 \mu_c$, since beyond this point the branch of large black holes ceases to exist.

\begin{figure}[h]
\centering
\includegraphics[width=0.45\textwidth]{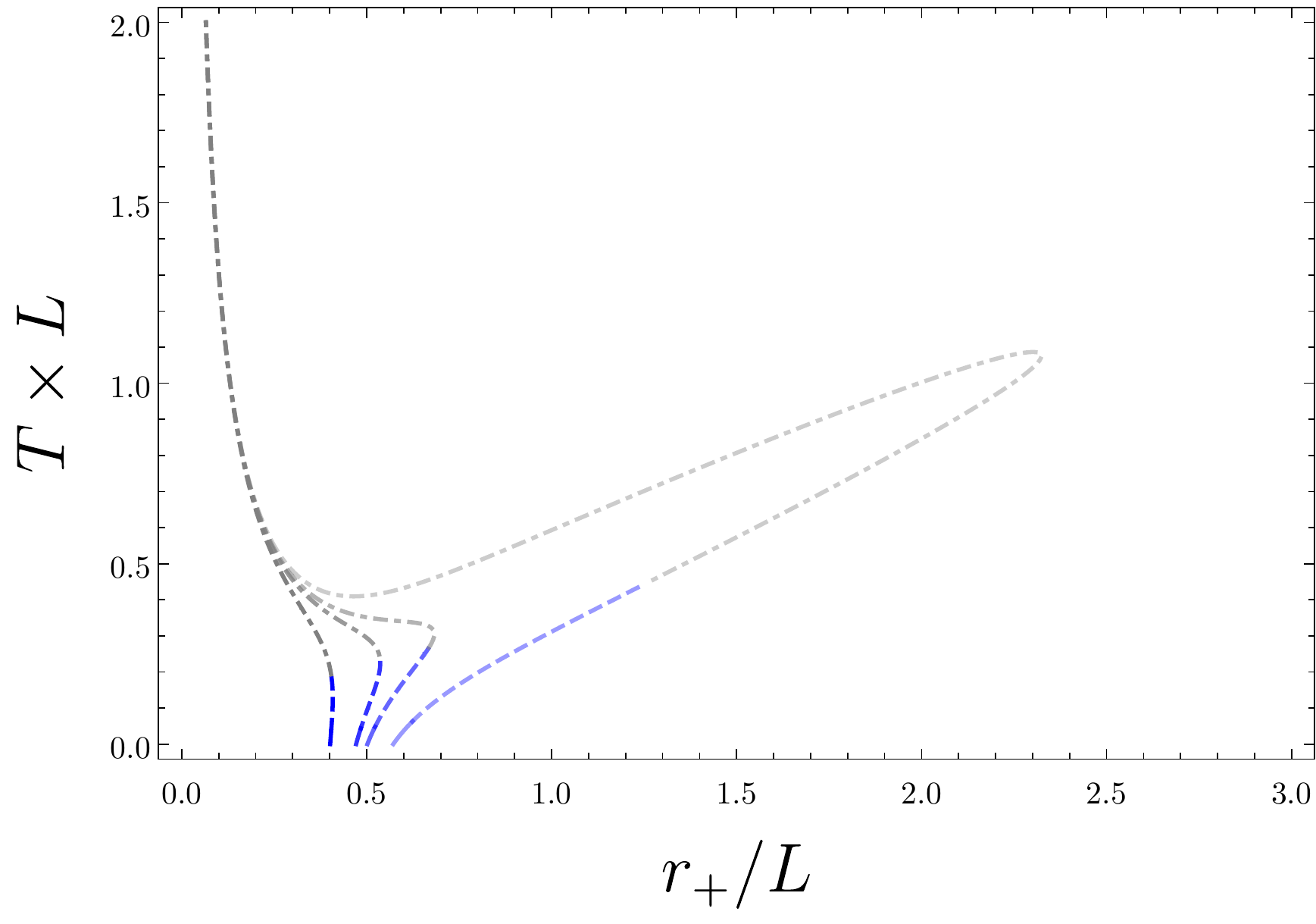}
\quad
\includegraphics[width=0.45\textwidth]{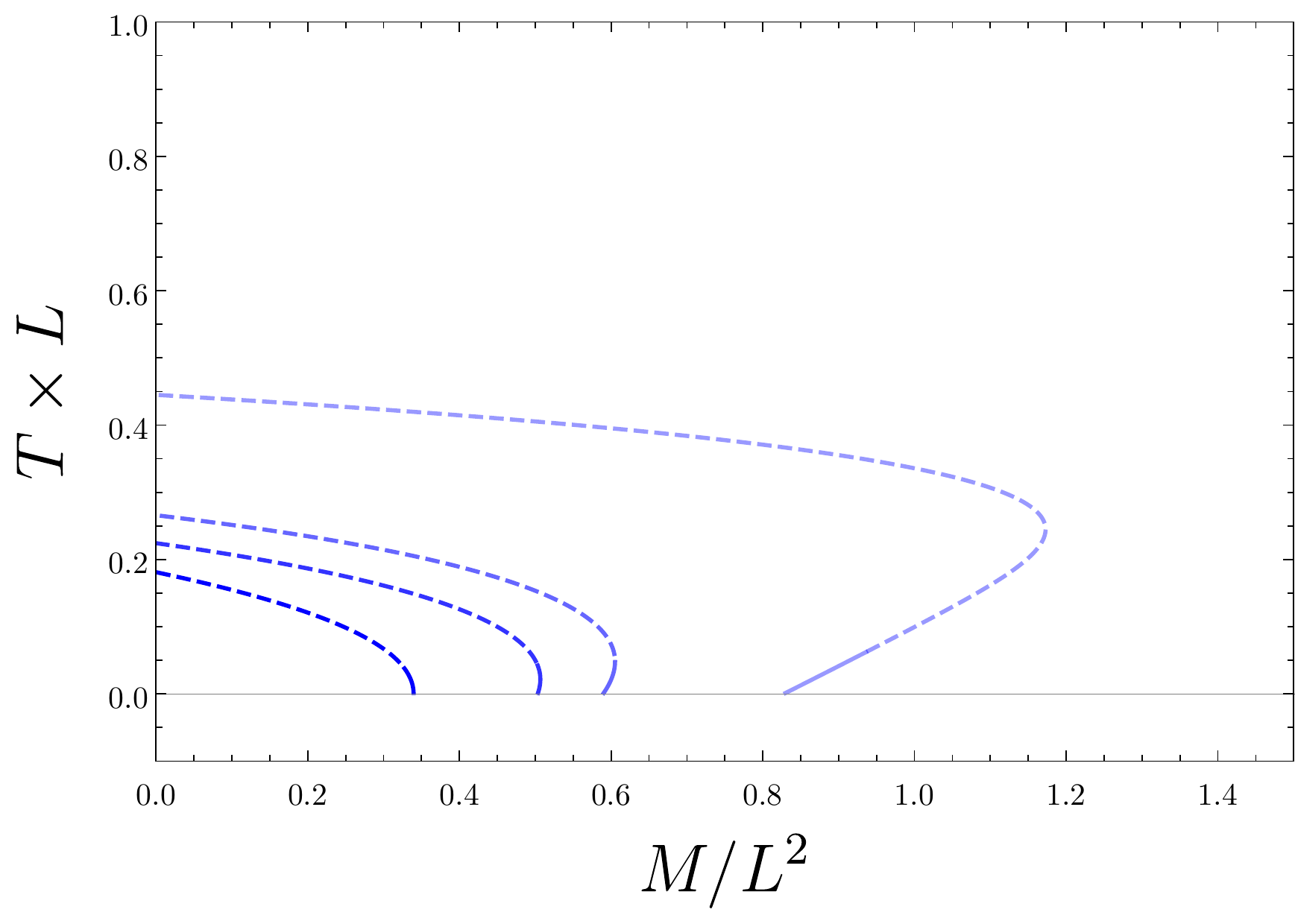}
\caption{{\bf Five-dimensional black hole properties in the grand canonical ensemble II}. Here we show plots of temperature vs.~horizon radius (left) and temperature vs.~mass (right) for $\mu/\mu_c = 0.3$, corresponding to $\mu < -79 L^4/1890$, providing an example of the behaviour when large black holes do not exist. The curves correspond to $\Phi = 0, 1.5, 1.7, 2$ in order of decreasing opacity (or left to right through a horizontal slice). The dashed blue curves indicate that the Wald entropy is negative, while the dot-dashed grey curves in the left plot indicate negative mass. }
\label{fig:5D-GC-Small}
\end{figure}

The value $\mu = -79 L^4/1890$ continues to mark a transition between the existence/nonexistence of large black holes. However, in the fixed potential ensemble, the structure is slightly different. We show some representative curves in Figure~\ref{fig:5D-GC-Small}.  For small values of $\Phi$, the behaviour is similar to that displayed in Figure~\ref{fig:5d-HP-small}. For larger values of $\Phi$, one can see the a protrusion begins to take shape in the profiles, pushing toward larger values of horizon radius. We can see that for much of the profiles the mass is negative, indicating that those parameters do not correspond to solutions with sensible asymptotics.

\begin{figure}[h]
\centering
\includegraphics[width=0.45\textwidth]{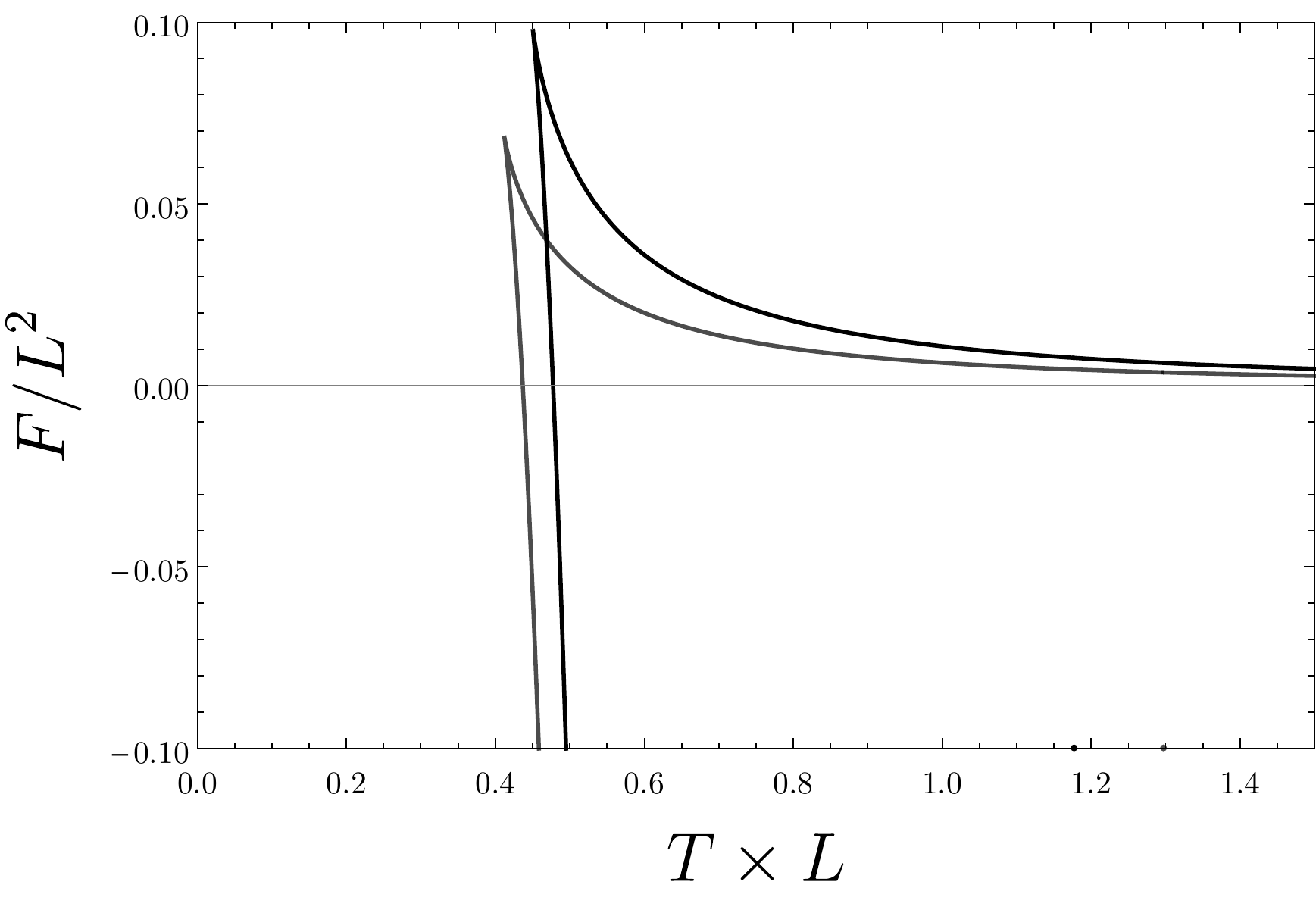}
\quad
\includegraphics[width=0.45\textwidth]{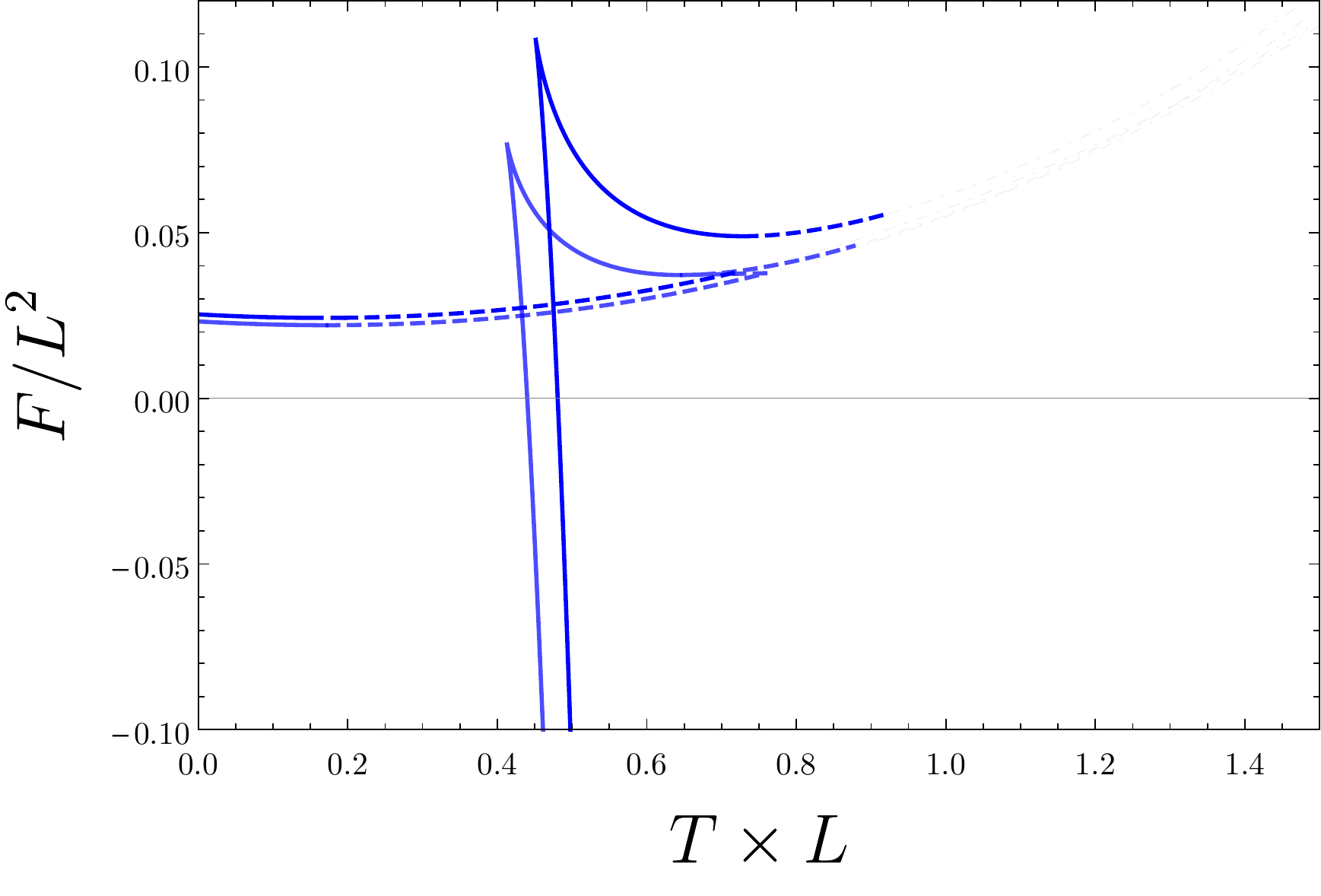}
\quad
\includegraphics[width=0.45\textwidth]{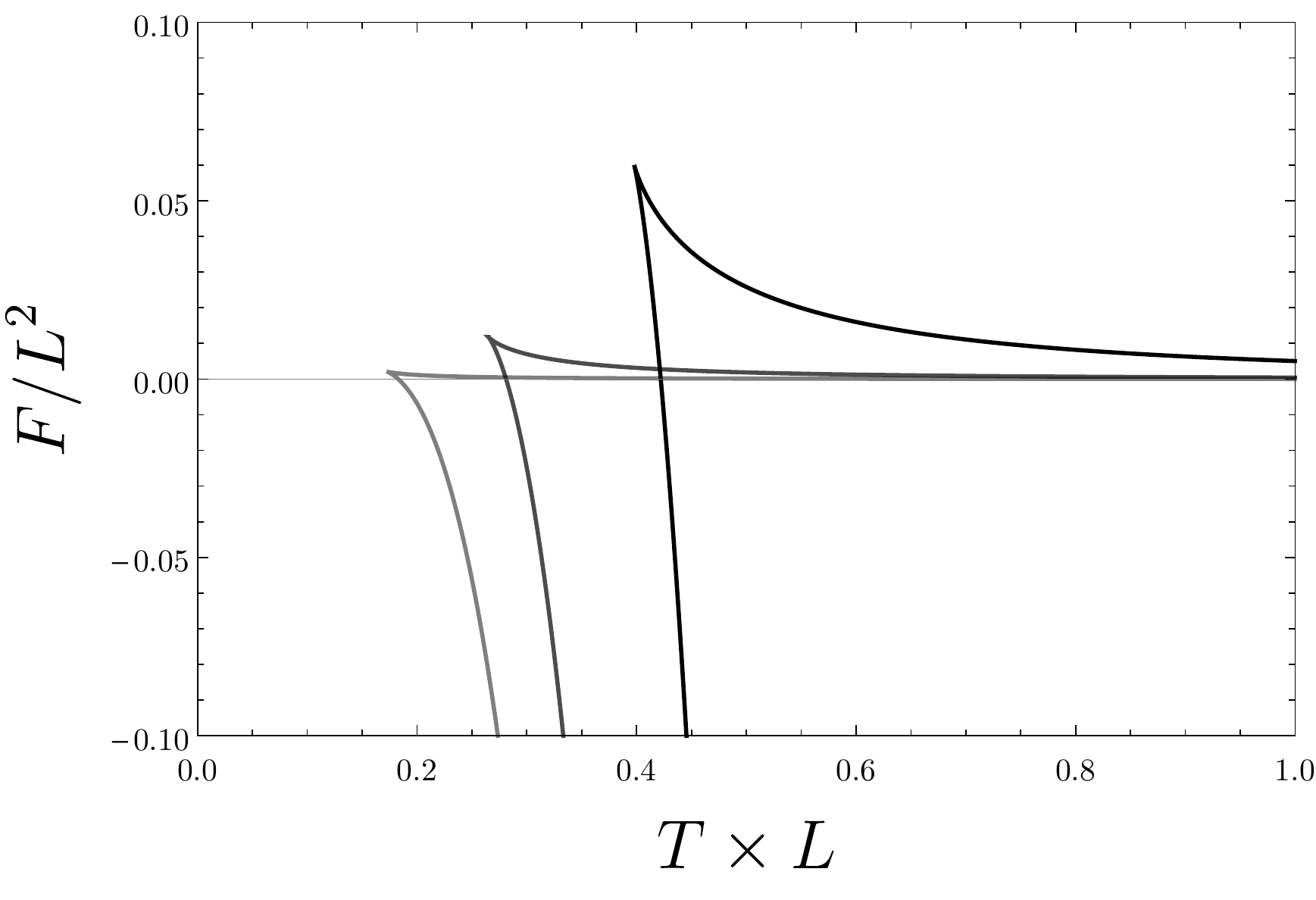}
\quad
\includegraphics[width=0.45\textwidth]{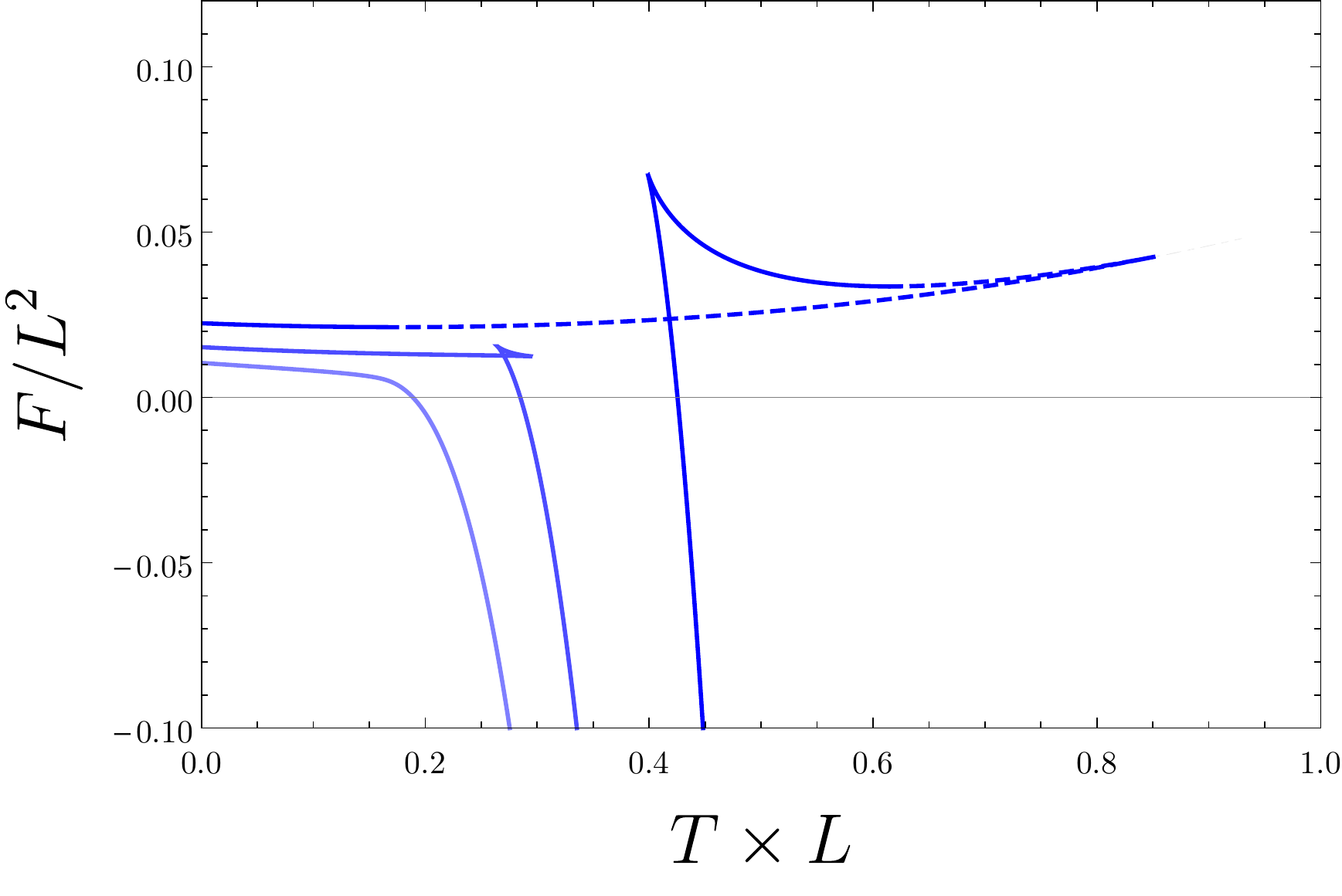}
\quad
\includegraphics[width=0.45\textwidth]{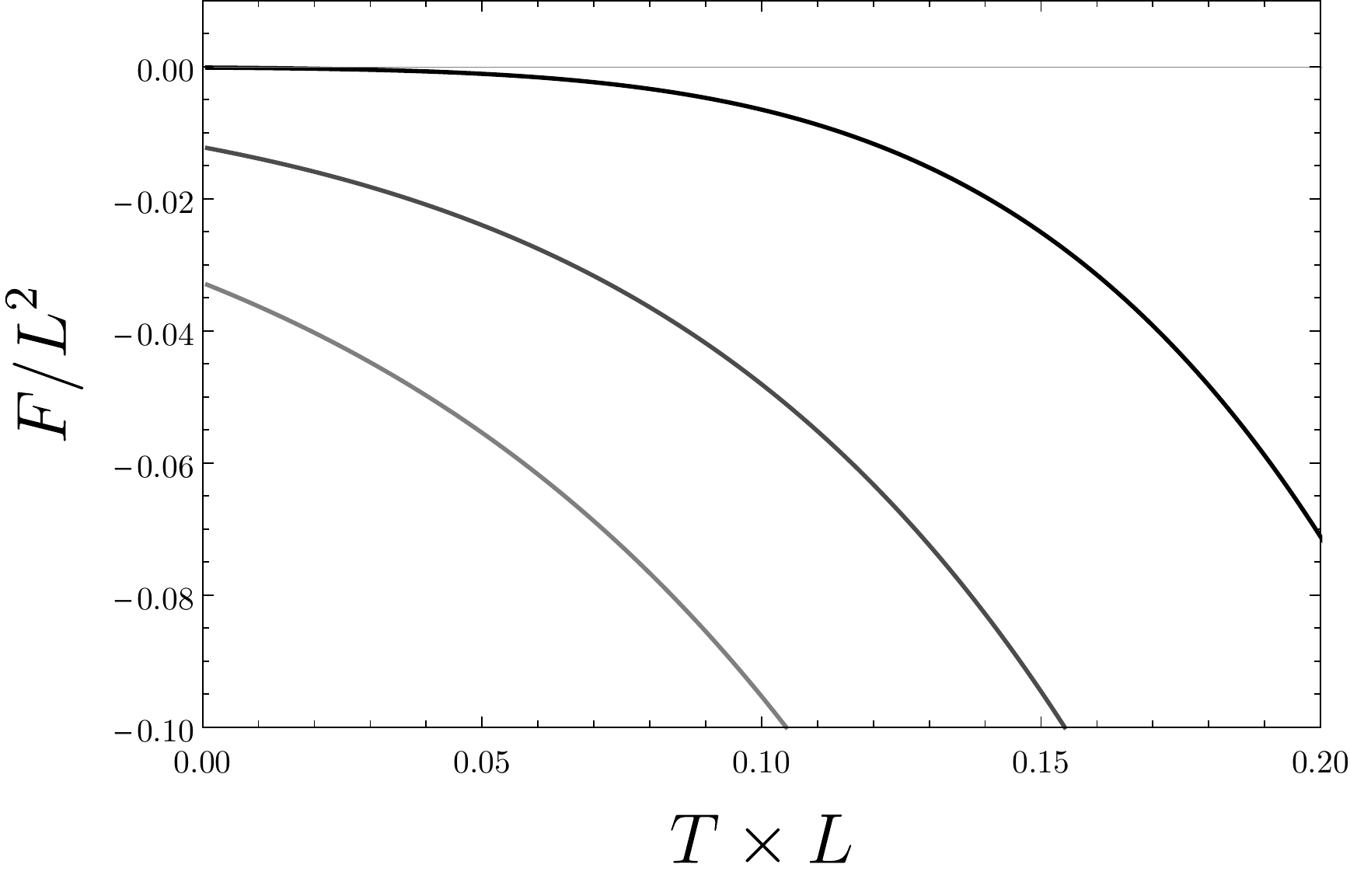}
\quad
\includegraphics[width=0.45\textwidth]{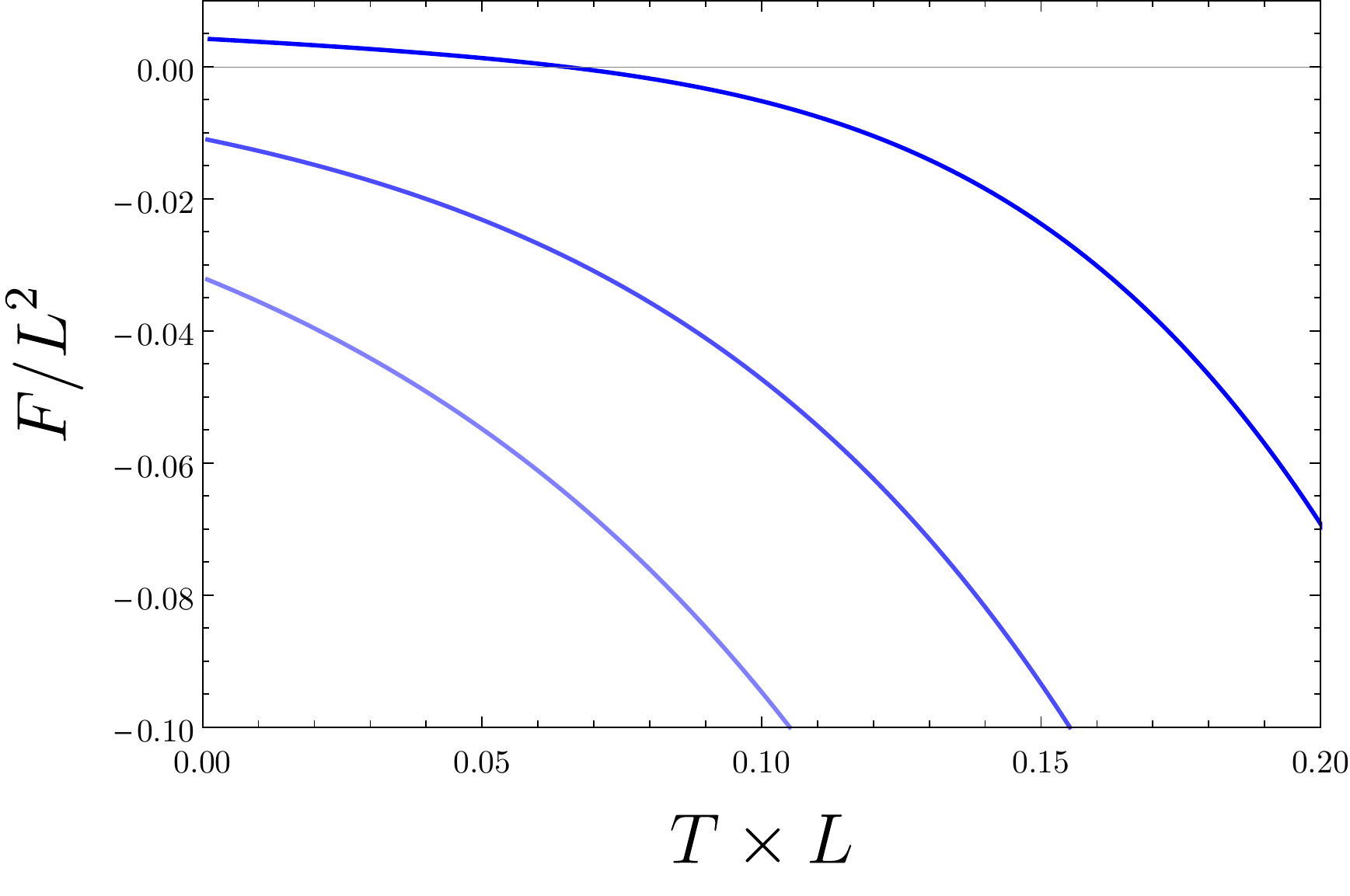}
\caption{{\bf Free energy: grand canonical ensemble in five dimensions}. {\it Top Row}: Here we show plots of the free energy for $\Phi = 0, 0.7$ (more to less opacity) for Einstein gravity (left) and the cubic theory with $\mu/\mu_c = 10^{-3}$ (right). {\it Middle Row}:  Here we show plots of the free energy for $\Phi = 0.81, 1.2, 1.6$ (more to less opacity) for Einstein gravity (left) and the cubic theory with $\mu/\mu_c = 10^{-3}$ (right).  {\it Bottom Row}:  Here we show plots of the free energy for $\Phi = 1.75, 1.9, 2.0$ (more to less opacity) for Einstein gravity (left) and the cubic theory with $\mu/\mu_c = 10^{-3}$ (right). In all cases the dashed portions of the blue curves indicate negative Wald entropy, and points where the blue curves simply terminate indicate that the mass has become negative.}
\label{fig:FvT-5d-GC}
\end{figure}

Finally, let us consider the free energy --- for the situation where $\mu < -79 L^4/1890$ we show representative plots in Figure~\ref{fig:FvT-5d-GC}. In the top row we show the results for small values of the potential. In both cases, when the potential is small the free energy has the same structure as in the uncharged case. For Einstein gravity this means that the free energy presents a cusp-like structure, with a Hawking-Page-like transition between thermal AdS and a large AdS black hole occurring at the point where the free energy vanishes.  For the cubic theory (shown on the right), the situation is much the same, exhibiting a phase transition between thermal AdS and a large black hole. Different parts of the free energy curve can have either negative entropy or mass. However, as in the uncharged case, this does not seem to pose a problem in the five dimensional case, as these cases are excluded due to the fact that they are not thermodynamically favoured. Note that the blue curve terminates at the point where $M=0$, since the cases with $M < 0$ do not exist as full solutions of the equations of motion.

As the potential is further increased, we enter into the regime where three branches of solutions emerge for the cubic theory. The value of $\Phi$ where this occurs is plotted in Figure~\ref{fig:PhiSpecial}, and representative free energy curves are shown in the middle row of Figure~\ref{fig:FvT-5d-GC}. In the Einstein case nothing of note changes. For the cubic case, we see a swallowtail emerge when $\Phi^2$ equals the value given in Figure~\ref{fig:PhiSpecial}. As $\Phi^2$ is further increased, the swallowtail shrinks, eventually terminating at what would be a critical point if it minimized the free energy\footnote{The critical exponents of this would-be critical point turn out to be the usual mean field theory values --- see Eq.~\eqref{eqn:mft}.}.  Of course, since the swallowtail occurs for positive values of the free energy, the usual first order phase transition it represents does not occur. Instead, what we observe is once again a Hawking-Page transition between thermal AdS and a large black hole at fixed potential.

As the value of $\Phi^2$ is further increased, we eventually reach a value for which there is only a single branch of solutions, which occurs for
\beq\label{eqn:5d-no-PT}
\Phi^2 > 3 + \frac{9}{2} \left(\frac{-122 \mu}{79 L^4} \right)^{1/3} \, .
\eeq
 Plots of free energy for $\Phi^2$ larger than $3$ are shown in the bottom row of Figure~\ref{fig:FvT-5d-GC}.  The upper-most curve in this these plots corresponds to $\Phi = 1.75$, for which we see that in Einstein gravity the black hole is thermodynamically preferred at all temperatures, while in the cubic theory a Hawking-Page transition continues to occur. The remaining curves correspond to values of $\Phi$ that satisfy the inequality given in Eq.~\eqref{eqn:5d-no-PT} --- the black hole is always thermodynamically preferred.  Due to the appearance of $\mu$ in Eq.~\eqref{eqn:5d-no-PT}, Hawking-Page transitions persist to larger values of $\Phi$ in the cubic theory than in Einstein gravity.

Lastly, let us note that in similar fashion to the uncharged case, if we adjust the entropy of the solutions so that they are always positive, this does nothing to change the interpretation of the phase structure described here, though it does push around the temperatures at which the phase transitions occur.

To close this section, let us make a few comments about what our results reveal about the black hole chemistry of these solutions. In both the uncharged and fixed potential cases, nothing qualitatively different is observed if one chooses to vary the cosmological constant. This can be seen, for example, just by considering the expressions for the critical point given in Eq.~\eqref{eqn:critPoint}. Notice that the critical pressure could be completely removed from the equation by redefining $\mu = x \mu_c$. Then the pressure simply serves as a relative scaling between the critical temperature and critical volume. Since the electric potential is dimensionless, in the uncharged and fixed potential ensembles, changing the pressure can only scale the results. In other words, by varying the pressure we can scale the points at which phase transitions and critical points occur, but we will not uncover any additional physics. The situation is a bit different in the canonical ensemble, since  there the electric charge (which is dimensionful) appears directly, and there is no natural analog of $\mu_c$ (i.e. a special value of the charge that relates it to the cosmological length scale) that occurs for the charge.

\section{Charged black holes: Canonical ensemble} \label{sec: thermoce}

 We now move on to consider thermodynamics in the canonical (fixed charge) ensemble. In this case, our aim will be to explore the critical points and phase behaviour working in the black hole chemistry framework.  A key difference between the thermodynamics in the canonical ensemble compared to the previous two sections is that here, due to conservation of charge, transitions to the vacuum are not possible. This means that, at fixed charge, we compare the free energy of all the black hole solutions, and that with the lowest free energy is the preferred phase.

%

Recall that the equation of state in general dimensions reads
\beqa
P&=&\frac{T}{v}-\frac{(d-3)}{\pi  (d-2)} \frac{k}{v^2}+\frac{e^2}{v^{2 d-4}}
+ \frac{2^8 (d-7)(d-4) (4 d^4-57 d^3+357 d^2-768 d+516 ) \mu k }{\pi  (d-2)^5 (4 d^4 - 49 d^3 + 291 d^2 - 514 d + 184)  v^6}
\nn\\
&&-\frac{3\times 2^8 (d-4)(d-6) \left(4 d^3-33 d^2+127 d-166\right) k^2  \mu T}{(d-2)^4 (4 d^4 - 49 d^3 + 291 d^2 - 514 d + 184) v^5}  \nonumber\\
&&\
+\frac{3\times 2^{12} \pi (d-5)  \left(d^2+5 d-15\right) k \mu T^2  }{(d-2)^3 (4 d^4 - 49 d^3 + 291 d^2 - 514 d + 184) v^4}
\nn\\
&&+ \frac{2^{11} \pi^2 (d-4) \left(d^2+5 d-15\right) \mu T^3 }{(d-2)^2 (4 d^4 - 49 d^3 + 291 d^2 - 514 d + 184) v^3}
\eeqa
where we note that the charge appears as $e^2$, and so
the same results hold for both positive and negative charge. The general idea for observing   phase transitions is to see whether the coefficients of different powers  of  $v$ in the equation of state have signs that allow for various
maxima and minima of $P$. The appearance/disappearance of distinct phases will generically be associated with \textit{critical points}. A necessary condition for a critical point to occur is that
\beqa
\frac{\partial P}{\partial v}=\frac{\partial^2 P}{\partial v^2}=0\label{dpd2p} \, .
\eeqa
which will generally have non-degenerate solutions. A free energy analysis is required to determine whether the critical point is physically realized in the system i.e., whether or not the critical point belongs to a minimizing branch of the free energy. Unfortunately it is difficult to make any very general statements about how many critical points occur and what their associated phase behaviour is. For this reason, we resort to a case-by-case analysis in four, five and six dimensions, presenting an essentially exhaustive analysis of the parameter space.  We close the section with a few brief remarks on the situation in general dimensions.

In what follows, we concentrate on several specific dimensions and investigate the thermodynamic behaviour in some detail.

\subsection{Critical behaviour in four dimensions}\label{four4}

The existence of critical points for four dimensional charged black holes has been previously pointed out in  Einstein gravity ($\mu=0$) \cite{Kubiznak:2012wp}. In four dimensions, the field equation for cubic generalized quasi-topological gravity reduces to that of Einsteinian cubic gravity;\footnote{In four-dimensions, the theory itself reduces to Einsteinian cubic gravity plus an additional term that does not contribute to the field equations of spherically symmetric black hole spacetimes~\cite{Hennigar:2017ego}.} the critical behaviour of black holes in Einsteinian cubic gravity have been previously studied~\cite{Hennigar:2016gkm} for the case of uncharged black holes (see also Section~\ref{sec:HawkingPage} above). Here we include an analysis of the charged case.

The equation of state~\eqref{eos0} takes the following relatively simple form:
\beqa
P=\frac{T}{v}-\frac{k}{2 \pi  v^2}+\frac{e^2}{v^4}-\frac{48 \pi \mu k T^2}{  v^4}
\eeqa
 Note that, at fixed temperature, the term arising from the electric charge and the term arising from the cubic correction both go like $v^{-4}$.  These terms dominate for small black holes and, due to them having the same fall-off behaviour, suggests there will be similarities between the cubic black holes and ordinary charged black holes in Einstein gravity.


Solving equation \reef{dpd2p}   we find  for general values of $\mu$ and $e^2$
that the critical temperature, volume and pressure are
\beqa
T^2_{c \pm}=\frac{3 \pi  e^2\pm  \sqrt{9 \pi^2 e^4- 64 k^4\mu  }}{288 \pi ^2 k\mu },\quad P_{c \pm}=\frac{3 \pi e^2\pm  \sqrt{9 \pi ^2 e^4-64  k^4\mu }}{512 \pi k^2\mu }
,\quad
v_{c \pm}=\frac{2 k}{3 \pi  T_{c \pm}}\nonumber\\ \label{Tpvc}
\eeqa
where the two choices result because the equation of state is quadratic in $T$.
 Under the restriction of negative coupling (which is required for the existence of sensible positive mass solutions), we can see that the term under the square root in the above expressions is always positive. However, by the same token we see that for $k=+1$ only $T_{c-}, P_{c-}$ and $v_{c-}$ are physically sensible, i.e. have all three critical values positive, while for $k=-1$ there is no physical solution. The end result then is that there are no ``new'' critical points introduced by the cubic theory in four dimensions. Effectively, the cubic correction shifts the critical quantities away from their Einstein gravity values, reducing the critical temperature and pressure, while increasing the critical volume.


We also find that the ratio of  critical quantities in \reef{Tpvc} is independent of the black hole parameters
\beqa
\frac{P_c v_c}{T_c}=\frac{3}{8}\label{ratio}
\eeqa
 and in this sense is universal.
Note that this ratio is independent of choice of spherical or hyperbolic geometry, though  in the latter case we do not have critical points since $p_{c-}$ and $v_{c-}$ are negative. In~\cite{Hennigar:2017umz} it was found that the van der Waals ratio differs from this value of $3/8$ for black branes, and so the ratio can be sensitive to the horizon topology. Remarkably, the ratio~\eqref{ratio} is precisely the same as that first observed for charged black holes in four dimensional Einstein gravity~\cite{Kubiznak:2012wp}; higher curvature corrections have not affected this universal value for spherical black holes.



It can be straight-forwardly confirmed that the various physical constraints are satisfied by the black holes at the critical point. That is, these black holes possess positive mass and the critical pressure is always less than the maximum pressure $P_{\rm max}$. We can also confirm that the entropy --- regardless if it has been shifted or not ---  is always positive at the critical point.    For the entropy from Eq. \reef{sratio} at critical point we obtain
\beqa
s_{c \pm}\propto\frac{16 \pi^2 k^4\mu +9 \pi ^4 e^4\pm 3 \pi ^3 e^2 \sqrt{9 \pi ^2 e^4-64 k^4\mu}}{48 \pi^2 k^4\mu}\label{ent4d} \, .
\eeqa
Noting that only the minus branch with $k=+1$ corresponds to a sensible critical point, some simple manipulations reveal that the Wald entropy is positive at the critical point.

The critical points are characterized by mean field theory critical exponents which,  for generic values of parameters and $k=1$ in the physical domain, are given by
\beqa
\alpha=0, \quad \beta=\frac{1}{2}, \quad \gamma=1 ,\quad \delta=3
\label{exponents}
\eeqa
and are obtained  by expanding the equation of state near the critical point \cite{Gunasekaran:2012dq}
\beqa
\frac{P}{P_{c \pm}}&=&1-\frac{1}{48 \pi^2 k^4{\mu}} \left(-160 \pi^2 k^4{\mu}+Y\right){\tau}
+\frac{4 }{48 \pi^2 k^4{\mu}} \left(-64 \pi^2 k^4{\mu}+Y\right){\phi} {\tau}-\frac{4}{3} {\phi}^3
\nn\\
&+&\cO(\tau \phi^2,\phi^4)\label{ppc}
\eeqa
with
\beqa
Y =9 \pi ^4 {e^4}\pm 3 \pi ^2  {e^2}\sqrt{9 \pi ^4 e^4- 64  \pi^2 k^4{\mu}}
\eeqa
and where we replaced the following terms for volume and temperature
\beqa
v = v_c(\phi + 1) \, , \quad T = T_c (\tau + 1) \, .
\eeqa
Since the prefactors multiplying the $\phi \tau$ and $\phi ^3$ terms are non-vanishing the the physical portion of parameter space,  the critical points given in \eqref{exponents} follow from this expansion.

\begin{figure*}[htp]
\centering
\begin{tabular}{cc}
\includegraphics[scale=.32]{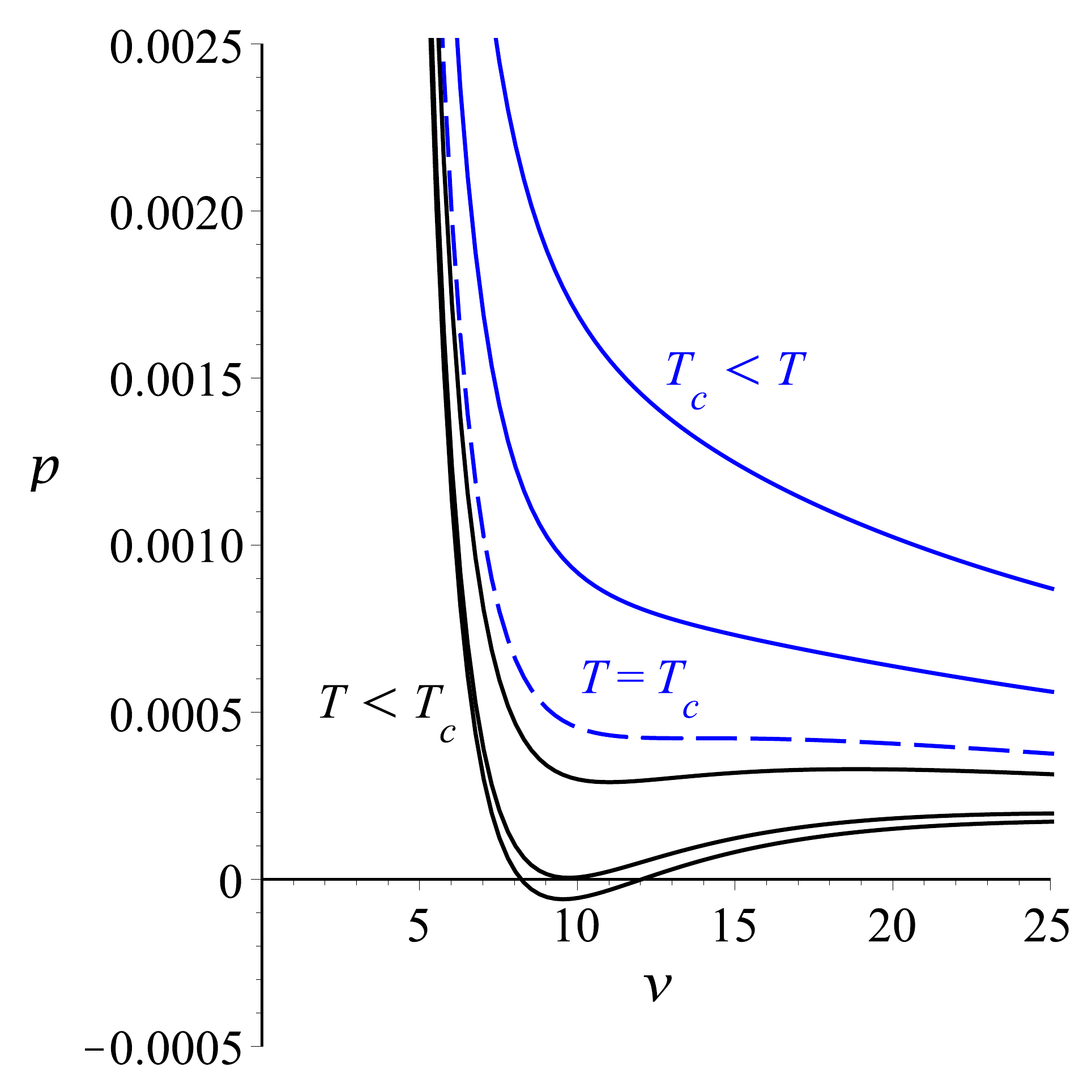}& \quad \quad
\includegraphics[scale=.32]{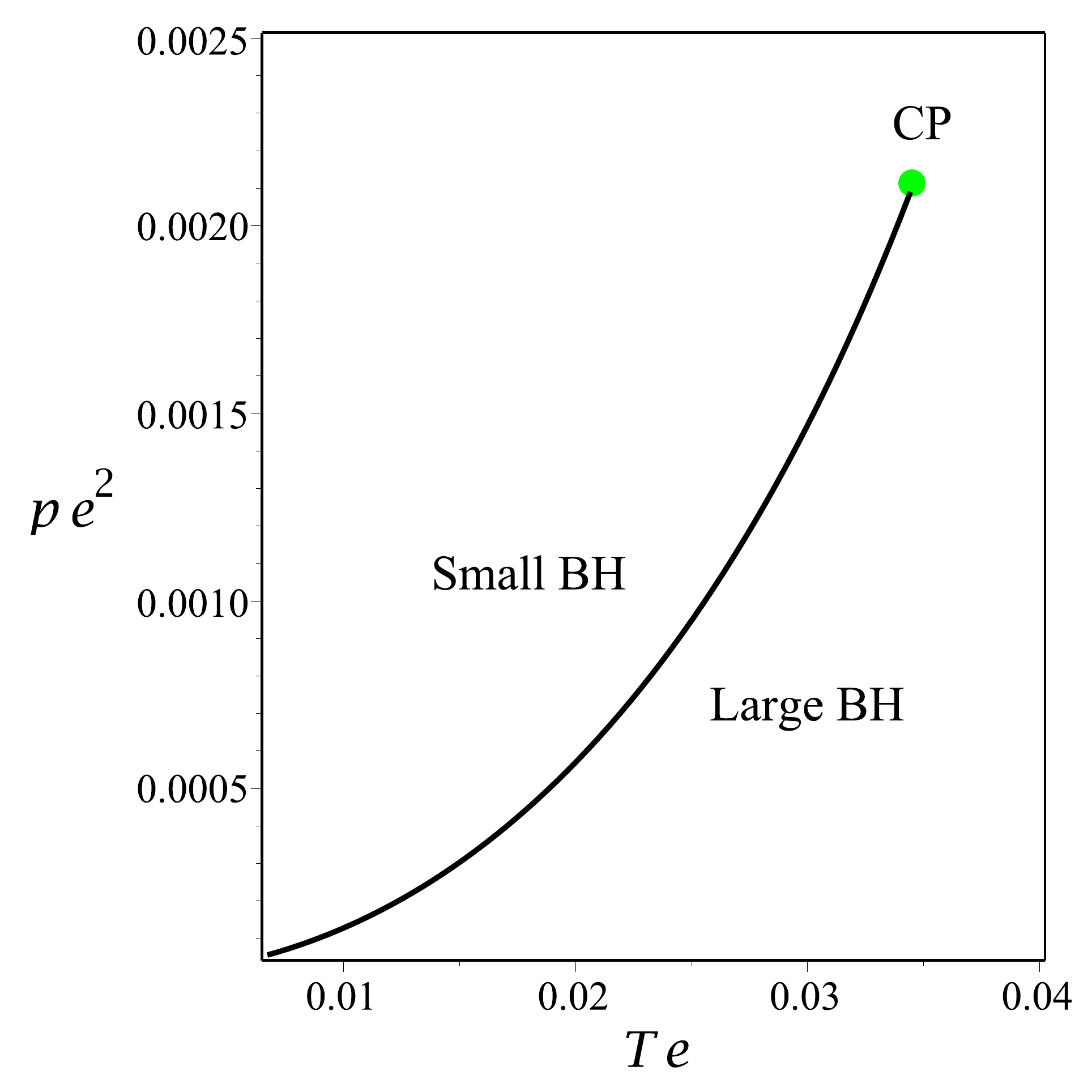}
\\
\end{tabular}
\caption{{\bf Critical behaviour in four dimensions}. \textit{Left}: the case for fixed charge, we show a $P-v$ graph that illustrates a first order phase transition with VdW behaviour in $d=4$ and with $k=1$.  The various curves correspond to different isotherms: at the critical point $T=T_c$ (dashed blue line), $T=0.9 T_c, 0.71 T_c, 0.67 T_c$ (solid black lines), and $T=1.3 T_c, 1.8 T_c$ (solid blue lines). Here we choose $\mu/e^4 \approx -0.00152$ with $T_c e\approx 0.03455$. \textit{Right}: Phase diagram in $P-T$ plane. The phase diagram for four dimensional charged black holes with $k=1$ is constructed  with $\mu/e^4 \approx -0.00152$, however the similar behaviour occurs for any other values. Note that here we are working in units of the electric charge.
}
\label{PT4d}
\end{figure*}

Considering the $P-v$ graph in Figure \ref{PT4d}, we observe two distinguishable (stable) phases
for $T<T_c$. These  merge at $T=T_c$ and then for $T>T_c$ they become indistinguishable,
the hallmark of a  standard Van-der-Waals (VdW) phase transition.  Note that for certain low temperature isotherms, portions of the $P-v$ curve can dip into negative pressure. A similar situation occurs already in Einstein gravity and, of course, negative pressure in this setup is unphysical. The solution to the problem is either that the negative pressure portion of the curve is excised via a Maxwell equal area prescription or, in some cases, it is just the case that these solutions are unphysical.

\begin{figure*}[htp]
\centering
\begin{tabular}{cc}
\includegraphics[scale=.32]{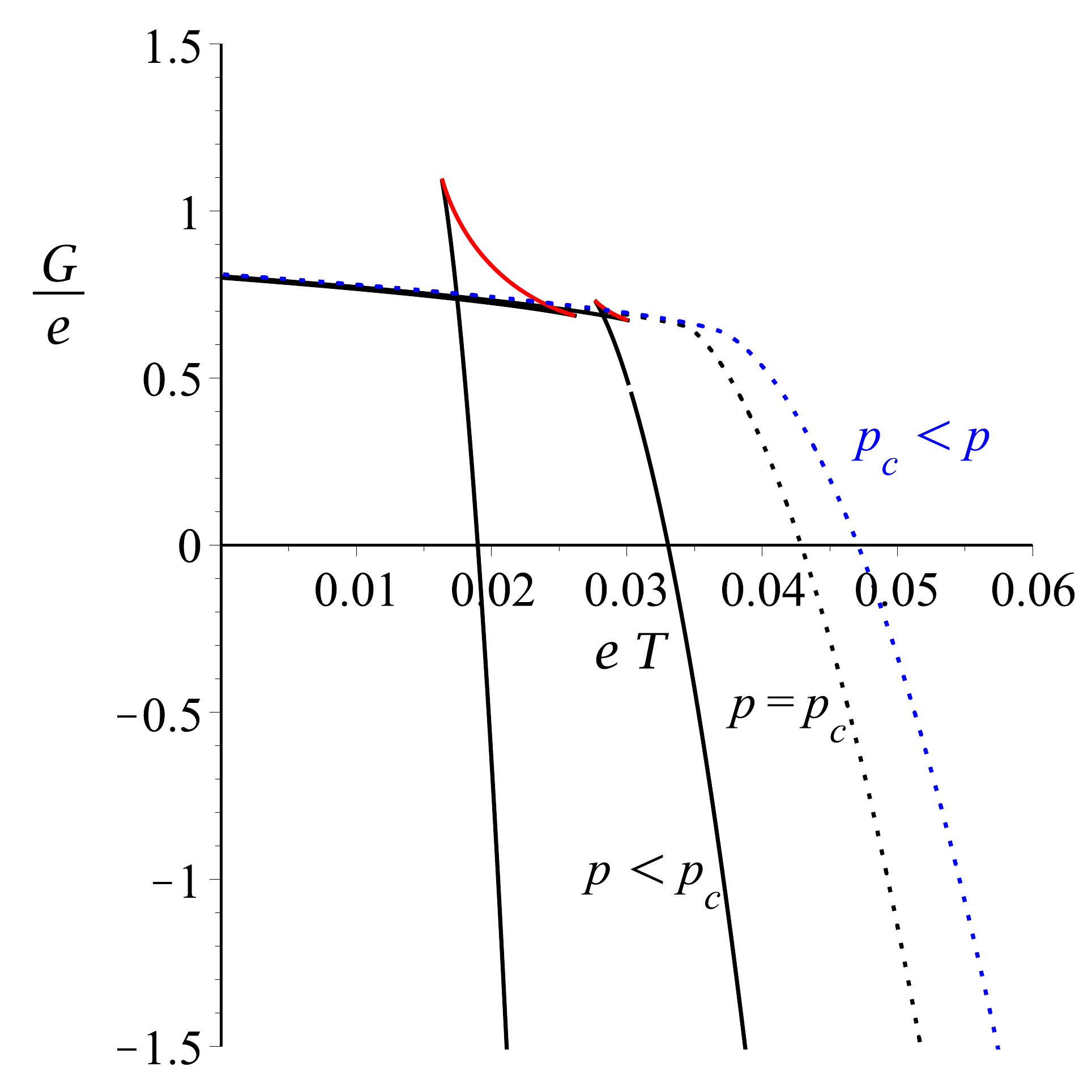}
&\quad\quad
\includegraphics[scale=.32]{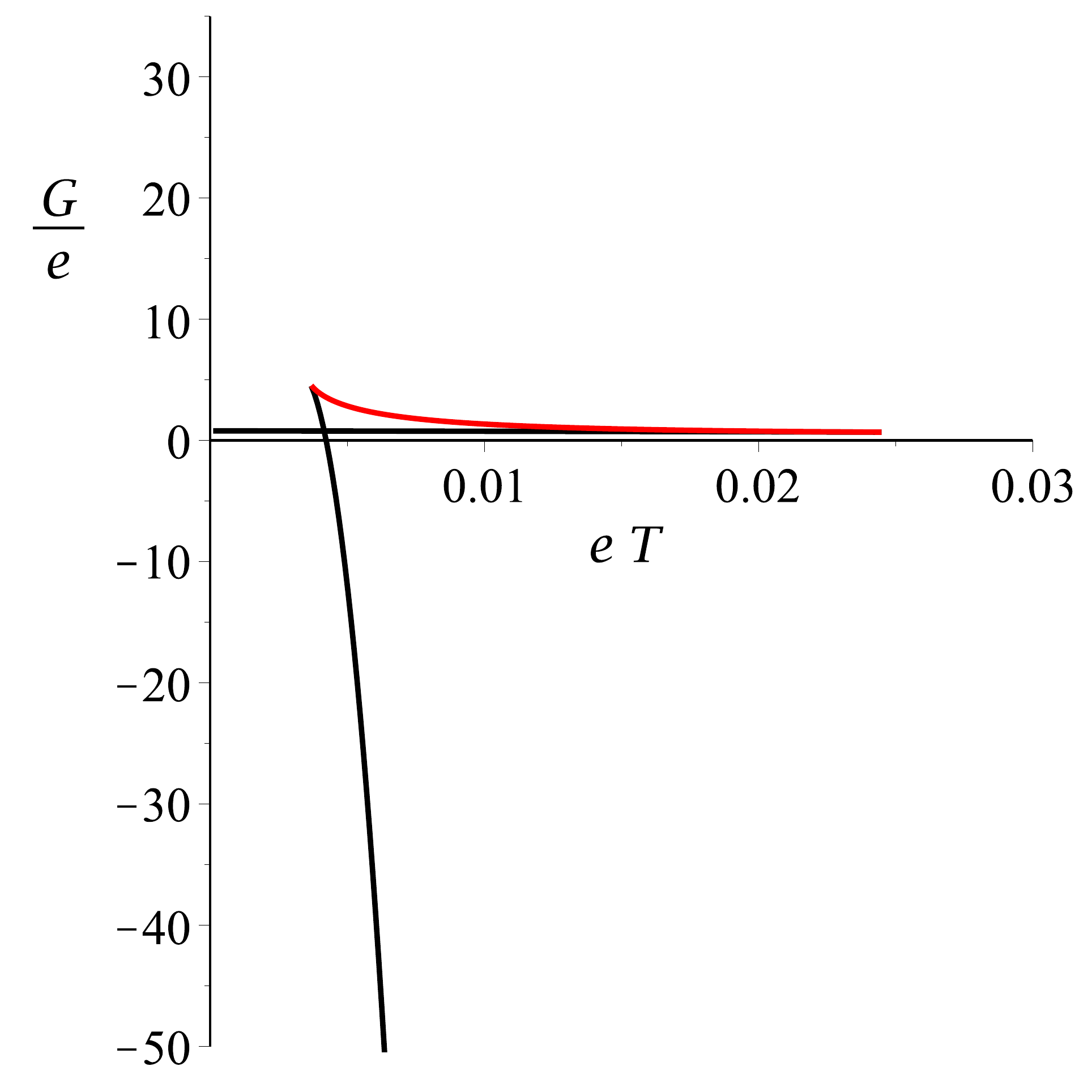}
\end{tabular}
\caption{\textbf{Free energy in four dimensions}. {\it Left}: Plot of Gibbs free energy versus temperature for $d=4$ and $k=1$, for $P=1.2 P_c$ (dotted, blue curve), for $P=P_c$ (dotted, black curve), for $P=0.6 P_c$ and $P=0.2 P_c$ (solid black and red lines). {\it Right}: Plot for  $P=0.01 P_c$.
In each plot, the red lines represent parts of the curves that the specific heat is negative. In all plots, $\mu/e^4 \approx -0.00152$  where physical conditions are satisfied with $P_c e^2\approx 0.00211$.
}
\label{GTd4}
\end{figure*}

The critical points correspond to the end point of a line of first order phase transitions, as shown in Figure~\ref{PT4d}. This line of coexistence demarcates phases of large and small black holes.
Furthermore, an analysis of the Gibbs free energy reveals typical van der Waals behaviour, shown in Figure~\ref{GTd4}.

For pressures larger than the critical value, there is only a single branch of black holes and no phase transition takes place.  For $P=P_c$, the free energy has a kink shape, characteristic of the diverging specific heat at the critical point and it is always stable ($C_p>0$). For pressures a bit less than the critical pressure, the Gibbs free energy demonstrates the swallowtail behaviour as expected from van der Waals manner. There are up to two branches of black holes that have positive specific heat (though only one ever minimizes the free energy), while the concave patch of the Gibbs free energy indicates negative specific heat,
\beqa
C_p = - T \frac{\partial ^2 G}{\partial T^2} \, .
\eeqa
We specify the negative specific heat in Figure~\ref{GTd4} by red lines. Note that all curves approximately converge to the same small domain as $T\rightarrow 0$ for different choices of pressure. Further decreasing the pressure, we observe a swallowtail. For very small values of pressure the swallowtail `grows'.

\subsection{Critical behaviour in five dimensions} \label{five5}

In five dimensions we obtain
\beqa
P=\frac{T}{v}-\frac{2 k}{3 \pi  v^2}+\frac{ 4480 \pi^2 \mu T^3}{711 v^3}+\frac{256  \mu k^2 T}{237  v^5}-\frac{31232 \mu k}{19197 \pi v^6}+\frac{e^2}{v^6}
\eeqa
for the equation of state. Using  \reef{dpd2p} and setting the first and second derivatives of $P$ with respect to $v$ to $0$,
 the general form of the critical temperature for given $k$ in terms of critical volume and other parameters reads
\beq
T_c = \frac{19197 \pi  e^2-31232 k \mu +1422 k v_c^4}{2133 \pi  v_c^5-11520 \pi  k^2 \mu  v_c}
\eeq
where $v_c$ satisfies
\begin{align}
0&=3 v_c^3 \left(79 v_c \left(4266 k v_c^4-95985 \pi  e^2+87040 k \mu \right)+40320 \pi ^3 \mu  T_c^3 \left(1280 k^2 \mu -237 v_c^4\right)\right)
	\nn\\
	&+1280 k \mu  \left(31232 \mu -19197 \pi  e^2 k\right)
\end{align}
Although the equations~\eqref{dpd2p} are non-linear in both $T$ and $v$, it is still possible to obtain the above explicit expression for $T_c$ in terms of $v_c$ and the other parameters by manipulating the equations~\eqref{dpd2p} to remove non-linear dependencies on the temperature. Note that these manipulations are possible only in four, five and six dimensions, since it is only in these cases that a single non-linear power of temperature appears in the equation of state.


Although it appears possible for the expression of the critical temperature to have a singularity for a particular combination of the specific volume and the coupling, this does not manifest as it would require positive coupling --- this is forbidden by the requirement of having sensible positive mass solutions.

%


\begin{figure*}[htp]
\centering
\includegraphics[width=0.45\textwidth]{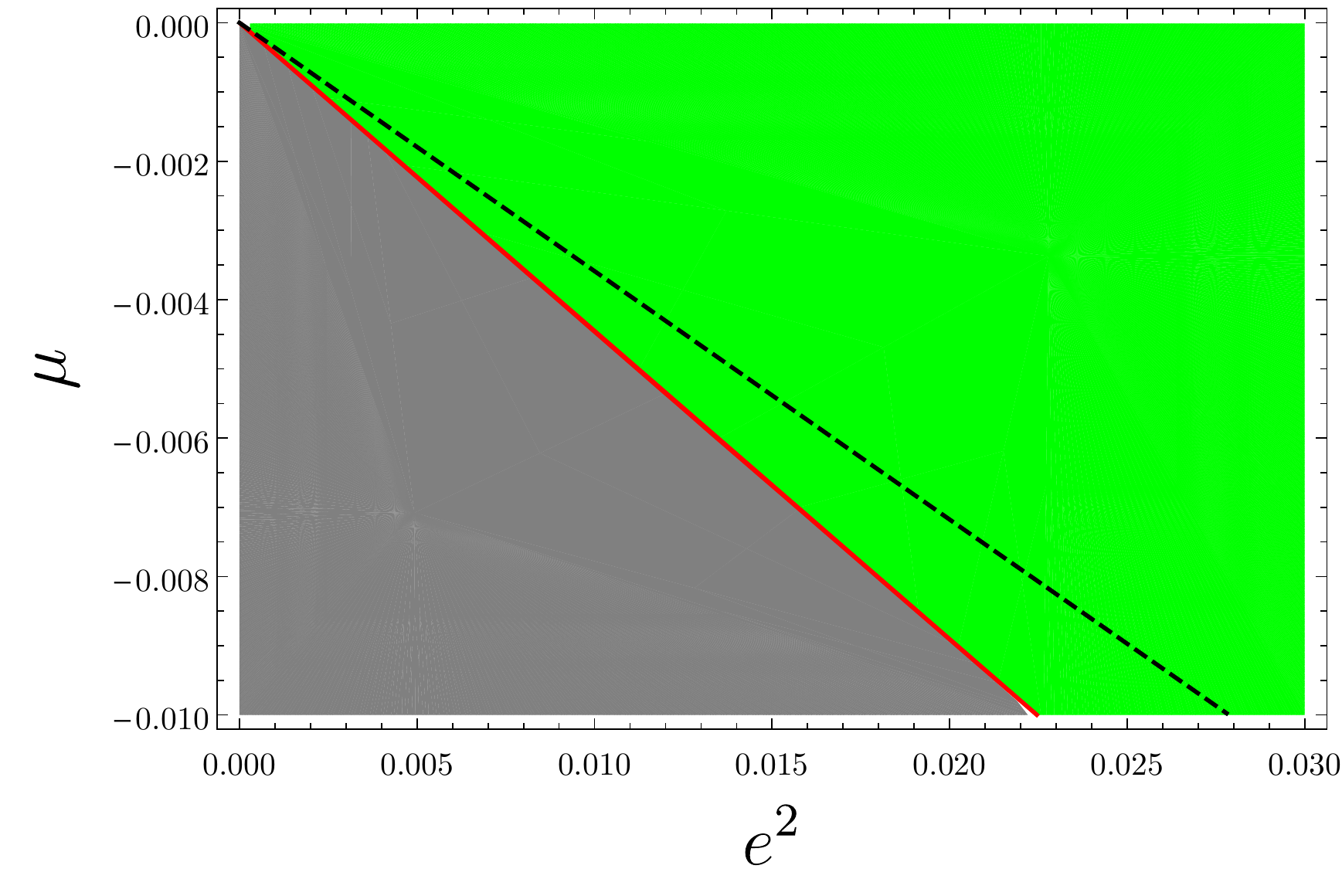}
\quad
\includegraphics[width=0.45\textwidth]{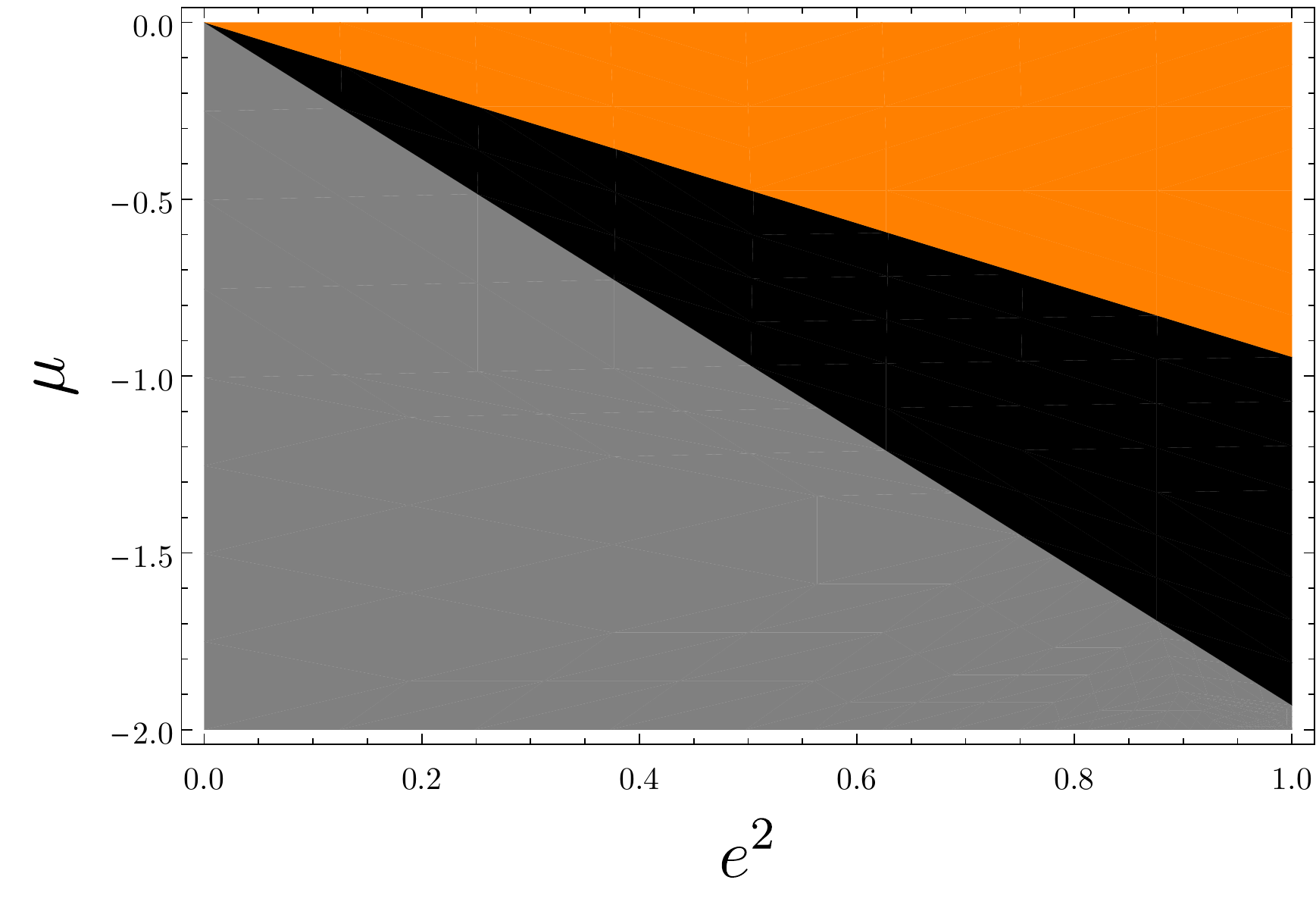}
\caption{\textbf{Phase space of constraints for $d=5$}.
The plot shows the different possibilities for
critical behaviour in $(\mu,e)$
parameter space for
 $k=1$ (left) and $k=-1$ (right).   Green regions denote single physical critical points with all physical constraints fulfilled. In grey regions there are no possible critical points. Blue regions depict single critical points with $\gamma^2<0$. The thin red line denotes two potential critical points.  In the left plot, the critical points that lie below the dashed black curve possess negative Wald entropy.  In the right plot, the black region indicates that the critical pressure exceeds the maximum pressure, while the orange region indicates that the potential critical point corresponds to a negative mass black hole.}
\label{domain5d}
\end{figure*}

 We now turn to an examination of the coupling/charge parameter space, with the relevant plot shown in Figure~\ref{domain5d}. We first focus on black holes with spherical horizons. In this case, we first note that when $\mu = 0$ there is a single physical critical point provided that the electric charge is non-zero. We find that when the cubic coupling is negative, and provided it satisfies the bound $\mu \lessapprox -0.446179 e^2$ then there are no possible critical points. When $-0.446179 e^2 \lessapprox \mu < 0$ the equations \eqref{dpd2p} admit two possible solutions for critical points. To determine if the black holes corresponding to these possible solutions are physical, we have to ensure that the various physical constraints are satisfied --- these have been incorporated into Figure~\ref{domain5d} directly. We find that one of the two possible critical points always possesses negative Wald entropy, while the second has negative Wald entropy only when $\mu \lessapprox -0.358799 e^2$, and otherwise has positive entropy. However, as mentioned in the previous sections, due to ambiguities in the definition of the entropy, it is unclear whether this alone means that the black holes are unphysical. More important is that the mass is positive, since it seems that the negative mass solutions do not exist. We find that for coupling in the range $-0.445201 e^2 \lessapprox \mu < 0$ there is a single critical point with positive mass, while in the interval $-0.446179 e^2 \lessapprox \mu \lessapprox -0.445201 e^2$ both of the critical points correspond to black holes with positive mass.

Let us now describe the phase behaviour in the various regions of parameter space. In the regime where there is a single physical critical point we find (unsurprisingly) Van der Waals type behaviour with the critical exponents coinciding with the mean field theory values. The plots that arise in this case are qualitatively similar to the four dimensional case, and so we do not present them here. This single physical critical point limits to the one in Einstein gravity as $\mu \to 0$; the effect of the higher curvature correction is to increase both the critical pressure and temperature, while decreasing the critical volume. We can examine the ratio of critical values numerically based on the data from Figure~\ref{domain5d}, and we find that it exhibits weak dependence on the cubic coupling constant. This dependence can be confirmed by solving Eqs.~\eqref{dpd2p} perturbatively in the coupling constant, giving the following for the leading order correction:
\beq
\frac{P_c v_c}{T_c} = \frac{5}{12} \left[1 - \frac{527872}{2399625} \left( \frac{\mu}{e^2} \right) + {\cal O} \left(\frac{\mu^2}{e^4} \right) \right] \, .
\eeq
The first term in the expansion is, of course, the result for five-dimensional charged black holes in Einstein gravity~\cite{Gunasekaran:2012dq}. Recalling that $\mu <0$, we see that the cubic correction increases the value of the ratio.

\begin{figure}[h]
\centering
\includegraphics[width=0.5\textwidth]{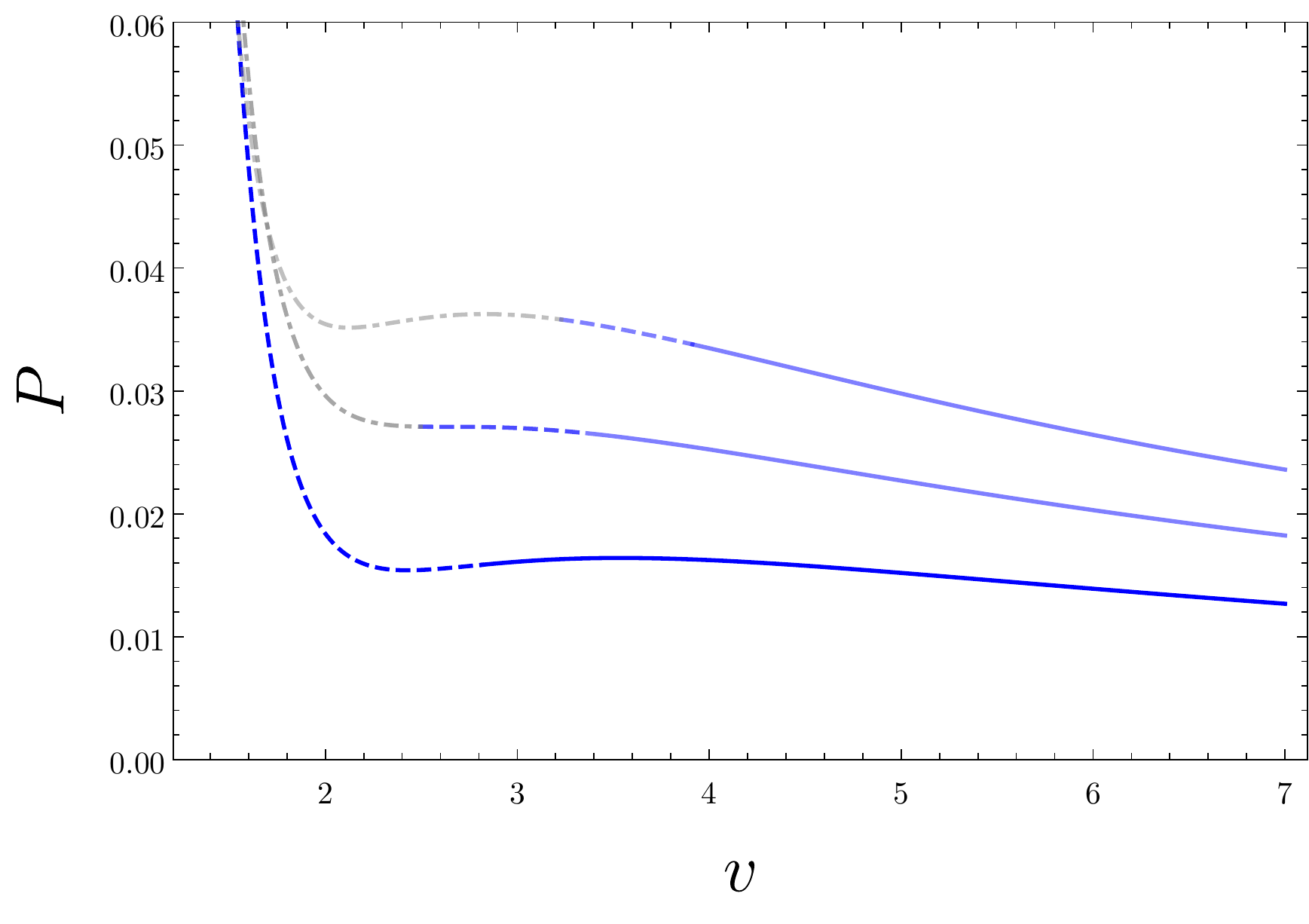}
\caption{{\bf Pressure vs.~volume plot depicting two critical points in $d=5$ for $k=+1$}.  This plot displays the situation for $\mu = -0.446 e^2$. There are two critical points with critical temperatures $T_{c_1} \approx 0.160234/e$ and $T_{c_2} \approx 0.163327/e $.  The isotherms shown correspond to $T \approx 0.75 T_{c_1}, T_{c_1}, 1.22 T_{c_2}$, from bottom to top (more to less opacity). Solid blue lines indicate positive Wald entropy and mass, dashed blue lines indicate positive mass but negative Wald entropy, dot-dashed gray lines indicate negative mass and hence that the corresponding black holes do not exist.}
\label{fig:5d-k1-PV2}
\end{figure}

As mentioned just above, when $-0.446179 e^2 \lessapprox \mu \lessapprox -0.445201 e^2$ there is an additional critical point that occurs for positive mass black holes (though they have negative Wald entropy). To illustrate the physics in this case, we refer to Figure~\ref{fig:5d-k1-PV2}, where the behaviour in the pressure volume plane is displayed for three different temperatures. The plot depicts three isotherms, corresponding to $T \approx 0.75 T_{c_1}, T_{c_1}, 1.22 T_{c_2}$. The behaviour can be understood as follows. For $T < T_{c_1}$, the system exhibits usual van der Waals type behaviour that terminates at the critical point $T_{c_1}$. For temperatures between $T_{c_1}$ and $T_{c_2}$, there is no interesting phase behaviour. (Curves with $T_{c_1} < T < T_{c_2}$ are not shown in Figure~\ref{fig:5d-k1-PV2} since they are too close together to distinguish.) When $T > T_{c_2}$, the system again exhibits van der Waals type oscillations, but with the caveat that these oscillations \textit{begin} at the critical point and then exist for arbitrarily large temperatures. However, it turns out that for much of the parameter space one of the possible phases possesses negative mass, and so there is no first order phase transition present.

Finally, let us make a few comments about the hyperbolic $k=-1$ case.  In this case we observe that the equations \eqref{dpd2p} admit a single solution provided that $ -1.93101 e^2 \lessapprox \mu < 0$. However, a further analysis reveals that the black holes corresponding to these critical points are unphysical.  For the coupling in the range $-0.946037 e^2 \lessapprox \mu < 0$, the black hole mass is negative at the critical point, while for $  -1.93101 e^2 \lessapprox \mu \lessapprox -0.946037 e^2$ the mass is positive but the value of the critical pressure exceeds the maximum allowable pressure $P_{\rm max}$.

\subsection{Critical behaviour in six dimensions}

\begin{figure*}[htp]
\centering
\includegraphics[width=0.65\textwidth]{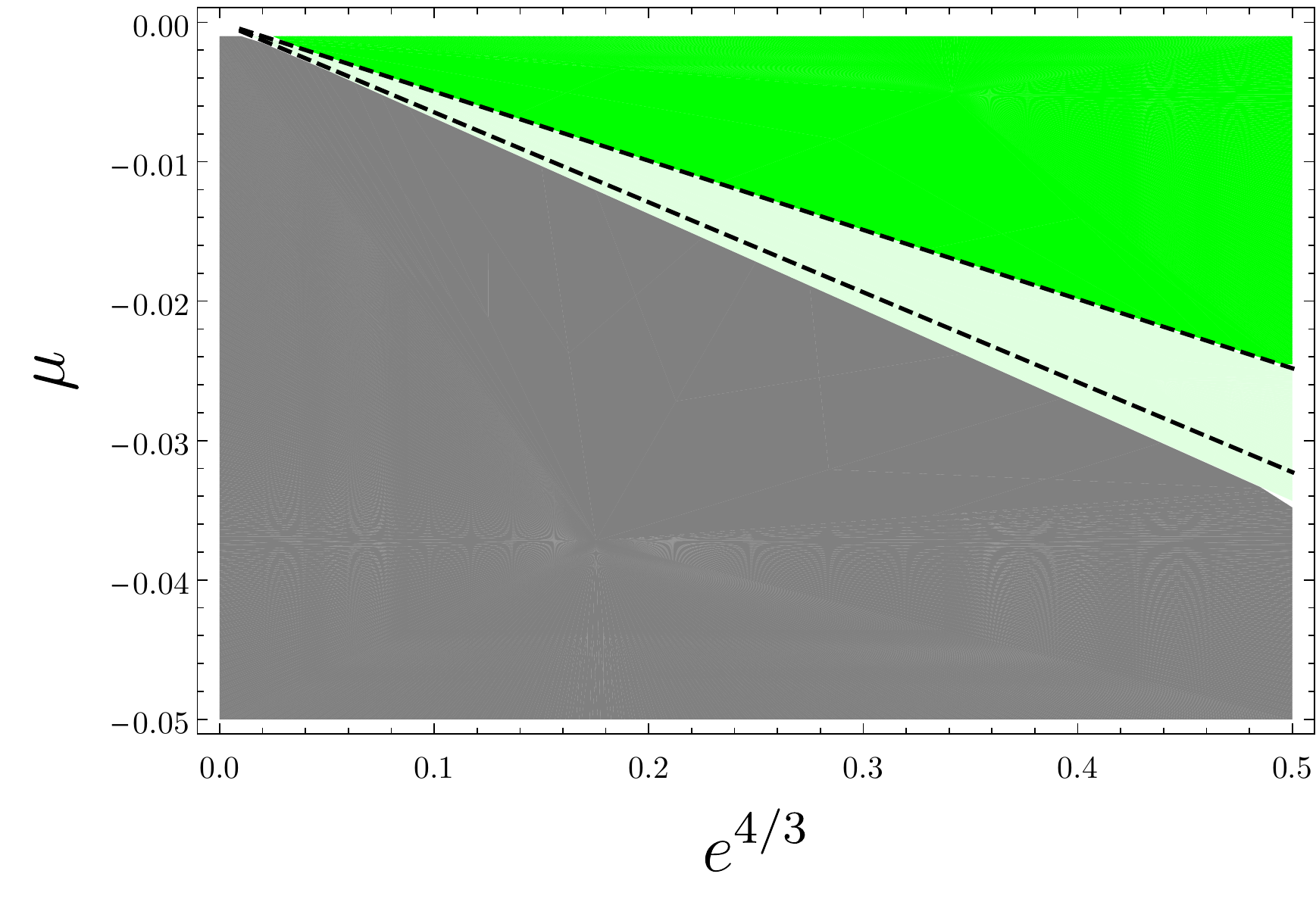}
\caption{\textbf{Phase space of constraints in six dimensions}.  Here we show the situation for spherical black holes in six dimensions. Green shaded areas represent a single physical critical point. Light green areas represent two physical critical points. In the gray regions, there are no solutions to the critical point equations~\eqref{dpd2p}.  In the region between the dashed black lines, the Wald entropy is negative for one of the two critical points.}
\label{domain6d}
\end{figure*}

For six dimensional Einstein metrics, the contributions of any cubic term to the linearized field equations vanishes~\cite{Bueno:2016xff}. This results in some simplification in this case, and we have $f_{\infty}=1$ by definition. From \reef{gamma2}, we obtain $\gamma^2=-\frac{2 }{9 \pi \mu P f_{\infty} m} $, which is positive provided $\mu<0$ (see Figure \ref{domain6d});  for $\mu>0$ we obtain $\gamma^2<0$. Therefore, in six dimensions, the pressure can be arbitrarily large.

In six dimensions, the equation of state becomes
\beqa
P=\frac{T}{v}-\frac{3 k}{4 \pi  v^2}+\frac{6 \pi^2 \mu T^3}{v^3}+\frac{9 \pi \mu  k T^2}{2  v^4}-\frac{3 \mu k}{8 \pi v^6}+\frac{e^2}{v^8} \, .
\eeqa
According to the analysis at the beginning of this section, there can be up to two critical points for the six dimensional spherical $(k=+1)$ black holes and three critical points for the hyperbolic ($k=-1$) ones. Applying Eq. \reef{dpd2p}
the critical temperature is related to the critical volume as
\beq
T_c =  -\frac{160 \pi  e^2 v_c^4-162 k \mu ^2 v_c^2+81 k \mu  v_c^6+6 k v_c^{10}+1152 \pi  e^2 \mu }{1440 \pi ^2 e^2 k \mu  v_c-243 \pi  \mu ^2 v_c^3-54 \pi  \mu  v_c^7-8 \pi  v_c^{11}}
\eeq
and $v_c$ satisfies the following relation
\beq
\pi ^2 v_c^2 \left(72 \pi ^2 k \mu  T_c^2 v_c^2+6 k v_c^4-8 \pi  T_c v_c^5-27 k \mu \right)+160 \pi ^3 e^2 = 0 \, .
\eeq
Once again, any apparent singularities of the above expression for the critical temperature actually do not occur within the physical parameter space.


%

A parameter space plot is shown in Figure~\ref{domain6d} for the case of $k=+1$. In this case we find that there are two solutions to Eqs.~\eqref{dpd2p} provided $-0.068658 e^{4/3} \lessapprox \mu < 0$, while there are no solutions for potential critical points when $\mu \lessapprox -0.068658 e^{4/3}$. To determine which (if any) of these potential critical points are physical, we must check the various physicality conditions. We find that when the coupling is in the range $0.049633 e^{4/3}  \lessapprox \mu < 0$ one of the potential critical points corresponds to a negative mass solution, while for $-0.068658 e^{4/3} \mu \lessapprox 0.049633 e^{4/3}$ both potential critical points have positive mass. We also find that in the interval $\-0.064541 e^{4/3} \lessapprox \mu \lessapprox 0.049633 e^{4/3}$ one of the two critical points possesses negative Wald entropy.

\begin{figure*}[htp]
\centering
\begin{tabular}{cc}
\includegraphics[scale=.32]{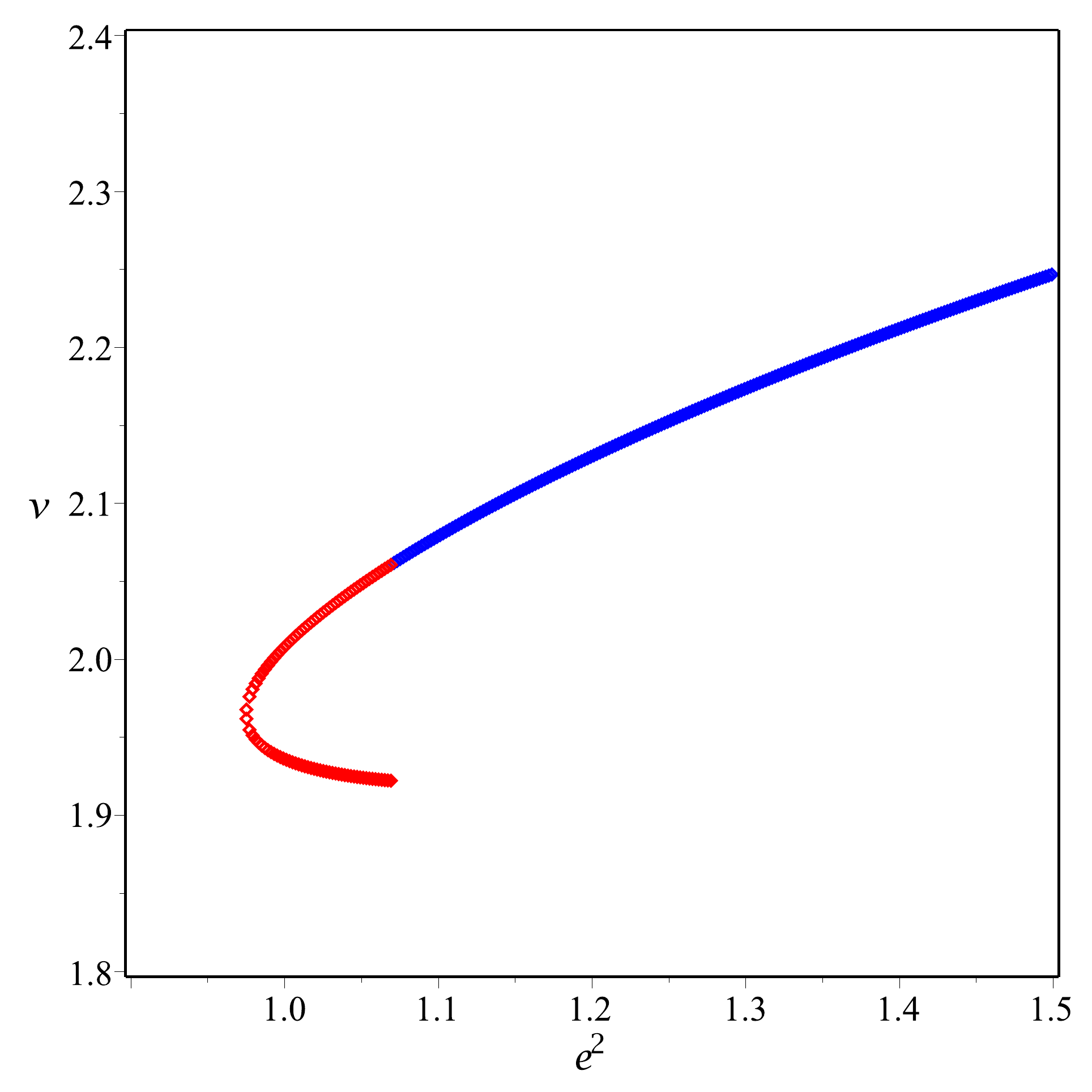}&\quad\quad
\includegraphics[scale=.32]{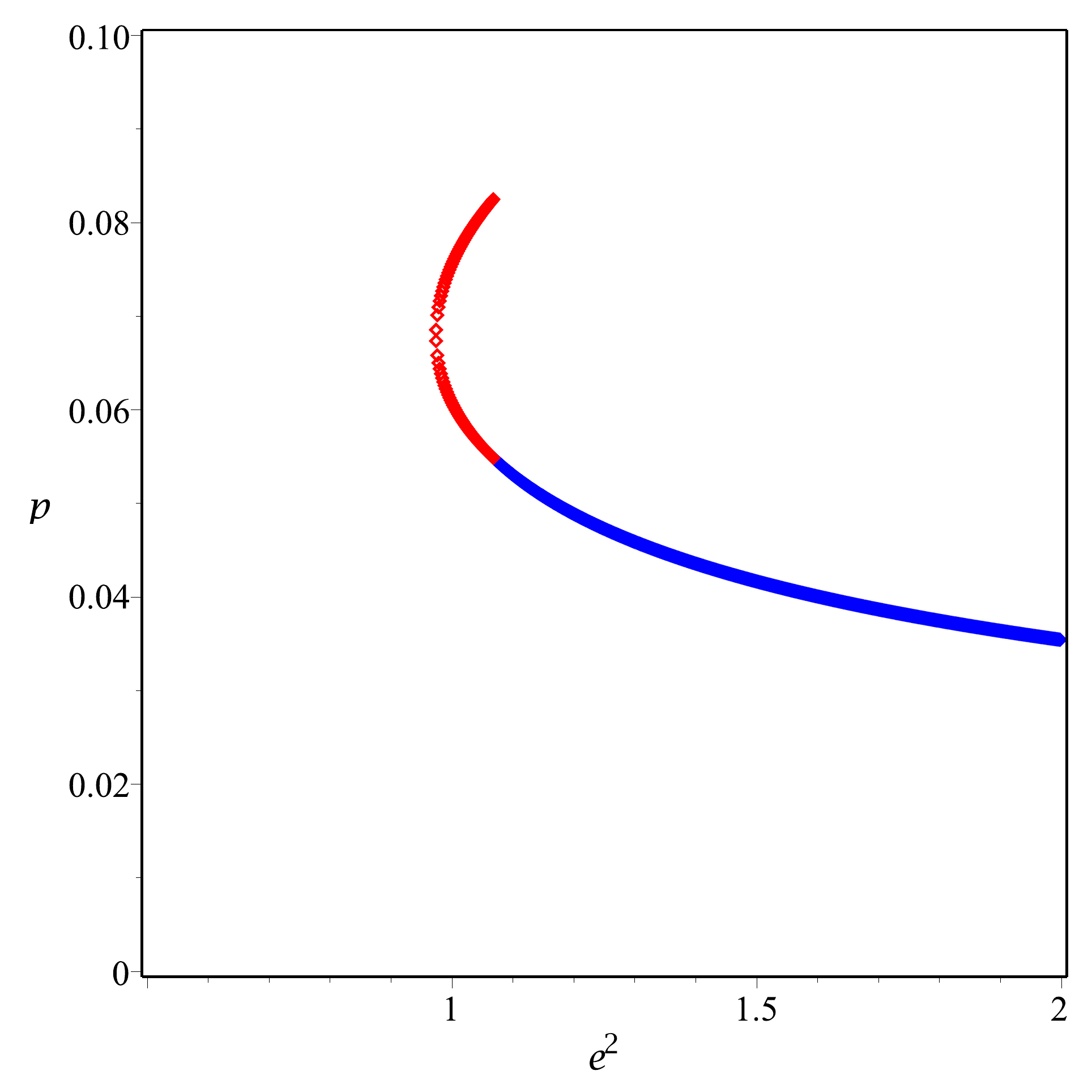}
\\
\end{tabular}
\caption{\textbf{Counting the number of critical points in six dimensions}. In six dimensions, The behaviour of critical volume (left) and critical pressure (right) versus electric charge. We set $\mu =-0.0675$  and $k=1$. The red curve shows in which region of electric charge two critical points exist.
}

\label{figPeVe6d}
\end{figure*}

\begin{figure*}[htp]
\centering
\begin{tabular}{cc}
\includegraphics[width=0.45\textwidth]{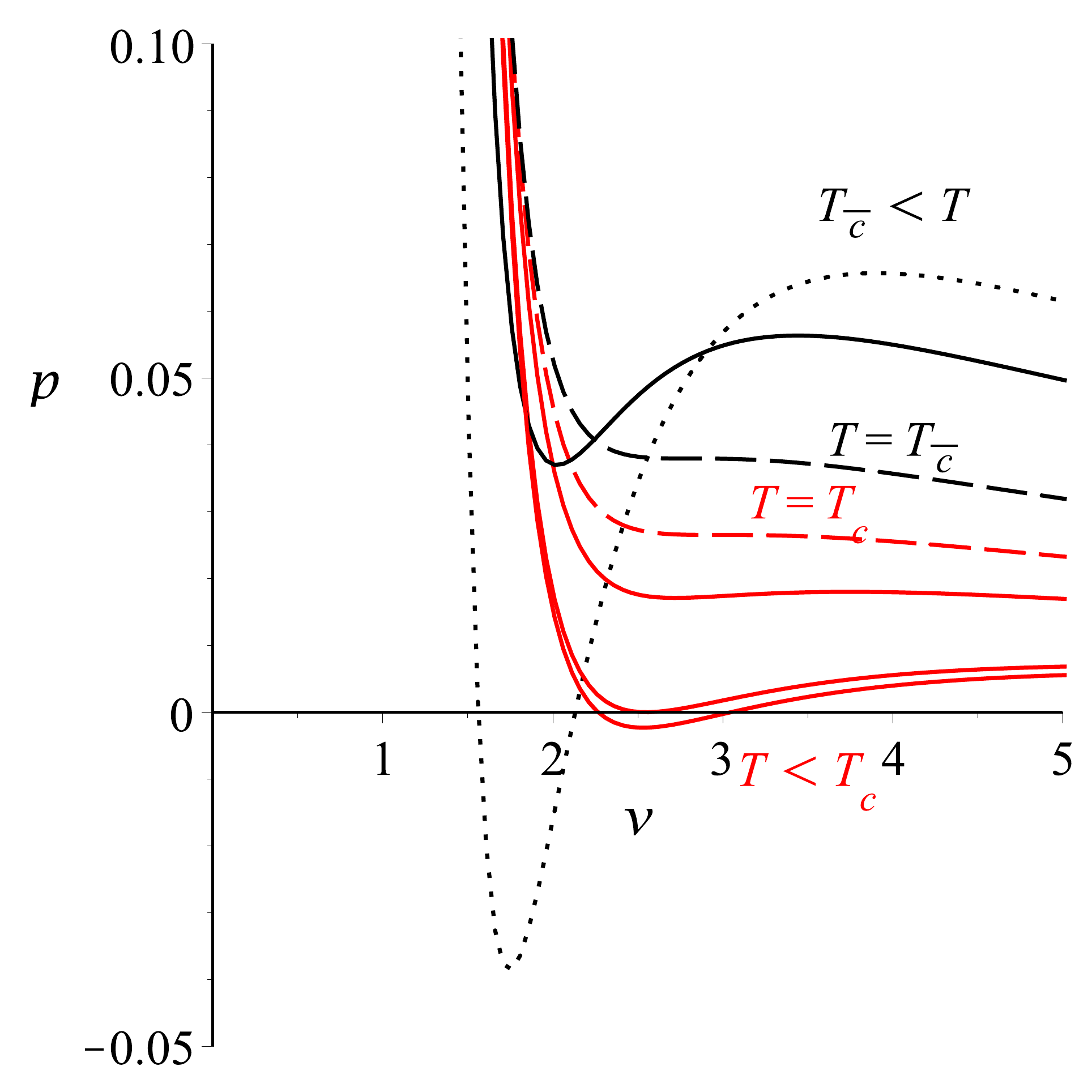}&
\\
\end{tabular}
\caption{\textbf{$p-v$ graph illustrating two first order phase transitions}. The behaviour of pressure versus volume for temperatures less than first critical point $T=T_c$ presented with dashed red line, $\ie$ $T=0.45 T_c, .48859 T_c,  0.8 T_c$ (red lines) and temperatures larger than second critical point $T=T_{\overline{c}}$ presented with dashed black line, $\ie$ $T=1.5 T_{\overline{c}},1.9 T_{\overline{c}}$ (black lines).  Here, we set $\mu/e^{4/3}=-0.0655$  giving $T_c e^{1/3} =0.24730$ and $T_{\overline{c}}e^{1/3}\approx 0.31746$.
}
\label{figPV6d}
\end{figure*}

 Let us discuss at greater length the situation in which there are two physical critical points since, as we will see, this leads to some interesting phase behaviour. The existence of these two solutions is dependent on the value of the electric charge and the coupling, as shown in Figure~\ref{figPeVe6d}. This plot shows the critical volume and pressure as a function of electric charge for particular choices of the coupling.  In Figure~\ref{figPV6d} we depict two separate first order transitions for two different critical points. The figure exhibits `double VdW' behaviour, in which a standard VdW transition takes place for cold temperatures, disappearing at a critical temperature $T_{c}$, and then reappearing once $T$ becomes greater than an even larger critical temperature $T_{\overline{c}}$. The intermediate region $T_c < T < T_{\overline{c}}$ is where both phases are indistinguishable,  and the associated isotherms are one-to-one functions $P(v)$. Note that, in Figure~\ref{figPV6d}, some of the isotherms dip below $P < 0$. Those portions of the curve are, of course, unphysical but are also naturally excluded via Maxwell's equal area law since the pressure at which the phase transition occurs is positive.




\begin{figure*}[htp]
\centering
\begin{tabular}{cc}
\includegraphics[scale=.32]{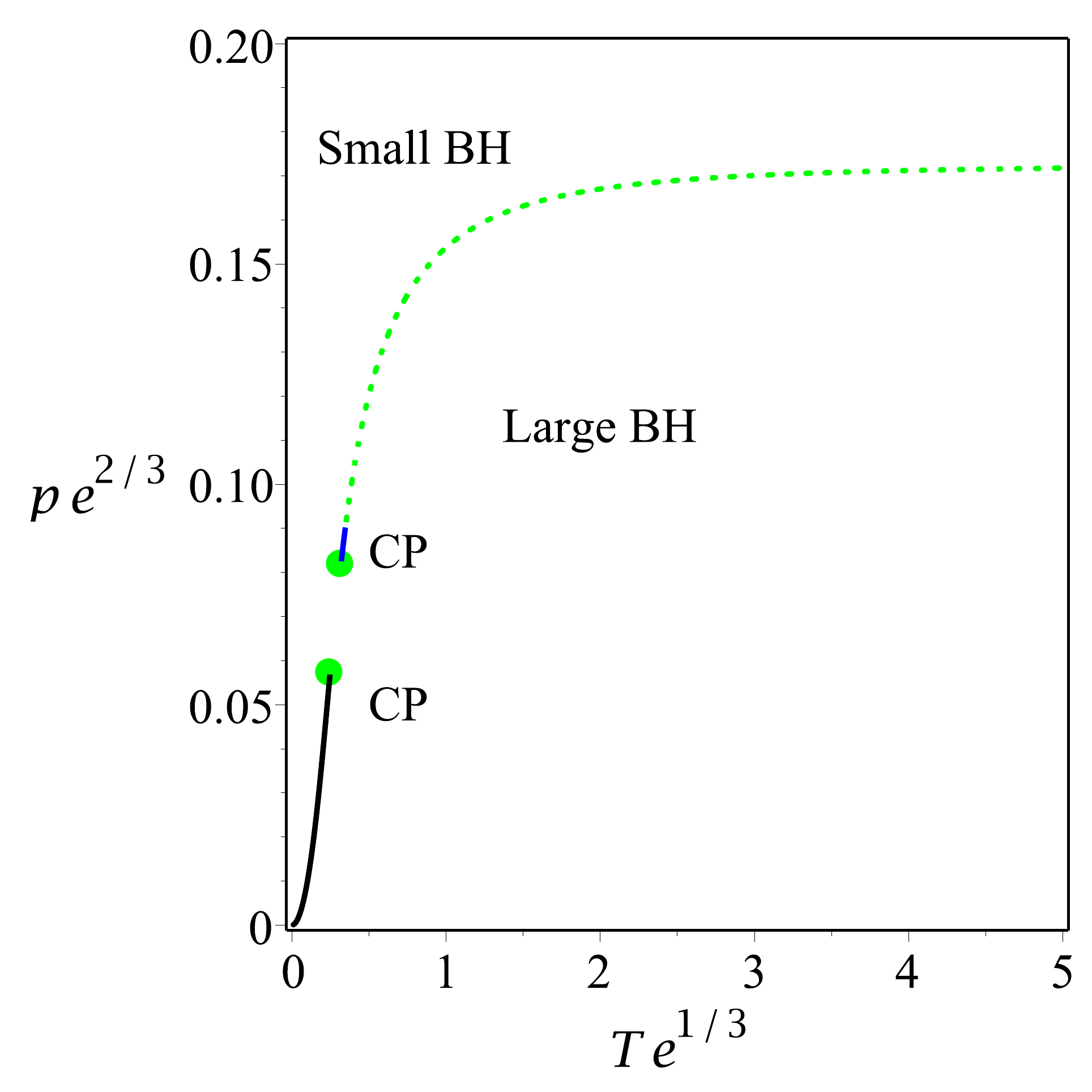}&\quad\quad
\includegraphics[scale=.32]{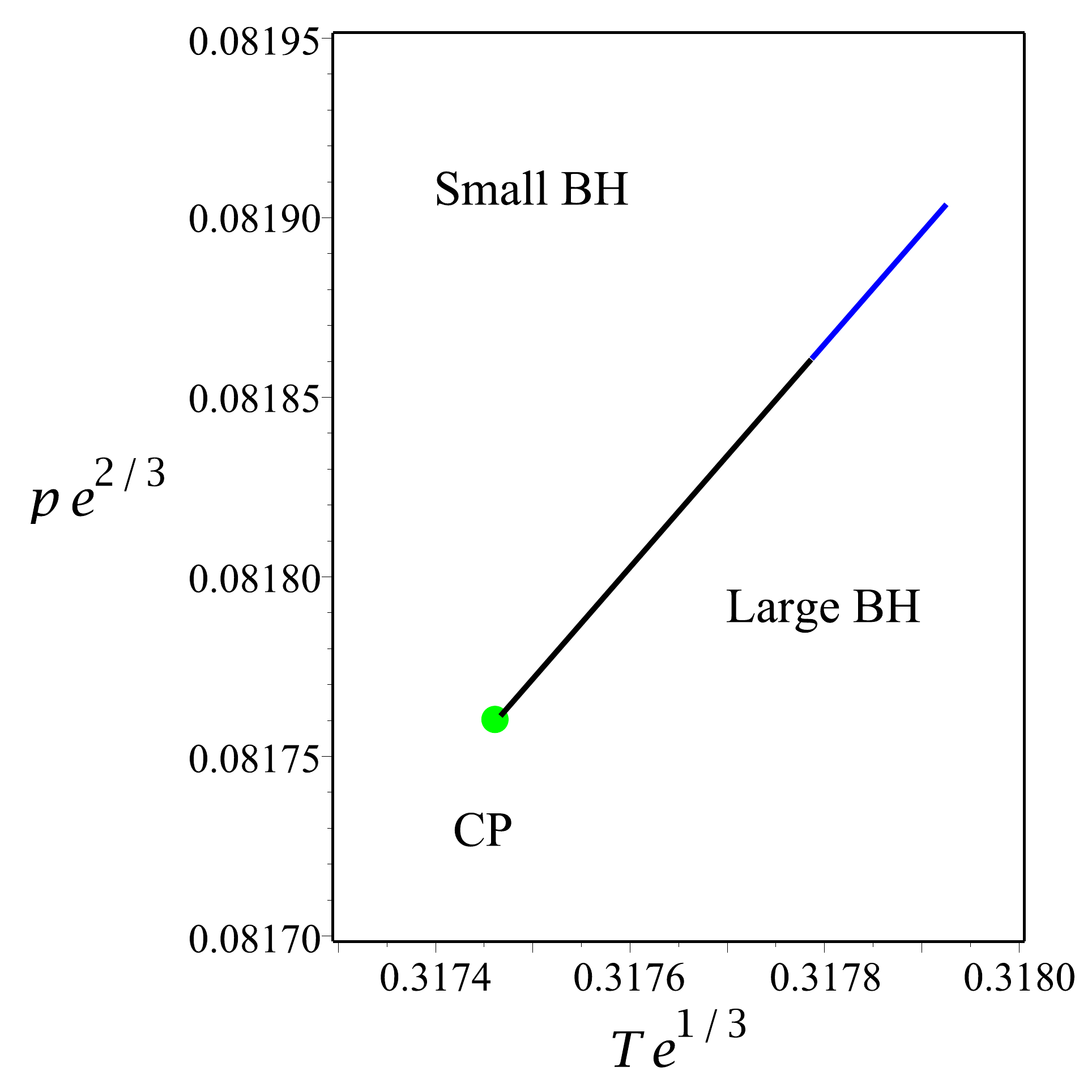}
\\
\includegraphics[scale=.32]{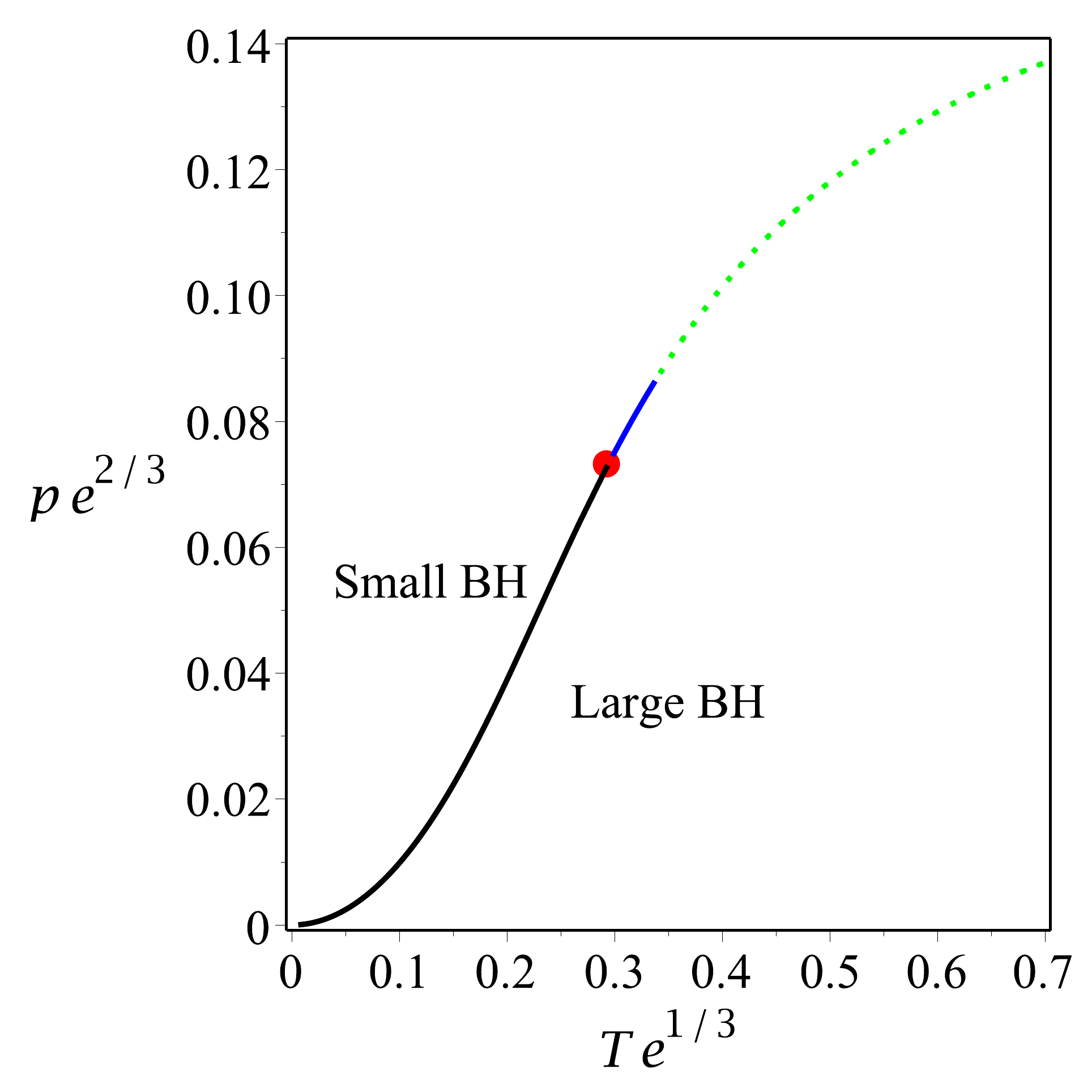}&\quad\quad
\includegraphics[scale=.32]{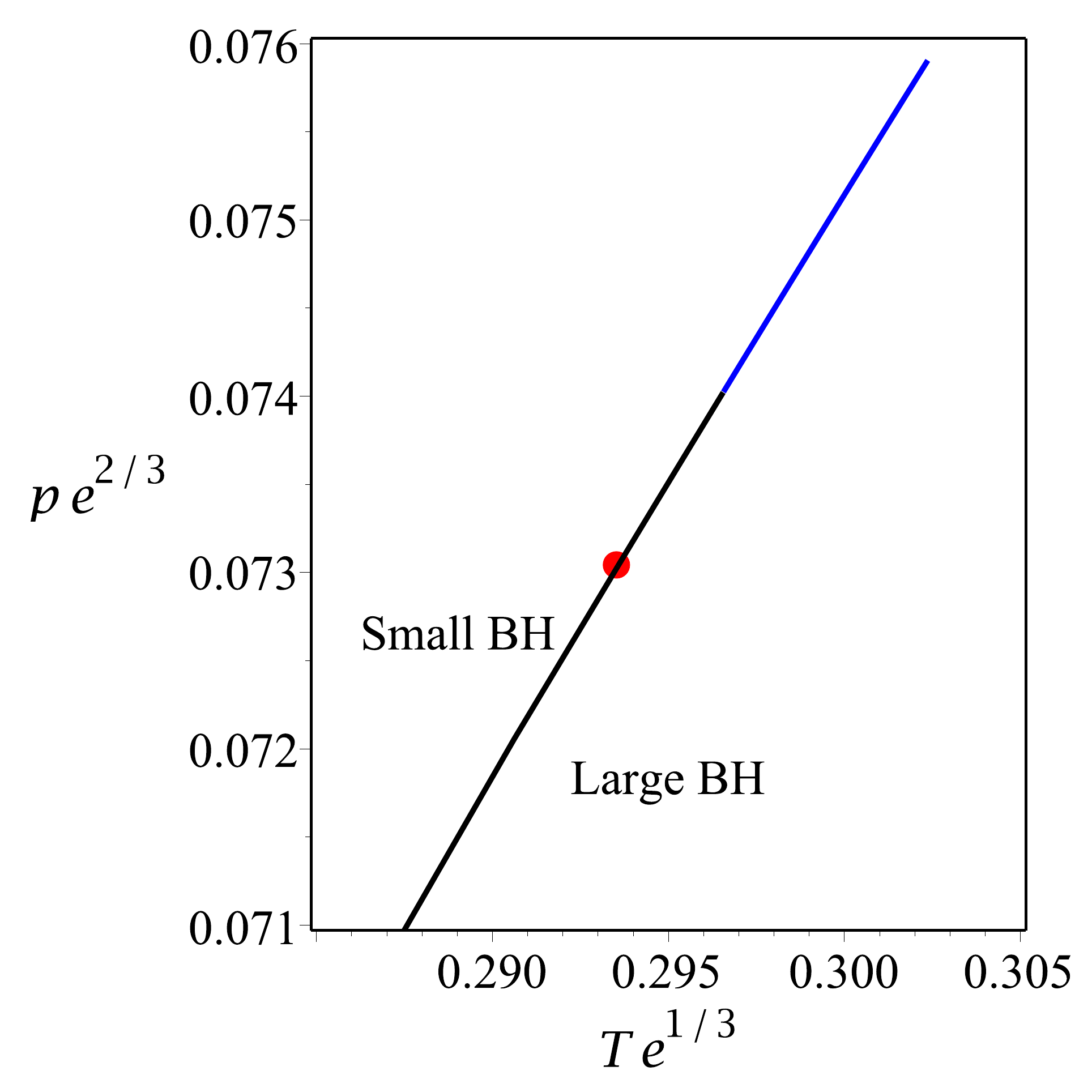}
\\
\end{tabular}
\caption{ \textbf{Phase diagrams depicting first and second order phase transitions in $d=6$}. \textit{Top left}:
Choosing  $\mu/e^{4/3}\approx -0.065487$ the critical quantities are $T_c e^{1/3} \approx 0.24730$ and $P_c e^{2/3} \approx 0.05712$, $T_{\overline{c}}e^{1/3}\approx 0.31746$ and $P_{\overline{c}}e^{2/3}\approx 0.08175$.  Green points denote   critical points and   black lines indicate a first-order phase transition. We
see that there is such a transition for  $T < T_c$ and another for   $T> T_{\overline{c}}$. Blue lines indicate negative entropy; green dotted lines indicate negative mass.
\textit{Top right }: A magnification of the region near the upper right critical point in the figure to the left, illustrating the existence of a small region with positive entropy (solid black line).
\textit{Bottom left}:  For  $\mu/e^{4/3}\approx-0.068658$ we obtain an isolated critical point (red point); the approximate values at the conjoined critical temperature and volume are $T_c e^{1/3} \approx
0.2766075924$ and $P_c e^{2/3} \approx 0.06725819565$. \textit{Bottom right}: A magnification of the bottom left plot close to the isolated critical point.
}
\label{PT6d}
\end{figure*}

For generic values of the coupling, each of the two critical points are described by mean field theory critical exponents. One marks the end point of a first order coexistence line, while the other marks the beginning of a  first order coexistence line, shown in Figure \ref{PT6d}.  The fact that the critical exponents are the mean field theory values can be deduced by examining the equation of state expanded near the critical point. Schematically, we obtain an expansion of the form
\beqa
\frac{P}{P_{c \pm}}&=&1+A \tau-B \tau \phi-C \phi^3+\cO(\tau \phi^2,\phi^4)\label{expcoeff}
\eeqa
where the coefficients $(A,B,C)$ are numerically determined from choices of the parameters.

For finely tuned values of the coupling, these two critical points merge into a single object known as an \textit{isolated critical point}.  Isolated critical points have been of interest since, in all known cases, they provide examples of critical exponents that deviate from the mean field theory values.  The first examples using Lovelock and quasi-topological gravity were discussed in \cite{Frassino:2014pha, Dolan:2014vba,Hennigar:2015esa} where the isolated critical points happen for hyperbolic horizons and massless black holes, and coincided with a thermodynamic singularity. For Lovelock and quasi-topological black holes with conformal scalar hair \cite{EricksonRobie, Dykaar:2017mba} isolated critical points were discussed in five and higher dimensions, providing first examples of isolated critical points for black holes of any mass and away from the thermodynamic singularity. Here we observe these points for the first time in six dimensions, and also for spherical horizons in pure gravity. Note also that these examples of isolated critical points do not correspond to any thermodynamic singularity, as the slope of the $P-T$ coexistence curve is non-zero.

We have confirmed that this isolated critical point associated with the parameters given in Figure \ref{PT6d} has positive mass, and therefore the associated black holes have sensible asymptotic structure.  Since the equation for finding the critical values of $T$ and $v$ \reef{dpd2p} is seventh  order in $v$ and third order in $T$, it is only feasible to solve these equations numerically.  From the numerics we can extract the form of the equation of state expanded near the critical point. We find that the coefficient $B$ in \reef{expcoeff} goes to zero as parameters approach those yielding an isolated critical point. The critical exponents corresponding to the isolated critical point are given according following prescription.

 To get the critical exponents, we follow the prescription outlined in \cite{Gunasekaran:2012dq}.  For the specific heat at constant volume
\beqa
C_v=T \frac{\partial s}{\partial T}\Big|_v = 0 \, .
\eeqa
we  find that the critical exponent $\tilde{\alpha}=0$,
despite the fact that the entropy  (naively) depends on temperature.  Using  \reef{expcoeff},   the fact that during the phase transition (between large/small black holes) the pressure remains constant, and Maxwell's area law  written in differential form as
\beqa
dP=-P_c(2 D \tau \omega+3 C \omega^2)d\omega
\eeqa
 we find   $\omega_{l,s}\propto \tau$.  Hence $\tilde{\beta}=1$. To evaluate the exponent $\tilde{\gamma}$ we compute the behaviour of the isothermal compressibility near criticality, finding
\beqa
\kappa_T=-\frac{1}{v}\frac{\partial v}{\partial P}\Big|_T=-\frac{1}{(1+\omega) P_c \left(-2 D \omega \tau-3 C \omega^2\right)}\propto \frac{1}{\tau^2}
\eeqa
 where we used the relation $\omega\propto \tau$ as mentioned above.  We thus obtain $\tilde{\gamma}=2$. These values for the exponents ($\beta,\gamma$) are different from the standard exponents  in \reef{exponents}  but match the non-standard critical exponents found in~\cite{Dolan:2014vba} for seven-dimensional Lovelock gravity.

Let us close this section by mentioning that there are no physical critical points in the hyperbolic case. We find that for any negative coupling the equations~\eqref{dpd2p} admit possible solutions, however these always correspond to negative mass black holes.

\subsection{Remarks on higher dimensions}

To close our considerations of the canonical ensemble, we present a few remarks on the situation in general dimensions. Rather than perform an exhaustive analysis --- which would require a case-by-case study --- here we limit the discussion to small values of the coupling and black holes with spherical horizons. This will allow us to understand how the cubic theory affects the critical behaviour already present in Einstein gravity.

Let us begin by recalling that in Einstein gravity charged black holes with spherical horizon present a single critical point in all dimensions with the critical values being given by~\cite{Gunasekaran:2012dq}
\beq
P_c^{(0)} = \frac{(d-3)^2}{(d-2)^2 \pi v_c^2} \, , \quad T_c^{(0)} = \frac{4 (d-3)^2}{(d-2)(2d-5)\pi v_c} \, , \quad v_c^{(0)} = \left[\frac{(d-2)^2(2d-5) \pi e^2}{d-3} \right]^{\frac{1}{2(d-3)}} \, .
\eeq
When the cubic coupling is turned on and is perturbatively small, the critical values given above become modified. The first order corrections are given by
\begin{align}
P_c^{(\mu)} &= P_c^{(0)} - \frac{256 (d-2)^{-(5d-9)/(d-3)}  \pi^{-d/(d-3)} }{(2 d-5)^3 \left(4 d^4-49 d^3+291 d^2-514 d+184\right)} \left( \frac{d-3}{2d-5}\right)^{\frac{3}{d-3}}
\nn\\
&\times \bigg(416 d^9+4424 d^8-200812 d^7+2129198 d^6-11437255 d^5+35957054 d^4
\nn\\
&-68280093 d^3+75654408 d^2-43205940 d+8802960 \bigg) \left(\frac{\mu}{|e|^{6/(d-3)}} \right)  + {\cal O}\left(\mu^2 \right) \, ,
	\\
T_c^{(\mu)} &= T_c^{(0)} - \frac{1536 (2 d-5)^{-3}  \pi^{(1-2d)/(2d-6)} }{(d-2)^5 \left(4 d^4-49 d^3+291 d^2-514 d+184\right)} \left(\frac{(d-2)^2(2d-5)}{d-3} \right)^{\frac{5}{2(3-d)}}	
	\nn\\
	&\times \bigg( 128 d^9+304 d^8-39908 d^7+464048 d^6-2604697 d^5+8471996 d^4
	\nn\\
	&-16664635 d^3 +19319904 d^2-11851020 d+2803440 \bigg) \left(\frac{\mu}{|e|^{5/(d-3)}} \right) + {\cal O}\left(\mu^2 \right) \, ,
	\\
v_c^{(\mu)} &= v_c^{(0)} + \frac{384 (d-2)^{-4} (2d-5)^{-3} \pi^{-3/(2d-6)} }{(d-3)^2\left(4 d^4-49 d^3+291 d^2-514 d+184\right)} \left(\frac{(d-2)^2(2d-5)}{d-3} \right)^{3/(6-2d)}	
	\nn\\
	&\times \bigg(32 d^9+7848 d^8-171340 d^7+1533478 d^6-7507951 d^5+21947526 d^4
	\nn\\
	&-38744053 d^3+39179688 d^2-19237140 d+2564880 \bigg) \left(\frac{\mu}{|e|^{3/(d-3)}} \right)  + {\cal O}\left(\mu^2 \right) \, .
\end{align}
Although it is not immediately obvious from these expressions, the effect of the higher-order coupling is different depending only on whether the spacetime dimension is four or higher. In four dimensions, the cubic coupling leads to an increase in the critical volume, while decreasing both the critical temperature and the critical pressure. In all higher dimensions, the effect is reversed: the critical volume is decreased, while the critical temperature and pressure are increased. One can readily check that for small values of the coupling, the critical points meet all physicality conditions. We can also compute the effect on the Van der Waals ratio:
\begin{align}
\frac{P_c v_c}{T_c} &= \frac{2d-5}{4d-8} \bigg[1 - \frac{256(d-4)(2d-5)^{-3} \pi^{-2/(d-3)} }{(d-3)^2(d-2)^4\left(4 d^4-49 d^3+291 d^2-514 d+184\right)}
	\nn\\
	&\times\left(\frac{d-3}{(d-2)^2(2d-5)} \right)^{\frac{2}{d-3}} \bigg(32 d^9+3720 d^8-84548 d^7+758186 d^6-3668673 d^5
	\nn\\
	&+10467583 d^4-17711283 d^3+16541067 d^2-6667680 d+106860 \bigg) \left(\frac{\mu}{|e|^{4/(d-3)}} \right)
	\nn\\
	&+ {\cal O}\left(\mu^2 \right) \bigg] \, .
\end{align}
In four dimensions the ratio is unaltered (as discussed above), while in all higher dimensions there is a correction dependent on the coupling and charge that serves to increase the ratio compared to its value in Einstein gravity. The conclusion, then, is that while the van der Waals ratio is a ``universal'' quantity in Einstein gravity, it is sensitive to the particular details of the solution in more general theories of gravity.

\section{Holographic hydrodynamics} \label{sec: holog}
\label{sec:holo_hydro}

\subsection{Review of black branes}

The thermodynamic properties of black branes in the cubic theory were studied in~\cite{Hennigar:2017umz}. Here we will review some of the properties of uncharged black branes in the cubic theory that will be useful in the following subsection.

When $k=0$, the near horizon equations of motion simplify dramatically:
\begin{align}
m &= r_+^{d-4} \bigg[ \frac{r_+^{3}}{L^2} + \frac{ 1024  \mu \pi^4 (d^2 + 5 d - 15)   T^3}{ (4 d^4 - 49 d^3 + 291 d^2 - 514 d + 184)}  \bigg]
\nn\\
0 &=  (d-1)\frac{r_+^2}{L^2} - 4 \pi r_+ T
	- \frac{512  (d-4)(d^2 + 5d - 15)  \mu \pi^3  T^3}{(4 d^4 - 49 d^3 + 291 d^2 - 514 d + 184) r_+}  \, .
\end{align}
In four dimensions, these equations imply that the temperature as a function of horizon radius is exactly the same in the cubic theory as it is in Einstein gravity. In higher dimensions there are corrections to this profile, and in both four and higher dimensions the mass receives corrections. Taking the discriminant of the second equation above we find that it changes from positive to negative when the coupling takes the value
\beq
\mu^* = -\frac{(4 d^4 - 49 d^3 + 291 d^2 - 514 d + 184)L^4}{54(d-4)(d-1)^2(d^2 + 5d - 15)} \, .
\eeq
When taking $d =5$, this reduces to $\mu^* = - 79 L^4/1890$ which we encountered earlier in Sections~\ref{sec:HawkingPage} and~\ref{sec:thermofpe}. In those cases we were interested in black holes with spherical horizon topology and the bound implied that large black holes simply do not exist when the coupling is smaller than this value. In  the present case, the bound applies \textit{for all values of the horizon radius} and we find that when the coupling exceeds this value there is no sensible solution for $T$ as a function of $r_+$.

The next important point we will note is that both of the near horizon equations are satisfied (with vanishing mass) for all values of the horizon radius when the temperature and coupling are given by
\beq
T_p = \frac{3(d-2) r_+}{8 \pi L^2} \, , \quad  \mu_p = - \frac{L^4(4d^4 - 49 d^3 + 291 d^2 - 514 d + 184)}{54 (d-2)^3(d^2 + 5d - 15)} \, .
\eeq
Interestingly these conditions also imply that the entropy of the black brane vanishes. This is the reason for the labels ``$p$'' since as we will see, in this limit the ratio of shear viscosity to entropy density has a pole. Comparing the above results, we notice that $\mu_p < \mu^*$ indicating that we reach the point where $M = 0$ and $S = 0$ before the point where solutions fail to exist. Further exploration reveals that for all $|\mu| > |\mu_p|$ the mass of the black holes is negative, indicating that the full solutions do not exist. This means that the point $\mu = \mu_p$ actually serves as the limit of sensible coupling for black branes in the cubic theory and we must constrain $\mu \in [\mu_p, 0]$. In this interval, we find that the mass and entropy of the black branes is always positive.  An interesting point is that in four-dimensions $\mu_p$ coincides with $\mu_c$ corresponding to the critical limit of the theory; however, in higher dimensions $\mu_p$ is always distinct from $\mu_c$.

While the equation determining the temperature as a function of horizon radius is a cubic, only a single branch of the solution is physical --- one gives negative temperature, while the other gives negative mass. In terms of $\mu_p$, the physical solution can be expressed quite simply as
\beq
T = \frac{3 r_+}{4 \pi L^2} \sqrt{\frac{(d-2)^3 \mu_p}{(d-4) \mu }} \cos \left(\frac{\theta + \pi}{3} \right)
\eeq
where
\beq
\cos\theta =  \frac{(d-1) \sqrt{(d-4) \mu/\mu_p}}{(d-2)^{3/2}} \, .
\eeq
The temperature therefore exhibits a linear dependence on the horizon radius, with the slope of the line depending on the spacetime dimension and the value of the coupling. In the limit $\mu \to 0$, the expression limits to $T = (d-1) r_+/(4\pi L^2)$, while when $\mu \to \mu_p$ it limits to $T = 3(d-2) r_+/(8\pi L^2)$.  These two lines bound all other curves.

We can also write an explicit expression for the entropy for the physical branch of black branes. This reads
\begin{align}
s &:= \frac{S}{L^{d-2} {\rm Vol} \left(\mathbb{R}^{d-2} \right)}
	\nn\\
	&= \frac{r_+^{d-2}}{4 L^{d-2}} \left[1 - \frac{4(d-2)}{d-4} \sin\left( \frac{1}{3} \arcsin \left( \frac{(d-1) \sqrt{(d-4) \mu/\mu_p}}{(d-2)^{3/2}} \right) \right)  \right] \, .
\end{align}
This entropy is vanishes only when $r_+ \to 0$ or when $\mu \to \mu_p$. This behaviour is precisely in line with what we would expect for the entropy, and so there is no need to be concerned with shifting it for the black branes. Indeed, any shift in the entropy would result in a non-zero entropy assigned to the spacetime when it does not contain a horizon. Let us now study the ratio of shear viscosity to entropy density.

\subsection{Computation of $\eta/s$}

As a step toward a full understanding of the generalized quasi-topological class of theories in the context of the AdS/CFT correspondence, we compute here the ratio of shear viscosity to entropy density $\eta/s$.

For field theories possessing Einstein gravity duals, the shear viscosity to entropy density ratio has a universal form $\eta/s = 1/(4\pi)$. It was conjectured by Kovtun, Son, and Starinets that this represents a universal lower bound for all substances~\cite{Kovtun:2004de}, i.e. $\eta/s \ge 1/ (4\pi)$ (the KSS bound). However, it was later discovered that the inclusion of higher derivative corrections can actually lead to violations of this bound~\cite{Brigante:2007nu}. Here we will compute $\eta/s$ for field theories dual to the cubic generalized quasi-topological theory in all dimensions and show that the KSS bound always holds. In fact, we show that the ratio $\eta/s$ takes on all real values $\eta / s \in [(4\pi)^{-1}, \infty)$ as a function of the coupling $\mu$.

For this computation, we are interested in the planar class of metrics,
\beq
ds^2 = \frac{r^2}{L^2} \left(-g(r) dt^2 + \sum_i dx_i^2 \right) + \frac{L^2 dr^2}{r^2 g(r)} \, .
\eeq
We transform the metric by introducing $z = 1 - r_+^2/r^2$, which compactifies the space outside the horizon. The transformed metric reads,
\beq
\label{eqn:zMetric}
ds^2 = \frac{r_+^2}{L^2 (1-z)} \left(-g(z) dt^2 + \sum_i dx_i^2  \right) + \frac{L^2}{4 g(z)} \frac{dz^2}{(1-z)^2}
\eeq
and $g(z)$ has a simple zero at $z=0$, and $g(1) = f_\infty$.      Near the horizon, we can expand $g(z)$ as,
\beq
g(z) = g_0^{(1)} z + g_0^{(2)} z^2 + g_0^{(3)} z^3 + \cdots \, .
\eeq
The field equations fix $g_0^{(i)}$ for $i \neq 2$. As mentioned earlier, the second derivative of the metric function near the horizon is undetermined, but is fixed by demanding that the numerical solution converges to the asymptotic solution without growing mode. It  is in this same way that $g_0^{(2)}$ must be determined.  Of course, the parameters $g_0^{(i)}$ are straightforwardly related to the parameters $a_i$ used in the near horizon expansion in Eq.~\eqref{eqn:nh_ansatz}. The relevant ones for our purposes below are,
\begin{align}
g_0^{(1)} &= \frac{2 \pi T L^2}{r_+} \, , \quad g_0^{(2)} = - \frac{L^2}{4 r_+} \left(2 \pi T - r_+ a_2 \right)
	\nn\\
	g_0^{(3)} &= -\frac{L^2}{8 r_+} \left(2 \pi T - r_+ a_2 - r_+^2 a_3 \right) \, .
\end{align}
We will need the expression for $a_3$ as a function of $a_2$. This can be obtained from the following equation, which is the ${\cal O}\left((r-r_+)^3\right)$ component of the near horizon field equations:
\begin{align}
0 &= \frac{16 \mu }{3 \mu_p} \frac{(4 \pi T)^2 L^6 }{(d-2)^3} a_3 + \frac{32 \mu}{9 \mu_p} \frac{4 \pi T L^6}{(d-2)^3} {a_2}^2 + \left(\frac{8 \mu}{9\mu_p} \frac{(5d-28) (4 \pi T)^2 L^6}{(d-2)^3 r_+} - 2 r_+ L^2 \right)a_2
	\nn\\
	&+ \frac{16 \mu}{27 \mu_p} \frac{(d-5)(d-10) (4 \pi T)^3 L^6}{(d-2)^3 r_+^2} + (d-1)(d-2) r_+ - 2 (d-3)(4 \pi T) L^2
\end{align}



Next, following~\cite{Paulos:2009yk} we perturb the metric~\eqref{eqn:zMetric} by the shift
\beq
dx_i \to dx_i + \epsilon e^{-i\omega t} dx_j \, .
\eeq
The perturbed metric is substituted into the Lagrangian and a small $\epsilon$ expansion is performed. The result gives,
\begin{align}
&\sqrt{-g}{\cal L} = \frac{1}{16 \pi} \bigg[ \cdots - \frac{\omega^2 \epsilon^2 r_+^{d-3}}{L^{d-4} g_0^{(1)} z}  \bigg\{1 + \frac{\mu}{L^4} \frac{48}{d-3} \bigg( \frac{(17 d^3 - 209 d^2 + 632 d - 566)}{(4d^4 - 49 d^3 + 291 d^2 - 514 d + 184)} (g_0^{(1)})^2
	\nn\\
	& -\frac{4(21d^3 - 289 d^2 + 740 d - 478)}{(4d^4 - 49 d^3 + 291 d^2 - 514 d + 184)} g_0^{(1)} g_0^{(2)}
- \frac{24 (21 d^2 - 62d + 38)}{(4d^4 - 49 d^3 + 291 d^2 - 514 d + 184)} g_0^{(1)}g_0^{(3)}
	\nn\\
	&- \frac{16(21 d^2 - 62 d + 38)}{(4d^4 - 49 d^3 + 291 d^2 - 514 d + 184)} (g_0^{(2)})^2 \bigg) \bigg\}  + {\rm Regular} \bigg]
\end{align}

Now, using the `time' formula, the shear viscosity is given by
\beq
\eta = - 8 \pi T \lim_{\omega,\epsilon \to 0} \frac{{\rm Res}_{z=0} \sqrt{-g}{\cal L}}{\omega^2 \epsilon^2}
\eeq
which we can read off to be
\begin{align}
\eta &= \frac{ T r_+^{d-3}}{8L^{d-4} g_0^{(1)}}  \bigg\{1 + \frac{48 \mu}{L^4 (d-3) (4d^4 - 49 d^3 + 291 d^2 - 514 d + 184)}
\nn\\
&\times  \bigg( (17 d^3 - 209 d^2 + 632 d - 566) (g_0^{(1)})^2
	\nn\\
	& -4(21d^3 - 289 d^2 + 740 d - 478) g_0^{(1)} g_0^{(2)}
- 24 (21 d^2 - 62d + 38) g_0^{(1)}g_0^{(3)}
	\nn\\
	&- 16(21 d^2 - 62 d + 38) (g_0^{(2)})^2 \bigg) \bigg\} \, .
\end{align}
Recalling that the entropy density (for planar black holes) takes the form,
\beq
s = \frac{S}{L^{d-2} {\rm Vol}\left(\mathbb{R}^{d-2}\right)} =  \frac{r_+^{d-2}}{4 L^{d-2}} \left[1 + \frac{384 \pi^2  (d-2)(d^2 + 5d - 15)}{4d^4 - 49 d^3 + 291 d^2 - 514 d + 184}  \frac{\mu T^2}{r_+^2}  \right]
\eeq
it is straightforward (if messy) to write down the ratio $\eta/s$.  Computing this ratio for arbitrary values of $\mu$ requires implementing a numerical scheme to determine the value of $a_2$ for a given black hole. However, insight can be easily gained by considering a small $\mu$ expansion of $\eta/s$. This can be performed analytically, and the result is
\beq
\frac{\eta}{s} = \frac{1}{4 \pi} \left[1 - \frac{12 \mu}{L^4} \frac{(d-1)^2(23 d^4 - 83 d^3 - 18 d^2 + 256 d - 136)}{(d-3)(4d^4 - 49 d^3 + 291 d^2 - 514 d + 184)} + {\cal O}(\mu^2) \right]  \, .
\eeq
The dimension dependent factor in the above is always positive (at least for $d \ge 4$), and since $\mu$ must be negative for sensible AdS asymptotics, this means that the KSS bound $\eta/s \ge 1/(4\pi)$ holds in all dimensions in the cubic generalized quasi-topological theories, at least when the coupling is small.

An interesting property of the generalized quasi-topological theories is that the entropy density of black branes is non-trivial~\cite{Hennigar:2017umz}.  It turns out that this actually leads to a pole in the ratio $\eta/s$ in all dimensions. Recall from above that for the special values
\beq
T_p = \frac{3(d-2) r_+}{8 \pi L^2} \, , \quad  \mu_p = - \frac{L^4(4d^4 - 49 d^3 + 291 d^2 - 514 d + 184)}{54 (d-2)^3(d^2 + 5d - 15)} \, ,
\eeq
the near horizon equations are satisfied identically and the entropy vanishes linearly as $\mu \to \mu_p$.
%
Meanwhile, our numerical investigations (see below) indicate that the shear viscosity is always strictly positive on the interval for $\mu \in (\mu_p, 0)$. Thus, in all dimensions there is a pole in the ratio of shear viscosity to entropy density. This is quite an interesting result -- since the ratio smoothly connects between $\eta/s = 1/(4\pi)$ (for $\mu = 0$) and $\eta/s = \infty$ (for $\mu = \mu_p$), a particular coupling can always be chosen to match $\eta/s$ for any fluid in nature.


The pole in $\eta/s$ is also universal in the following sense. If we were to include cubic quasi-topological or Lovelock terms into the action, these terms would not disturb our result. This is because quasi-topological and Lovelock terms do not modify the black hole entropy from its Einstein gravity value~\cite{Hennigar:2017umz}, and it is the vanishing of $s$ that gives rise to the pole.  It would be interesting to see if this behaviour persists at higher order in the curvature in four and higher dimensions.

\begin{figure}[h]
\centering
\includegraphics[width=0.7\textwidth]{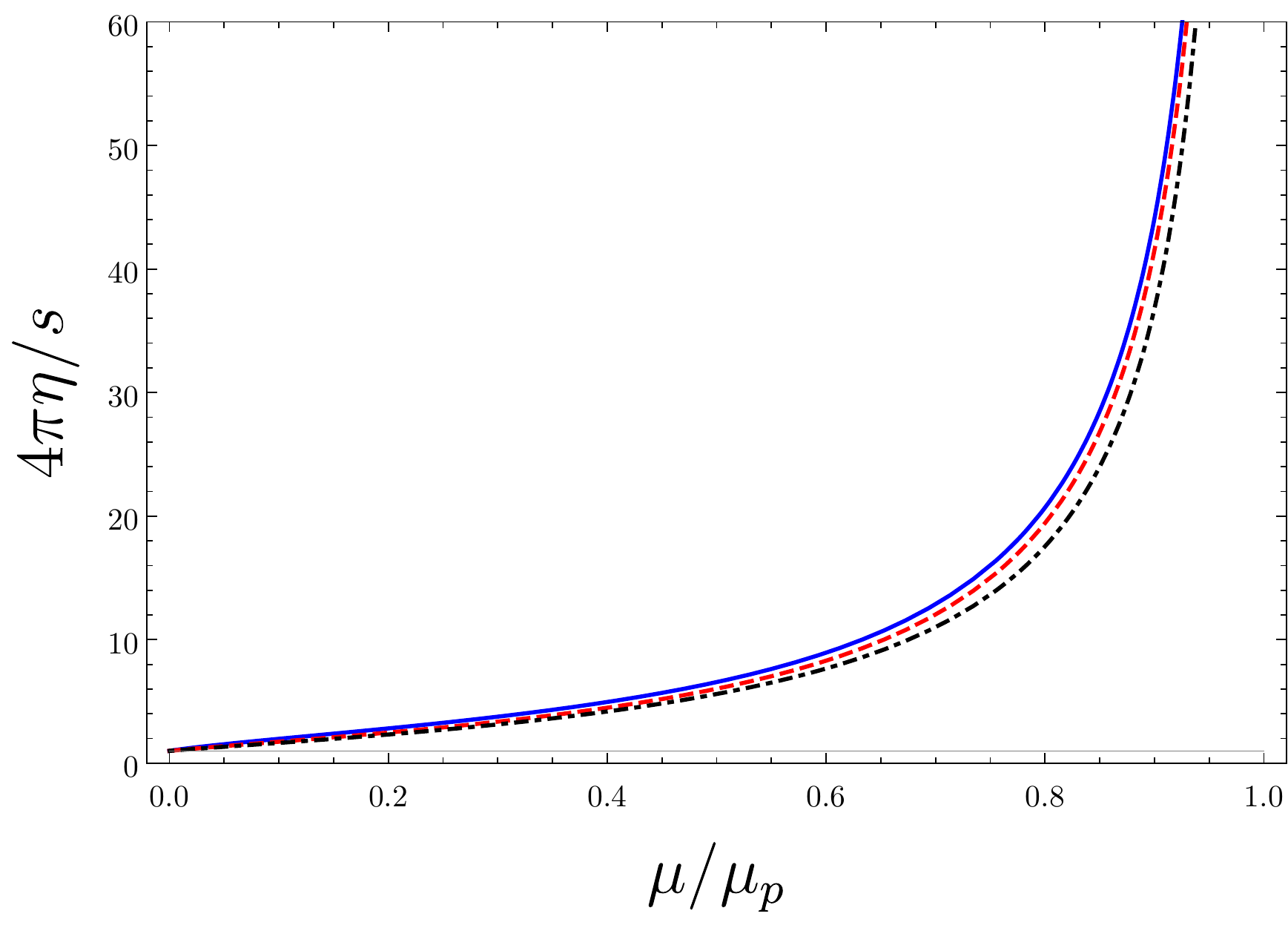}
\caption{{\bf Ratio of shear viscosity to entropy density}: Plots of the ratio $\eta/s$ in four (blue, solid), five (red, dashed) and six (black, dot-dashed) dimensions. In all cases, the thin grey line represents the universal Einstein gravity value of $\eta/s  = 1/(4\pi)$. In all cases, a $[7|7]$ order Pad\'e approximant was used for $a_2$ (see Appendix~\ref{sec:pade}). }
\label{fig:etaOvers}
\end{figure}

To see the explicit $\mu$ dependence of $\eta/s$, we must resort to numerical techniques to determine the parameter $a_2$, or the Pad\'e approximant method outlined in the appendix. Either is computationally costly, and therefore we present only a few example dimensions in Figure~\ref{fig:etaOvers}.  In these plots, we see the same characteristic structure: for $\mu = 0$, $\eta/s$ begins at $1/(4\pi)$ and then monotonically increases as $\mu \to \mu_p$.  Increases the spacetime dimension shifts the curves down slightly, but the overall structure is the same in all dimensions.

Of course, there is no good reason to believe that sensible CFT duals will exist over the whole range $\mu \in (\mu_p, 0)$.  Indeed,  it was found in~\cite{Bueno:2018uoy} that the putative CFT dual of the four-dimensional theory is consistent only for $\mu > -100489 L^4/64314864$, which is a tighter constraint than that imposed by $\mu > \mu_p$. Determining the equivalent constraints for the higher-dimensional versions of the theory is an interesting --- and important --- task that would require a careful analysis of causality constraints and positivity of energy flux in the dual CFT (see, for example,~\cite{Hofman:2008ar, Myers:2010jv,Camanho:2014apa}). We hope to come back to these issues in subsequent work.

\section{Discussion}

We have studied electrically charged static AdS black holes in cubic generalized quasi-topological gravity. These black holes are characterized by a single metric function, and our study considered spherical, planar, and hyperbolic base manifolds. The full field equations due not admit an analytic solutions. We have constructed a number of numeric solutions to the full field equations, finding that the black holes are non-hairy generalizations of the usual Schwarzschild solutions, characterized by their mass and charge alone. We have found that increasing the electric charge has the effect of increasing the horizon radius. We have found that imposing that the theory admits solutions with sensible asymptotics requires that $\mu M < 0$, i.e. that the cubic coupling constant is negative for positive mass solutions.

While the field equations cannot be solved exactly, evaluating them near the horizon of a black hole simplifies them dramatically. In solving the equations order-by-order near the horizon, we find that the two lowest order equations  involve powers of the temperature and horizon radius (along with the coupling constant and electric charge). Thus, the thermodynamic properties of the solutions can be studied exactly by solving these polynomial equations. We verified the extended first law and Smarr relation for the solutions, working in the framework of black hole chemistry, treating the cosmological constant as a thermodynamic pressure.

In both the charged and uncharged cases, the most dramatic differences between the cubic theory and Einstein gravity arise for small black holes. For example, in the four-dimensional case, small uncharged black holes are thermodynamically stable in the cubic theory, while unstable in Einstein gravity. In the five-dimensional case at fixed charge and coupling, the cubic theory does not admit arbitrarily small black hole solutions as eventually the mass becomes negative. Further, if the magnitude of the cubic coupling is made too large in five dimensions, the theory no longer admits large black hole solutions.

We have studied the phase structure of these black holes in the uncharged case. The four-dimensional case is dramatically different: ensuring that the black holes possess positive entropy, the usual Hawking-Page transition is replaced by a first order small/large black hole transition that terminates at a critical point, exhibiting van der Waals type behaviour that is typically seen only for charged solutions. In the five dimensional case, we see the usual Hawking-Page transition with only slight modifications, e.g. the transition temperature is larger for the cubic theory than in Einstein gravity.

Qualitatively, a similar behaviour is observed in the grand canonical (fixed potential) ensemble. In four dimensions for the cubic theory, black holes are thermodynamically preferred for all values of the electric potential and we observe a small/large black hole phase transition provided that $\Phi^2  < 4$. In four-dimensional Einstein gravity, we see a Hawking-Page type transition provided that $\Phi^2 < 4$ while black holes are thermally preferred for large values of the potential, but exhibit no further phase structure. In the five dimensional version of the cubic theory, the phase structure qualitatively the same as in five-dimensional Einstein gravity: at small values of the potential there is a Hawking-Page type transition, while at larger values of the potential black holes are always preferred. The difference is that, in five dimensions Hawking-Page transitions persist to larger values of $\Phi$ in the cubic theory than in Einstein gravity.

Most of our study of the thermodynamics has focused on the canonical (fixed charge) ensemble. Here we find that in four and five dimensions, physical critical points exist only for the five-dimensional solutions. In these cases, we find a variety of interesting phase structure, including Van der Waals type behaviour. Six dimensions is somewhat special in this ensemble, as there is the possibility for two physical critical points. By tuning the charge and coupling it is possible to merge these generically distinct critical points. At the point where they merge, we find that the critical exponents change resulting in what is known as an isolated critical point.

We have taken a first step toward holographic studies of these theories by studying the ratio of shear viscosity to entropy density $\eta/s$ in all dimensions. Interestingly, we find that in all dimensions the KSS bound is upheld in these theories, subject only to the constraint that the solutions possess well-behaved asymptotics. This extends the observation made in~\cite{Bueno:2018xqc} to all dimensions.

Finally, we close with a brief discussion of questions and issues raised by our study.

{\bf Negative entropy}. From the perspective of statistical mechanics, negative entropy would seem to make little sense. However, it has long been known that the Wald entropy of black holes in higher curvature theories can --- and often is --- negative. Ambiguities in the definition of the Wald entropy permit one to shift it by an arbitrary constant by, for example, adding a total derivative to the action. These ambiguities are present even in the context of four-dimensional Einstein gravity, since there one could add a Gauss-Bonnet contribution to the action, which would shift the black hole entropy by a constant proportional to the coupling while having no effect on the solutions themselves. While this ambiguity has no serious implications in the context of the canonical ensemble (since there one compares different branches of black hole solutions amongst themselves), it can have serious implications for thermodynamics of uncharged black holes and thermodynamics in the fixed potential ensemble (since there one compares the free energy of the black holes to the free energy of the vacuum). Since the entropy of the vacuum is unaffected by adding such a total derivative term, it is possible to, for example, completely eliminate the Hawking-Page transition or drastically alter the temperature at which it occurs through the addition of such a constant to the black hole entropy. It would seem that the most natural way to deal with this issue is by choosing the constant in the entropy so that $S \to 0$ when the spacetime does not contain a horizon. This would avoid any subtleties to do with the order of limits that would arise for arbitrary shifts in the entropy.  However, this choice may not always be possible (for example, if multiple branches of black holes exist) and is certainly not the only option. Quite frequently the negative entropy solutions are thrown out as unphysical, and in the context of Gauss-Bonnet gravity it has been argued that some of these may be unstable~\cite{Cvetic:2001bk}. While beyond the scope of this work, it seems clear that the role of negative entropy and its connection with any instabilities/pathologies requires further investigation.

{\bf Holography}. In this work, we have limited our holographic analysis to the study of holographic hydrodynamics.  This already shows that some of the interesting behaviour observed in four-dimensions extends to all dimensions. As this class of theories provides sensible holographic toy models in all dimensions four and larger, it would be beneficial to further extend the holographic dictionary for them. For example, computing the parameters characterizing the stress tensor three-point function would allow one to constrain the range of couplings for which the theory could describe sensible CFT duals, and would help establish further evidence for the conjectured relationship between derivatives of the embedding function $h(f_\infty)$ and the stress tensor correlators made in~\cite{Bueno:2018yzo}. It would also be interesting to extend holographic considerations to higher dimensions and higher-orders in curvature. Since the generalized quasi-topological theories are non-trivial in dimensions where both Lovelock and quasi-topological gravities are trivial, these models can help fill the gaps, providing toy models allowing for calculations non-perturbative in the higher-order couplings in all dimensions.

{\bf Generalized quasi-topological theories}. Our work also suggests a number of future directions concerning the generalized quasi-topological theories themselves. One avenue would involve considering how the properties of black hole solutions in higher-dimensions are affected by additional curvature terms. In four dimensions, it has been observed in the asymptotically flat case that properties of black holes in the cubic theory persist for an infinite family of essentially unique theories~\cite{Bueno:2017qce, GeometricInflation}. It is natural to wonder if this holds also in higher dimensions. For example, one may wonder if the non-hairy properties of the black holes persist in higher dimensions if multiple generalized quasi-topological terms --- like those constructed in~\cite{Ahmed:2017jod} --- are included in the action. One noteworthy observation is that in six and higher dimensions, there appear to be three families of solutions based on an analysis of the metric near $r = 0$, while it seems that only one of these represents a black hole with regular horizon.  It would be interesting to see what the full geometry of the remaining solutions represents. Additionally, in this work we have observed that there are qualitative differences between the behaviour of the black hole solutions in four dimensions and in all higher dimensions. It would be interesting to investigate if this is a feature of all such theories in higher dimensions, or if it is a peculiar property of the cubic representative we have focused on in this work.

On the front of thermodynamics, it is clear that black holes in cubic GQG have a richer thermodynamic structure than do their counterparts in Einstein gravity. Explorations beyond this -- into higher curvature GQG theories, or black holes with more features (rotation, non-linear electromagnetic couplings, etc.) remain to be undertaken.

\section*{Acknowledgements}

We thank Pablo Cano and Hugo Marrochio for helpful discussions. M. M. appreciates the hospitality of the University of Waterloo where this work was completed. The work of R.~A.~H.~is supported by the Natural Sciences and Engineering Research Council of Canada through the Banting Postdoctoral Fellowship programme. The work of R.~B.~M.~was supported in part by the Natural Sciences and Engineering Research Council of Canada. The work of Jamil Ahmed is supported by the Higher Education Commission of Pakistan under its Project No. 20-2087 which is gratefully acknowledged.

\appendix

\section{The Gauss-Bonnet term and black hole entropy}\label{sec:GBent}

Here, in an aim for completeness, we discuss the effect of the Gauss-Bonnet term on the entropy of four-dimensional black holes. Similar conclusions would follow for the higher-order Euler densities in higher even dimensions. We consider a metric of the form
\beq
\label{eqn:4dsphere}
ds^2 = - f(r) dt^2 + \frac{dr^2}{f(r)} + r^2 d \Omega^2_2 \, .
\eeq
For the sake of example, we will take the gravitational action to be the Einstein-Hilbert term along with the Gauss-Bonnet density:
\beq
{\cal I} = \frac{1}{16 \pi G} \int d^4 x \sqrt{-g} \left[ -2 \Lambda + R + \alpha \left(R_{abcd}R^{abcd} - 4 R_{ab}R^{ab} + R^2 \right) \right] \, .
\eeq
Of course, in four-dimensions the Gauss-Bonnet term is a total divergence and so makes no contribution to the field equations. However, we will see that the term does indeed make a contribution to the entropy of black holes.

The derivative of the Lagrangian density with respect to the Riemann tensor gives
\beq
P^{ab}_{cd} := \frac{\partial {\cal L}}{\partial R_{ab}^{cd}} = \frac{1}{16 \pi G} \left[\frac{1}{2} \left(\delta^a_c \delta_d^b - \delta_d^a \delta_c^b \right) + 2 \alpha \left(R_{cd}^{ab} + G_c ^b \delta^a_d - G_c^a \delta^b_d + R^a_d \delta^b_c - R^b_d \delta^a_c \right) \right] \, .
\eeq
The Wald entropy is given by
\beq
S_{\rm Wald} = -2 \pi \int d^2 x \sqrt{\gamma} P_{ab}^{cd} \hat{\epsilon}_{cd}\hat{\epsilon}^{ab} \, ,
\eeq
where the integration is carried out over a $(t,r)$ constant hypersurface. Strictly speaking, this should be taken at the horizon $r = r_+$, but for the moment let us just perform the computation without that assumption. The result of the computation is
\beq
S_{\rm Wald} = \frac{\pi r^2}{G} - \frac{4 \pi \alpha \left( f(r) - 1 \right)}{G} \, .
\eeq
Clearly, when we take $r = r_+$ where $f(r_+) = 0$, we obtain the usual Bekenstein-Hawking entropy plus the contribution $4 \pi \alpha / G$. The Gauss-Bonnet contribution in the action shifts the entropy by a constant.  However, if we consider the metric~\eqref{eqn:4dsphere} to describe a maximally symmetric vacuum then it is clear that both the usual Bekenstein-Hawking part of the entropy and the Gauss-Bonnet contribution vanish as we push the $r = constant$ surface toward $r = 0$. In other words, adding an explicit Gauss-Bonnet contribution to the action shifts the entropy of a black hole, but leaves the entropy of the vacuum unaffected.

So does this mean that the black hole entropy is completely ambiguous? The fact that methods to shift the entropy do not affect the entropy of the vacuum provides a natural way to select the ``correct'' correction. Namely, it seems natural to (when possible) choose the correction $\alpha$ so that $S \to 0$ as $r_+ \to 0$, ensuring that there is no ambiguous limit when $r_+ \to 0$.

\section{Thermodynamics from the Euclidean action}
\label{sec:EucAct}

In this appendix, we compute the thermodynamic quantities for the black hole solutions to the cubic theory using the Euclidean action approach. We use the method described in~\cite{Bueno:2018xqc}, which is much simpler than the usual approach, provided that the solutions are asymptotically maximally symmetric. According to this, the usual Gibbons-Hawking-York boundary term of Einstein gravity \cite{York:1972sj,Gibbons:1976ue}, along with the counterterms that ensure a finite on-shell action for Einstein gravity, appear modified through an overall factor proportional to $a^*$.
\begin{equation}\label{assd}
{\cal I}_E=-\int_{\mathcal{M}} d^d x \sqrt{g}\mathcal{L}(g^{ef},R_{abcd})-\frac{2a^*}{\Omega_{(d-2)}\tilde{L}^{d-2}}\int_{\partial \mathcal{M}}d^{d-1} x \sqrt{h}\, \Big[ K+ \text{counterterms} \Big]\, ,
\end{equation}
where $\Omega_{d-2}\equiv 2\pi^{(d-1)/2}/\Gamma((d-1)/2)$ is the area of the unit sphere $\mathbb{S}^{d-2}$, $\tilde{L}$ is the AdS radius, and $a^*$ is the charge appearing in the universal contribution to the entanglement entropy across a spherical entangling surface $\mathbb{S}^{d-3}$ in the dual CFT. This quantity is related, for any higher-curvature theory of gravity, to the on-shell Lagrangian of the theory on pure AdS through \cite{Imbimbo:1999bj,Schwimmer:2008yh,Myers:2010tj,Myers:2010xs,Bueno:2018xqc}
\begin{equation}
a^*=-\frac{\Omega_{d-2} \tilde{L}^{d}}{2(d-1)}\left. \mathcal{L}\right|_{\text{AdS}}\,.
\end{equation}
The explicit counterterms that ensure finite on-shell action in Einstein gravity depend on the spacetime dimension. The first few read
\begin{align}
{\rm counterterms} =&\, -\frac{(d-2)}{\tilde L}-\frac{\tilde L}{2(d-3)}\mathcal{R}
	\nn\\
	&-\frac{\tilde L^3\theta(d-6)}{2(d-3)^2(d-5)}\left(\mathcal{R}_{ab}\mathcal{R}^{ab}-\frac{d-1}{4(d-2)}\mathcal{R}^2\right)+\ldots\, ,
\end{align}
where $\theta(x)$ is the Heaviside step function, and the dots indicate additional counterterms that would be required when $d > 7$. In these expressions, ${\cal R}_{abcd}$ and its contractions denote the intrinsic curvature of the boundary. In the present case, we will also need to add a boundary term for the Maxwell part of the action when working in the canonical ensemble. This boundary term reads
\beq
\label{eqn:MxWBdry}
{\cal I}_{\partial {\cal M}}^{\rm EM} = - \frac{1}{16 \pi G} \int d^{d-1} x \sqrt{h} F^{\mu\nu}n_\mu A_\nu \, ,
\eeq
and ensures that the electric charge is fixed on the boundary. Let us now discuss the computation in more detail.

The gravitational part of the Lagrangian  is given by
\begin{align}
{\cal L} =& \frac{1}{16\pi G} \left[ \frac{(d-1)(d-2)}{L^2}+R+\frac{12(2d-1)(d-2) \mu \cS_{3,d}}{(d-3)(4d^4 - 49 d^3 + 291 d^2 - 514 d + 184)} \right] \, .
\end{align}
Evaluating this on an AdS space with curvature scale $\tilde{L} = L/\sqrt{f_\infty}$ we find the following result for $a^*$:
\beq
a^* = \frac{ \Omega_{d-2} L^{d-2} }{32 \pi G f_\infty^{d/2}}  \left[2 - d(1-f_\infty) - d(d-2) f_\infty^3 \frac{\mu}{L^4} \right] \, .
\eeq

To evaluate the Lagrangian for the static and spherically symmetric ansatz of interest, it is helpful to note that the Riemann tensor can be written in the following form:
\beq
R_{ab}^{cd} = 2 \left[2 f'' \tau^{[c}_{[a} \rho^{d]}_{b]} + \frac{2 f'}{r} \left(\tau^{[c}_{[a} \sigma^{d]}_{b]} + \rho^{[c}_{[a} \sigma^{d]}_{b]}  \right)   + \frac{(k-f)}{r^2} \sigma^{[c}_{[a} \sigma^{d]}_{b]}  \right] \, ,
\eeq
where $\tau$, $\rho$ and $\sigma$ are projection tensors satisfying the following relations~\cite{Deser:2005pc}:
\beq
\tau_a^b\tau_b^c = \tau_a^c \, , \quad \rho_a^b\rho_b^c = \rho_a^c \, , \quad \sigma_a^b\sigma_b^c = \sigma_a^c \, , \quad \tau_a^b\rho_b^c = \tau_a^c \sigma_b^c = \rho_a^c \sigma_b^c = 0 \, ,
\eeq
with traces ${\rm tr} \, \tau = 1$, ${\rm tr} \, \rho  = 1$ and ${\rm tr} \, \sigma = d-2$.

Evaluating the bulk part of the gravitational action, we find that on-shell it is a total derivative. This allows us to express it in the following (relatively) simple way:
\beq
{\cal I}^{\rm grav}_{\cal M} = - \frac{\Sigma_{d-2, k} \beta}{16 \pi G} H(r) \bigg|_{r_+}^{R_0}
\eeq
where $R_0$ is a large-$r$ cutoff and
\begin{align}
H(r) =& \, (d-2) \left((k-f) + \frac{r^2}{L^2} \right) r^{d-3} - f' r^{d-2}  - \frac{4 (d-2)\mu r^{d-7}}{4d^4 - 49 d^3 + 291 d^2 - 514 d + 184}
	\nn\\
	&\times \bigg[(d-4)\left(d^4 - \frac{57}{4} d^3 + \frac{357}{4} d^2 - 192 d + 129\right)(k-f)^3
	\nn\\
	&+ 2 r^3(d^2 + 5d - 15) f'^3 + 12 r^2 (d^2 + 5 d - 15)(k-f) f'^2
	\nn\\
	& - 3 r (d-4) \left(d^3 - \frac{33}{4} d^2 + \frac{127}{4} d - \frac{83}{2} \right) (k-f)^2 f' \bigg] \, .
\end{align}
To evaluate the generalized boundary and counterterms, we note that the trace of the extrinsic curvature reads
\beq
K = (d-2) \frac{\sqrt{f}}{r} + \frac{f'}{2 \sqrt{f}}\, ,
\eeq
while the intrinsic Riemann tensor of the boundary reads
\beq
{\cal R}_{ab}{}^{cd} = \frac{2 k }{r^2} \sigma_{[a}^{[c} \sigma_{b]}^{d]} \, .
\eeq
Using these results, the boundary and counterterms valid up to $d = 7$ take the following explicit form:
\begin{align}
\int_{\partial \mathcal{M}} &d^{d-1} x \sqrt{h}\, \Big[ K+ \text{counterterms} \Big] = \Sigma_{d-2, k} \beta r^{d-2} \bigg[ \frac{(d-2) f}{r} + \frac{f'}{2}
	\nn\\
	&+ (d-2) \sqrt{f} \bigg(-\frac{ \sqrt{f_\infty}}{L} - \frac{  k L  }{2\sqrt{f_\infty} r^2 }  + \frac{L^3 k^2  \theta(d-6)}{8 f_\infty^{3/2} r^4}  \bigg) + \cdots \bigg]
\end{align}
where the expression is to be evaluated at $r = R_0$ and the dots indicate additional terms that would be present beyond $d = 7$.

To evaluate both the bulk and boundary/counterterms at $R_0$, we make use of the asymptotic expansion of the metric function, which reads:
\beq
f = k + \frac{f_\infty r^2}{L^2} + \frac{m}{h'(f_\infty) r^{d-3}} - \frac{q^2}{h'(f_\infty) r^{2d-6}} + \cdots \, .
\eeq
Plugging this into the expression for $H(r)$, we obtain the following contribution:
\begin{align}
H(R_0) =& -\frac{R_0^{d-1}}{L^2} \left(2 - d(1-f_\infty) - d(d-2) f_\infty^3 \frac{\mu}{L^4} \right)
	\nn\\
	&- \frac{m}{h'(f_\infty) } \left(1 - \frac{3 (d-2) \mu f_\infty^2}{L^4} \right) + {\cal O}(R_0^{1-d})
\end{align}
By plugging the same asymptotic expansion into the boundary/counterterm part of the action for the cases $d = 4,5,6,7$ it is possible to extract the general pattern:
\begin{align}
\int_{\partial \mathcal{M}} d^{d-1} x \sqrt{h}\, \Big[ K+& \, \text{counterterms} \Big] = \beta \Sigma_{d-2, k} \bigg[ \frac{f_\infty R_0^{d-1}}{L^2} + \frac{m}{2 h'(f_\infty)}
	\nn\\
	&- \frac{(d-2)!!^2}{(d-1)!} \frac{(-k)^{(d-1)/2} L^{d-3}}{f_\infty^{(d-3)/2}} + \cdots \bigg]
\end{align}
where the dots represent terms that vanish in the limit $R_0 \to \infty$. The last term above is present only when $d$ is odd, and is related to the Casimir energy.

Adding together the bulk and boundary/counterterm contributions at $R_0$ we can immediately see that the leading divergence cancels. The constant term is slightly more subtle, but combining terms leads to the nice form:
\begin{align}
{\cal I}^{{\rm grav}, R_0}_{\rm E} =  \frac{\beta \Sigma_{d-2, k}}{16 \pi G} \bigg[&\frac{(d-2) m}{f_\infty h'(f_\infty)}  \left(1 - f_\infty + \frac{(d-6) \mu f_\infty^3}{L^4} \right)
	\nn\\
	&+ \frac{32 \pi G (d-2)!!^2 a^* }{ \Omega_{d-2} (d-1)!  }    \frac{(-k)^{(d-1)/2} \sqrt{f_\infty} }{L} \bigg] \, .
\end{align}
We can now recognize that the term multiplying the mass parameter is nothing other than the embedding function $h(f_\infty)$, which must vanish. Thus, we have the following final result in the limit $R_0 \to \infty$:
\begin{align}
{\cal I}^{{\rm grav}, R_0}_{\rm E} =  - \frac{2 (d-2)!!^2 \beta \Sigma_{d-2, k} a^*}{ \Omega_{d-2} (d-1)! }  \frac{(-k)^{(d-1)/2} \sqrt{f_\infty} }{L} \, ,
\end{align}
and we once again emphasis that this term is present only for odd $d$.

The bulk action at the horizon can be evaluated by expanding $f(r)$ as a near horizon power series. The result for $H(r_+)$ is
\begin{align}
H(r_+) =& \,  (d-2) \left(k + \frac{r_+^2}{L^2} \right) r_+^{d-3} - 4 \pi T r_+^{d-2} - \frac{(d-2) \mu r_+^{d-7}}{4 d^4-49 d^3+291 d^2-514 d+184}
	\nn\\
	&\times\bigg[512 ( d^2 + 5d  -15) \pi^3 T^3 r_+^3  + 768 (d^2 + 5d - 15 ) k \pi^2 T^2  r_+^2
		\nn\\
		&- 12 (d-4)(4d^3 - 33 d^2 + 127 d - 166) k^2 \pi T r_+
		\nn\\
		&+ (d-4)(4d^4 - 57 d^3 + 357 d^2 - 768 d + 516) k^3 \bigg] \, .
\end{align}
We can then write the full Euclidean gravitational action for these solutions as,
\begin{align}
{\cal I}^{\rm grav}_{\rm E} =&\, \frac{\Sigma_{d-2, k} \beta}{16 \pi G} \bigg\{
(d-2) \left(k + \frac{r_+^2}{L^2} \right) r_+^{d-3} - 4 \pi T r_+^{d-2} - \frac{(d-2) \mu r_+^{d-7}}{4 d^4-49 d^3+291 d^2-514 d+184}
	\nn\\
	&\times\bigg[512 ( d^2 + 5d  -15) \pi^3 T^3 r_+^3  + 768 (d^2 + 5d - 15 ) k \pi^2 T^2  r_+^2
		\nn\\
		&- 12 (d-4)(4d^3 - 33 d^2 + 127 d - 166) k^2 \pi T r_+
		\nn\\
		&+ (d-4)(4d^4 - 57 d^3 + 357 d^2 - 768 d + 516) k^3 \bigg]
		\nn\\
		&+  \frac{32 \pi G (d-2)!!^2  a^*}{ \Omega_{d-2} (d-1)! }  \frac{(-k)^{(d-1)/2} \sqrt{f_\infty} }{L}
 \bigg\} \, .
\end{align}

All that remains now is to take the Maxwell field into account. The bulk part of the Maxwell action is easily evaluated and gives the following result:
\beq
{\cal I}_{\cal M}^{\rm EM} = - \frac{1}{16 \pi G} \int_{\cal M} d^dx \sqrt{-g} \frac{F_{\mu\nu}F^{\mu\nu}}{4}  = - \frac{(d-2)\beta \Sigma_{d-2, k} q^2}{16 \pi G r_+^{d-3}} \, .
\eeq
When working in the fixed charge ensemble, we must also add to the action the boundary term given in Eq.~\eqref{eqn:MxWBdry}. This leads to the following contribution:
\beq
{\cal I}_{\partial {\cal M}}^{\rm EM} = - \frac{1}{16 \pi G} \int d^{d-1} x \sqrt{h} F^{\mu\nu}n_\mu A_\nu =  \frac{(d-2) \beta \Sigma_{d-2, k} q^2}{8 \pi G r_+^{d-3}} \, .
\eeq
This result can be obtained in the following way. The form of the vector potential quoted in Eq.~\eqref{eqn:vecPot} is singular on the horizon when written in an orthonormal frame, but can be brought into a non-singular form via a gauge transformation $A_\mu \to A_\mu - A_\mu\big|_{r_+}$. Using this regular potential, the boundary term evaluated at the surface $R_0$, followed by taking the limit $R_0 \to \infty$ gives the result quoted above.

Having all of the ingredients at hand, we can express the full Euclidean action as
\begin{align}
{\cal I}_{\rm E} =&\, \frac{\Sigma_{d-2, k} \beta}{16 \pi G} \bigg\{
(d-2) \left(k + \frac{r_+^2}{L^2} \right) r_+^{d-3} - 4 \pi T r_+^{d-2} - \frac{(d-2) \mu r_+^{d-7}}{4 d^4-49 d^3+291 d^2-514 d+184}
	\nn\\
	&\times\bigg[512 ( d^2 + 5d  -15) \pi^3 T^3 r_+^3  + 768 (d^2 + 5d - 15 ) k \pi^2 T^2  r_+^2
		\nn\\
		&- 12 (d-4)(4d^3 - 33 d^2 + 127 d - 166) k^2 \pi T r_+
		\nn\\
		&+ (d-4)(4d^4 - 57 d^3 + 357 d^2 - 768 d + 516) k^3 \bigg]
		\nn\\
		& +  \frac{32 \pi G (d-2)!!^2  a^*}{ \Omega_{d-2} (d-1)! }  \frac{(-k)^{(d-1)/2} \sqrt{f_\infty} }{L}
  -  (1 +2 \alpha) \frac{(d-2) q^2}{r_+^{d-3}}  \bigg\} \, ,
\end{align}
where $\alpha = 0$ for the fixed potential ensemble and $\alpha = 1$ for the fixed charge ensemble.

The standard statistical mechanical argument relates the free energy to the Euclidean action
\beq
F = -T \log Z = T {\cal I}_{\rm E} .
\eeq
When working in the fixed potential ensemble the free energy would be identified as $F = M' - TS' - \Phi Q'$, while in the fixed charge ensemble it is just $F = M' - TS'$. Here we use primes to allow for the possibility that the mass, entropy and charge defined via the Euclidean action could differ from those calculated in Section~\ref{sec:thermo}.  These identifications allow one to compute the energy, entropy and charge from the Euclidean action. For example, in the fixed potential ensemble the identities would read
\beq
M' = \left(\partial_\beta {\cal I}_{\rm E} \right)_\Phi - \frac{\Phi}{\beta} \left(\partial_\Phi {\cal I}_{\rm E} \right)_\beta\, , \quad S' = \beta \left( \partial_\beta {\cal I}_{\rm E} \right)_\Phi - {\cal I}_{\rm E} \, , \quad Q' = - \frac{1}{\beta} \left( \partial_\Phi {\cal I}_{\rm E} \right)_\beta \, .
\eeq
A somewhat tedious but straight-forward computation making use of the near horizon equations of motion~\eqref{nearsol} reveals that the thermodynamic parameters defined by the Euclidean action match those defined in Section~\ref{sec:thermo}. The only subtlety arises for the mass, which reads
\beq
M' = M +  \frac{2 \Sigma_{d-2, k} (d-2)!!^2 (-k)^{(d-1)/2}  }{ \Omega_{d-2} (d-1)! }  \frac{ a^* }{\tilde L}
\eeq
where the second contribution is present only in odd $d$. This is, of course, just the Casimir energy associated with AdS, and if we take AdS to be the zero of action and energy (as done in the bulk of this paper), this contribution is just subtracted.\footnote{It is a bit intriguing that $a^*$ appears in the Casimir energy in the cubic theory (recall $a^*$ is the charge appearing in the universal contribution to the entanglement entropy across a spherical entangling surface $\mathbb{S}^{d-3}$ in the dual CFT.) Since this expression does not make explicit reference to the gravitational theory under consideration, one may expect that it holds in general for Einstein-like higher-order gravities. } The agreement between the results in Section~\ref{sec:thermo} and those presented here provides a non-trivial check of the thermodynamic quantities presented here, and illustrates the utility of the method for evaluating the Euclidean action presented in~\cite{Bueno:2018xqc}.

\section{Using Pad\'e approximants to determine the shooting parameter} \label{sec:pade}

It was discussed in the appendix of~\cite{Hennigar:2018hza} that it is possible to derive an useful analytic approximation for the shooting parameter $a_2$ that appears in the near horizon expansion. In that work, the focus was asymptotically flat black holes, but the technique works as well for the AdS case. Here we will discuss the method in the context of black branes, which provides useful insight into the results of Section~\ref{sec:holo_hydro}.  We write $\lambda = - \mu$ just for convenience.

Near the horizon we have,
\beq
f(r) = 4 \pi T (r-r_+) + a_2 (r-r_+)^2 + \sum_{i=3}^{\infty} a_i (r-r_+)^3
\eeq
Recall that the field equations fix $a_i$ for $i \ge 3$ in terms of $a_2$ and the other parameters of the black hole, but $a_2 = f'(r_+)/2$ is left undetermined by the field equations. The parameter can be determined numerically by demanding that the numerical solution joins smoothly onto the asymptotic solution at large $r$. An alternate method was outlined in~\cite{Hennigar:2018hza}. The idea is to write,
\beq
a_2 = g(\lambda)
\eeq
and then demand that $a_i$ for $i\ge 2$ joins smoothly onto the Einstein solution as $\lambda \to 0$.  This fixes the derivatives of $g(\lambda)$ by demanding that no terms like $\lambda^{-n}$ appear in these expansions. As an example, the four dimensional case, we have the first few terms,
\begin{align}
a_3  &= \frac{C_1 L^4 g(0) }{\lambda r_+} + \frac{L^6 g'(0) - 6048 L^4 g(0)^2 + 36288 L^2 g(0) - 54432}{27216 r_+ L^2} + \cdots \, ,
	\nn\\
a_4 &= \frac{C_1 L^8 g(0)}{r_+^2 \lambda^2} + \frac{C_2 L^2  \left(L^6 g'(0) - 48384 L^4 g(0)^2 + 172368 L^2 g(0) - 81648 \right)}{r_+^2 \lambda} + {\rm finite} \, ,
\end{align}
where $C_i$ are large constants that are irrelevant for the present discussion. Clearly, demanding a finite limit for $a_3$ fixes $g(0) = 0$. This then cures the $1/\lambda^2$ divergence in $a_4$ and we must select $g'(0) = 81648/L^6$ to cure the $1/\lambda$ divergence in $a_4$. Note also that removing the $1/\lambda$ pole in $a_4$ then ensures that the finite part of $a_3$ matches the Einstein value of $a_3 = 1/(r_+ L^2) = f^{(3)}(r_+)/6$. This pattern continues to arbitrary high order: $g^{(n)}(0)$ is determined by ensuring that $a_{n+3}$ has a smooth $\lambda \to 0$ limit, and this choice of $g^{(n)}(0)$ also ensures that $a_{n+2}$ limits to the value from Einstein gravity as $\lambda \to 0$.

Carrying out this procedure, in general it is found that that coefficients of the derivatives grow without bound. Thus, an ordinary Taylor series is not a good approximation since it has a vanishing radius of convergence. That is, $a_2$ is not a real analytic function of the coupling. However, a very good result can be obtained by matching the $g^{(n)}(0)$ terms to a Pad\'e approximant. The reason for the diverging  coefficients of the Taylor series is the existence of a pole at negative $\lambda$ (positive $\mu$). This is a consequence of the fact that the derivatives $g^{(n)}(0)$ implicitly contain derivatives of the temperature (treated as a function of the coupling), and while the temperature has a closed form, it is not a real analytic function.

\begin{figure}[htp]
\centering
\includegraphics[width=0.45\textwidth]{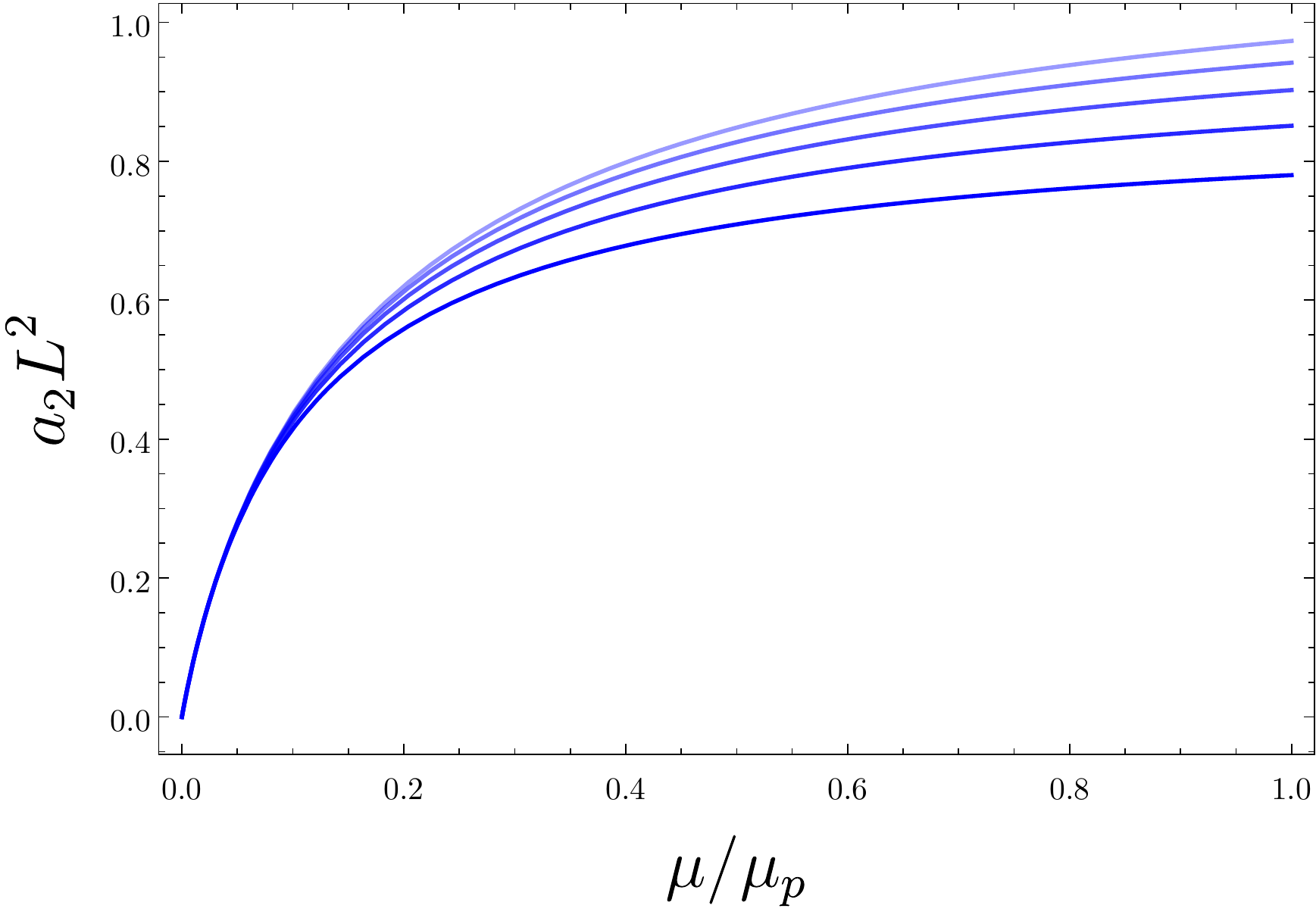}
\quad
\includegraphics[width=0.45\textwidth]{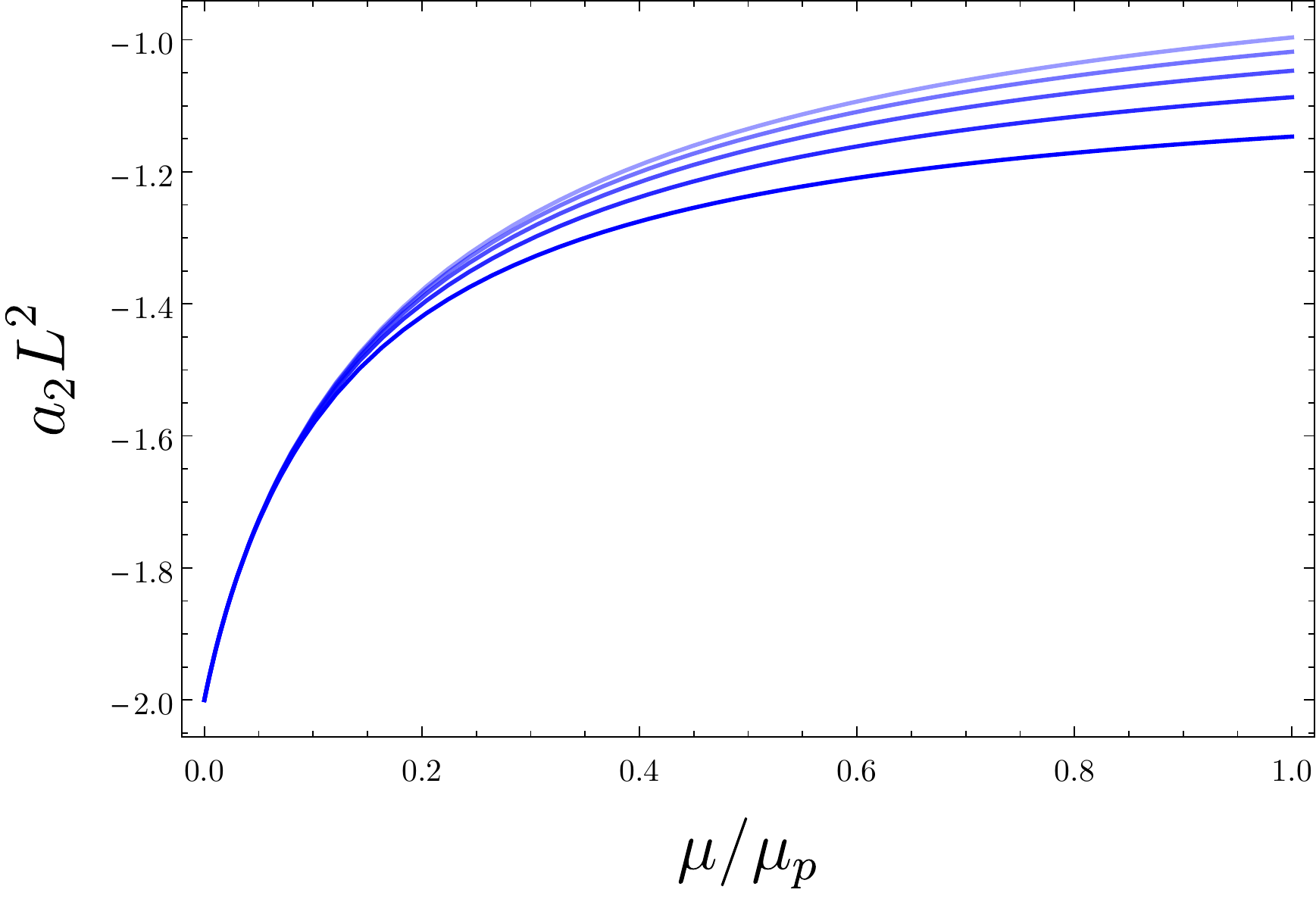}
\caption{{\bf Shooting parameter vs. coupling}: $a_2$ for black branes in four (left) and five (right) dimensions. In all cases, the curves illustrate $[3/3]$ to $[7/7]$ order Pad\'e approximants (more to less opacity, respectively). }
\label{fig:shooting_parameters}
\end{figure}

The expressions for the Pad\'e approximants are quite complicated at high order. However, as can be seen in Figure~\ref{fig:shooting_parameters}, for small coupling even a low order Pad\'e approximant gives consistent results. Here we list $[2|2]$ Pad\'e approximants in a few sample dimensions. In each case, $x = \mu/\mu_p$,
\begin{align}
a_2^{d = 4} L^2 &= \frac{3 x  (167636 x +2463)}{\left(687616 x ^2+70383 x +821\right) } \, ,
	\nn\\
a_2^{d = 5} L^2 &=	-\frac{2 \left(349748159973199286272 x ^2+72904557105141027840 x +1236978094606448985\right)}{105 \left(5731035367572926464 x ^2+742938823413991872 x +11780743758156657\right) } \, ,
	\nn\\
a_2^{d = 6} L^2  &= 	-\frac{10 \left(15347758658125 x ^2+2397935458800 x +44287344864\right)}{3 \left(12293400435625x ^2+1646464906800 x +29524896576\right)  } \, .
\end{align}

\bibliography{LBIB}
\bibliographystyle{JHEP}

\end{document}